\documentclass[notitlepage,12pt]{article}
\usepackage{amsfonts}
\usepackage{amsmath}
\usepackage{amssymb}
\usepackage{tikz}
\usepackage{makeidx}
\usepackage{graphicx}
\usepackage{subcaption}
\usepackage{natbib}
\usepackage[margin=1in]{geometry}
\usepackage{color}
\usepackage{url}
\usepackage[pdftex, 	colorlinks=true, linkcolor=blue, citecolor=blue, urlcolor=blue]{hyperref}
\usepackage{titlesec}
\usepackage{setspace}

\titleformat{\paragraph}
{\normalfont\normalsize\bfseries}{\theparagraph}{1em}{}
\titlespacing*{\paragraph}
{0pt}{3.25ex plus 1ex minus .2ex}{1.5ex plus .2ex}

\newtheorem{theorem}{Theorem}

\newtheorem{claim}{Claim}

\newtheorem{definition}{Definition}
\newtheorem{example}{Example}

\newtheorem{lemma}{Lemma}

\newtheorem{proposition}{Proposition}

\newtheorem{comment}{Remark}

\newenvironment{proof}[1][Proof]{\noindent \textbf{#1.} }{\  \rule{0.5em}{0.5em}}
\input{tcilatex}

\begin{document}

\title{\textbf{Maskin Meets Abreu and Matsushima}\thanks{%
We owe special thanks to a coeditor and four anonymous referees for their
insightful comments as well as to Kim-Sau Chung, Eddie Dekel, and Phil Reny
for long discussions which led to major improvements of the paper. We also
thank Soumen Banerjee, Olivier Bochet, Matthew Jackson, Hitoshi Matsushima,
Stephen Morris, Daisuke Nakajima, Hamid Sabourian, Roberto Serrano, Bal\'{a}%
zs Szentes, Xiangqian Yang, and participants at various seminars and
conference presentations for helpful comments. Part of this paper was
written while the four authors were visiting Academia Sinica, and we would
like to thank the institution for its hospitality and support. }}
\author{Yi-Chun Chen\thanks{%
Department of Economics and Risk Management Institute, National University
of Singapore. Email: ecsycc@nus.edu.sg} \and Takashi Kunimoto\thanks{%
School of Economics, Singapore Management University. Email:
tkunimoto@smu.edu.sg} \and Yifei Sun\thanks{%
School of International Trade and Economics, University of International
Business and Economics. Email: sunyifei@uibe.edu.cn} \and Siyang Xiong%
\thanks{
Department of Economics, University of California, Riverside. Email:
siyang.xiong@ucr.edu}}
\maketitle

\begin{abstract}
\begin{spacing}{1.2}
The theory of full implementation has been criticized for using
integer/modulo games which admit no
equilibrium (\cite{jackson92}). To address the critique, we revisit the
classical Nash implementation problem due to \cite{maskin99} but allow for
the use of lotteries and monetary transfers as in Abreu and Matsushima (1992,
1994). We unify the two well-established but somewhat orthogonal approaches
in full implementation theory. We show that Maskin monotonicity is a
necessary and sufficient condition for (exact) mixed-strategy Nash
implementation by a finite mechanism. In contrast to previous papers, our
approach possesses the following features: finite mechanisms (with
no integer or modulo game) are used; mixed strategies are handled
explicitly; neither undesirable outcomes nor transfers occur in equilibrium;
the size of transfers can be made arbitrarily small; and our mechanism is
robust to information perturbations. Finally, our result can be extended to
infinite/continuous settings and ordinal settings.
\end{spacing}
\end{abstract}

\baselineskip=20pt\textit{\ }

\section{Introduction}

Implementation theory can be seen as reverse engineering game theory.\
Suppose that a society has decided on a social choice rule -- a recipe for
choosing the socially-optimal alternatives alternatives on the basis of
individuals' preferences over alternatives. The individuals' preferences
vary across states and the realized state is common knowledge among the
agents but unknown to a social planner/mechanism designer. To (fully)
implement the social choice rule, the designer chooses a mechanism so that
at each state, the equilibrium outcomes of the mechanism coincide with the
outcomes designated by the social choice rule.

We study Nash implementation by a finite mechanism where agents report only
their preferences and preference profiles. We focus on the monotonicity
condition (hereafter, Maskin monotonicity) which Maskin shows is necessary
and \textquotedblleft almost sufficient" for Nash implementation. We aim to
implement \textit{social choice functions} (henceforth, SCFs) that are
Maskin-monotonic in mixed-strategy Nash equilibria without making use of the 
\textit{integer game }or the\textit{\ modulo game} which prevails in the
full implementation literature.

In the integer game, each agent announces some integer and the person who
announces the highest integer gets to name his favorite outcome. When the
agents' favorite outcomes differ, an integer game has no pure-strategy Nash
equilibria. The questionable feature is also shared by modulo games. The
modulo game is considered a finite version of the integer game in which
agents announce integers from a finite set. The agent who makes the integer
announcement matching the modulo of the sum of the integers gets to name the
allocation. In order to \textquotedblleft knock out" undesirable equilibria
in general environments, most constructive proofs in the literature,
following \cite{maskin99}, have either taken advantage of the fact that the
integer/modulo game has no solution in pure strategies and/or restricted
attention to pure-strategy Nash equilibria.

Instead of invoking integer/modulo games, we study Nash implementation in a
restricted domain where the designer can invoke both lotteries and
(off-the-equilibrium) transfers in designing the implementing mechanism. We
study a finite environment in which a finite mechanism is to be anticipated.%
\footnote{%
More precisely, the implementing mechanism which we construct is finite as
long as both the set of agents and the set of states are finite. In Section %
\ref{infinite}, we consider infinite environments in which we construct an
infinite yet \textquotedblleft well-behaved\textquotedblright\ implementing
mechanism to achieve the same goal.} Finite mechanisms are also bounded in
the sense of \cite{jackson92} and have no aforementioned questionable
features. Indeed, \citet[Example
4]{jackson92} shows that when no domain restriction on the environment is
imposed, some Maskin-monotonic SCF is not implementable in mixed-strategy
Nash equilibria by any finite mechanism. It raises the question as to
whether \emph{every} Maskin-monotonic SCF is mixed-strategy Nash
implementable with domain restrictions imposed by lotteries and transfers.%
\footnote{%
Another direction one can take is to characterize, without making any domain
restriction, the subclass of Maskin-monotonic SCFs which can be implemented
in mixed-strategy Nash equilibria in a finite mechanism. For that goal, our
exercise serves as a clarification of whether in certain environments, the
class of implementable SCFs is as permissive as it can be.}




Our main result (Theorem \ref{Nash}) shows that when the designer can make
use of lotteries and transfers off the equilibrium,\ Maskin monotonicity is
indeed a necessary and sufficient condition for mixed-strategy Nash
implementation by a finite mechanism. In our finite mechanism, each agent is
asked to report only his preference and a preference profile. That is, we
replace the integer announcement in Maskin's mechanism with an announcement
of each agent's own preference. The preference announcement plays the same
role as an integer in knocking out unwanted equilibria, albeit in a
different manner. Following the idea of \cite{AM94}, we design the mechanism
so that whenever an \textquotedblleft unwanted
equilibrium\textquotedblright\ occurs, the agents' reports must be
\textquotedblleft truthful,\textquotedblright\ namely, they announce their
own preference and preference profile as prescribed under the true state.
That in turn implies, through cross-checking the (truthful) preferences and
the preference profiles reported by the agents, that the unwanted
equilibrium could not have happened. In our finite mechanism, a
pure-strategy (truth-telling) equilibrium exists, and all mixed-strategy
equilibria achieve the desirable social outcome in each state.

We also provide several extensions of our main results. First, we show that
our implementation is robust to information perturbations. Second, we extend
Theorem \ref{Nash} to cover social choice correspondences (i.e.,
multi-valued social choice rules), studied in \cite{maskin99} as well as in
many subsequent papers. Formally, we show that when there are at least three
agents, every Maskin-monotonic social choice correspondence is
mixed-strategy Nash implementable (Theorem \ref{NashC}). Moreover, as long
as the social choice correspondence is finite-valued, our implementing
mechanism remains finite. Third, we show that if there are at least three
agents and the SCF satisfies Maskin monotonicity in the restricted domain
without any transfer, then it is implementable in mixed-strategy Nash
equilibria by a finite mechanism in which the size of transfers remains zero
on the equilibrium and can be made arbitrarily small off the equilibrium
(Theorem \ref{Thm:small-transfer}).Fourth, we consider an infinite setting
in which the state space is a compact set, and the utility functions and the
SCF are all continuous. In this setting, we show that Maskin monotonicity is
a necessary and sufficient condition for mixed-strategy Nash implementation
by a mechanism with a compact message space, a continuous outcome function,
and a continuous transfer rule (Theorem \ref{infiNash}). The extension
covers many applications and verifies that our finite settings approximates
settings with a continuum of states. To our knowledge, such an extension to
an infinite setting has not appeared in the literature. Moreover, a compact
and continuous mechanism allows each player to find a best response to every
(pure or mixed) strategy profiles of the other players. Indeed, in
discussing the notion of \textquotedblleft well-behaved\textquotedblright\
mechanisms, \citet[footnote 8]{AM92} also regard compactness and continuity
as \textquotedblleft plausible necessary desiderata\textquotedblright .

Finally, the extension to an infinite setting yields another interesting
extension. Specifically, in proving Theorem \ref{Nash}, we have assumed that
each agent is an expected utility maximizer with a fixed cardinal utility
function over pure alternatives. This raises the question as to whether our
result is an artifact of the fixed finite set of cardinalizations. To answer
the question, we study the concept of \textit{ordinal} Nash implementation
proposed by \cite{MR12}. The notion requires that the implementing mechanism
achieve mixed-strategy Nash implementation for \emph{every} cardinal
representation of preferences over lotteries. By making use of our
implementing mechanism in the infinite setting, we show that ordinal almost
monotonicity, as defined in \cite{sanver06}, is a necessary and sufficient
condition for ordinal Nash implementation (Theorem \ref{ord}). This
extension also verifies that our implementation result does not suffer from
the critique raised by \citet[p. 490]{JPS94} to the dependence of the result
of \cite{AM94} on a finite set of cardinalizations.

The rest of the paper is organized as follows. In Section \ref{RL}, we
position our paper in the literature. In Section \ref{pre}, we present the
basic setup and definitions. Section \ref{NE} proves our main result. We
discuss the extensions of our main result in Section \ref{extensions} and
conclude in Section \ref{conclusion}. The appendix contains all proofs,
which are omitted from the main text.

\section{Related Literature}

\label{RL}

\cite{maskin99} proposes the notion of Maskin monotonicity and implements a
Maskin-monotonic social choice correspondence by constructing an infinite
mechanism with integer games. While integer games are useful in achieving
positive results in general settings, the hope has been that for more
specific environments, more realistic mechanisms, or mechanisms without the
\textquotedblleft questionable features,\textquotedblright\ may suffice. The
research program has been proposed by \cite{jackson92}. One such class of
specific environments is the one with lotteries and transfers which our
paper, as well as the partial implementation literature, focuses on.

In environments with lotteries and transfers, Abreu and Matsushima (1992,
1994) obtain permissive full implementation results using finite mechanisms
without the aforementioned questionable features. Like our implementing
mechanisms, their mechanisms also make use of only payoff relevant messages,
such as preferences or preference profiles. However, Abreu and Matsushima
(1992, 1994) do not investigate Nash implementation but rather appeal to a
different notion of implementation: virtual implementation (in \cite{AM92})
or exact implementation under iterated weak dominance (in \cite{AM94}).%
\footnote{%
Iterated weak dominance in \cite{AM94} also yields the unique undominated
Nash equilibrium outcome. For undominated Nash implementation by
\textquotedblleft well-behaved\textquotedblright\ mechanisms, see also \cite%
{JPS94} and \cite{sjostrom94}.}

Virtual implementation means that the planner contents herself with
implementing the SCF with arbitrarily high probability. In contrast, by
studying \emph{exact} Nash implementation in the specific setting, we unify
the two well-established but somewhat orthogonal approaches to
implementation theory which are due to \cite{maskin99} and to Abreu and
Matsushima (1992, 1994). Our exercise is directly comparable to \cite%
{maskin99} and highlights the pivotal trade-off between domain restrictions
and the feature of implementing mechanisms. We consider it to be one step in
advancing the research program proposed by \cite{jackson92}.

An alternative approach to handling mixed-strategy equilibria is to resort
to refinements such as undominated Nash equilibria or subgame-perfect
equilibria. With such refinements, essentially every SCF, whether
Maskin-monotonic or not, is implementable in a complete-information
environment; see, for example, \cite{MR88} and \cite{AM94}. However,
according to \cite{CE03} and \cite{AFHKT}, if we were to achieve exact
implementation in these refinements which are also robust to a \emph{small
amount of incomplete information},\ then Maskin monotonicity would be
restored as a necessary condition. Those permissive implementation results,
which are driven by the lack of the closed-graph property of the
refinements, cast doubt on the success of taking care of
non-Maskin-monotonic SCFs by resorting to equilibrium refinements. In
contrast, our Theorem \ref{Nash} establishes exact and \emph{robust}
implementation in mixed-strategy Nash equilibria to the maximal extent that
every Maskin-monotonic SCF is implementable (Proposition \ref{prop:ne-bar}).%
\footnote{\cite{harsanyi73} shows that a mixed-strategy Nash equilibrium
outcome may occur as the limit of a sequence of pure-strategy Bayesian Nash
equilibria for \textquotedblleft nearby games" in which players are
uncertain about the exact profile of preferences. Hence, ignoring
mixed-strategy equilibria would be particularly problematic if we were to
achieve implementation which is robust to information perturbations.}

\cite{OP17} study a full implementation problem using transfers both on and
off the equilibrium. Specifically, Theorem 2 of \cite{OP17} provides a
sufficient condition restricting agents' beliefs via moment conditions under
which their notion of robust full implementation is possible in a direct
mechanism. Their notion of robustness is a \textquotedblleft global
notion\textquotedblright\ which accommodates arbitrary information
structures consistent with a fixed payoff environment.\footnote{%
See \cite{ollar2021network} for a further extension of the approach of \cite%
{OP17}.} In contrast, our paper follows the classical implementation
literature in dealing with the specific belief restriction of complete
information, and our notion of robustness (in Section \ref{perturb})
accommodates only perturbations around the complete-information benchmark.
With the specific belief restriction, we prove that Maskin monotonicity is
both necessary and sufficient for implementation in mixed-strategy Nash
equilibria in a finite, indirect mechanism with only off-the-equilibrium
transfers.

\section{Preliminaries}

\label{pre}

\subsection{Environment}

Consider a finite set of agents $\mathcal{I}=\{1,2,...,I\}$ with $I\geq 2$;
a finite set of possible states $\Theta $; and a set of pure alternatives $A$%
. We consider an environment with lotteries and transfers. Specifically, we
work with the space of allocations/outcomes $X\equiv \Delta \left( A\right)
\times 
\mathbb{R}
^{I}$ where $\Delta (A)$ denotes the set of lotteries on $A$ that have a
countable support, and $%
\mathbb{R}
^{I}$ denotes the set of transfers to the agents. We identify $a\in A$ with
a degenerate lottery in $\Delta (A)$.

For each $x=\left( \ell ,\left( t_{i}\right) _{i\in \mathcal{I}}\right) \in
X $, agent $i$ receives the utility $\tilde{u}_{i}(x,\theta )=v_{i}(\ell
,\theta )+t_{i}$ for some bounded expected utility function $v_{i}(\cdot
,\theta )$ over $\Delta \left( A\right) $. That is, we work with an
environment with a transferable utility (TU) on agents' preferences, which 
\cite{maskin99} does not impose. We abuse notation to identify $\Delta
\left( A\right) $ with a subset of $X$, i.e., each $\ell \in \Delta \left(
A\right) $ is identified with the allocation $\left( \ell ,0,...,0\right) $
in $X$.

We focus on a \textit{complete-information} environment in which a true
state $\theta $ is common knowledge among the agents but unknown to a
mechanism designer. The designer's objective is specified by a \textit{%
social choice function} $f:\Theta \rightarrow X$, namely, if the state is $%
\theta $, the designer would like to implement the social outcome $f\left(
\theta \right) $.

\subsection{Mechanism and Solution}

Let $\mathcal{M}=\left( (M_{i},\tau _{i}\right) _{i\in \mathcal{I}},g)$ be a
finite mechanism where $M_{i}$ is a nonempty finite \textit{set of messages}
available to agent $i$; $g:M\rightarrow X$ (where $M\equiv \times
_{i=1}^{I}M_{i}$) is the \textit{outcome function}; and $\tau
_{i}:M\rightarrow \mathbb{R}$ is the \textit{transfer rule} which specifies
the payment to agent $i$. At each state $\theta \in \Theta $, the
environment and the mechanism together constitute a \textit{game with
complete information} which we denote by $\Gamma (\mathcal{M},\theta )$.
Note that the restriction of $M_{i}$ to a finite set rules out the use of
integer games \`{a} la \cite{maskin99}.

Let $\sigma _{i}\in \Delta (M_{i})$ be a mixed \textit{strategy} of agent $i$
in the game $\Gamma (\mathcal{M},\theta )$. A strategy profile $\sigma
=(\sigma _{1},\ldots ,\sigma _{I})\in \times _{i\in \mathcal{I}}\Delta
(M_{i})$ is said to be a mixed-strategy \textit{Nash equilibrium} of the
game $\Gamma (\mathcal{M},\theta )$ if, for all agents $i\in \mathcal{I}$
and all messages $m_{i}\in $supp$\left( \sigma _{i}\right) $ and $%
m_{i}^{\prime }\in M_{i}$, we have 
\begin{eqnarray*}
&&\sum_{m_{-i}\in M_{-i}}\prod_{j\not=i}\sigma _{j}(m_{j})\left[ \tilde{u}%
_{i}(g(m_{i},m_{-i}),\theta )+\tau _{i}(m_{i},m_{-i})\right] \\
&\geq &\sum_{m_{-i}\in M_{-i}}\prod_{j\not=i}\sigma _{j}(m_{j})\left[ \tilde{%
u}_{i}(g(m_{i}^{\prime },m_{-i}),\theta )+\tau _{i}(m_{i}^{\prime },m_{-i})%
\right] \text{.}
\end{eqnarray*}%
A pure-strategy Nash equilibrium is a mixed-strategy Nash equilibrium $%
\sigma $ such that each agent $i$'s mixed-strategy $\sigma _{i}$ assigns
probability one to some $m_{i}\in M_{i}$. For any message profile $m\in M$,
let $\sigma (m)\equiv \prod_{j\in \mathcal{I}}\sigma _{j}(m_{j})$.

Let $NE(\Gamma (\mathcal{M},\theta ))$ denote the set of mixed-strategy Nash
equilibria of the game $\Gamma (\mathcal{M},\theta )$. We also denote by $%
\mbox{supp}\ (NE(\Gamma (\mathcal{M},\theta )))\ $the set of message
profiles that can be played with positive probability under some
mixed-strategy Nash equilibrium $\sigma \in NE(\Gamma (\mathcal{M},\theta )$%
, i.e., 
\begin{equation*}
\mbox{supp}\ (NE(\Gamma (\mathcal{M},\theta )))=\{m\in M:%
\mbox{there exists
$\sigma \in NE(\Gamma(\mathcal{M}, \theta))$ such that $\sigma(m) > 0$}\}.
\end{equation*}%
We now define our notion of Nash implementation.

\begin{definition}
\label{def-nash}An SCF $f$ is \textbf{implementable in mixed-strategy Nash
equilibria by a finite mechanism} if there exists a finite mechanism $%
\mathcal{M}=\left( (M_{i},\tau _{i}\right) _{i\in \mathcal{I}},g)$ such that
for every state $\theta \in \Theta $, (i) there exists a pure-strategy Nash
equilibrium in the game $\Gamma (\mathcal{M},\theta )$; and (ii) $m\in %
\mbox{supp}\ (NE(\Gamma (\mathcal{M},\theta )))\Rightarrow g\left( m\right)
=f\left( \theta \right) $ and $\tau _{i}\left( m\right) =0$ for every $i\in 
\mathcal{I}$.
\end{definition}

Our definition is adapted from mixed-strategy Nash implementation in \cite%
{maskin99} to (1) require that the implementing mechanism be finite; and (2)
accommodate our quasilinear environments with transfers. In particular,
Condition (ii) requires that transfers be imposed only off the equilibrium. 
\cite{MR12} propose another definition of Nash implementation that keeps
Condition (ii) but weakens Condition (i) in requiring only the existence of
a mixed-strategy Nash equilibrium, which, by Nash's theorem, is guaranteed
in a finite mechanism. In their sufficiency result, however, \cite{MR12}
construct an infinite mechanism with integer games.

\subsection{Maskin Monotonicity}

\label{sec:mono}

We now restate the definition of Maskin monotonicity that \cite{maskin99}
proposes for Nash implementation.

\begin{definition}
\label{mm}An SCF $f$ satisfies \textbf{Maskin monotonicity} if, for every
pair of states $\tilde{\theta}$ and $\theta $ with $f(\tilde{\theta}%
)\not=f\left( \theta \right) $, some agent $i\in \mathcal{I}$ and some
allocation $x\in X$ exist such that 
\begin{equation}
\tilde{u}_{i}(x,\tilde{\theta})\leq \tilde{u}_{i}(f(\tilde{\theta}),\tilde{%
\theta})\text{ and }\tilde{u}_{i}(x,\theta )>\tilde{u}_{i}(f(\tilde{\theta}%
),\theta )\text{.}  \label{maskin-m}
\end{equation}
\end{definition}

To illustrate how the idea of Maskin monotonicity is applied, suppose that
the SCF $f$ is implemented in Nash equilibria by a mechanism. When $\tilde{%
\theta}$ is the true state, there exists a pure-strategy Nash equilibrium $%
m\in M$ in $\Gamma (\mathcal{M},\tilde{\theta})$ which induces $f(\tilde{%
\theta})$. If $f(\tilde{\theta})\neq f\left( \theta \right) $ and $\theta $
is the true state, then $m$ cannot be a Nash equilibrium, i.e., there exists
some agent $i$ who has a profitable deviation. Suppose that the deviation
induces outcome $x$, i.e., agent $i$ strictly prefers $x$ to $f(\tilde{\theta%
})$ at state $\theta $. Since $m$ is a Nash equilibrium at state $\tilde{%
\theta}$, such a deviation cannot be profitable in state $\tilde{\theta}$;
that is, agent $i$ weakly prefers $f(\tilde{\theta})$ to $x$ at state $%
\tilde{\theta}$. In other words, $x$ belongs to agent $i$'s lower contour
set at $f(\tilde{\theta})$ of state $\tilde{\theta}$, whereas it belongs to
the strict upper-contour set at $f(\tilde{\theta})$ of state $\theta $.
Therefore, Maskin monotonicity is a necessary condition for Nash
implementation; in fact, it is a necessary condition even for Nash
implementation that restricts attention to pure-strategy equilibria (i.e.,
to require that condition (ii) of Definition \ref{def-nash} hold only for
pure-strategy Nash equilibria).

\section{Main Result}

\label{NE}

In this section, we present our main result, which shows that Maskin
monotonicity is necessary and sufficient for mixed-strategy Nash
implementation. We formally state the result as follows.

\begin{theorem}
\label{Nash}An SCF $f$ is implementable in mixed-strategy Nash equilibria by
a finite mechanism if and only if it satisfies Maskin monotonicity.
\end{theorem}

In the rest of this section, we will establish Theorem \ref{Nash} and
discuss the issues regarding the theorem. Section \ref{M} details how our
implementing mechanism is constructed. In Section \ref{proof}, we prove
Theorem \ref{Nash} by making use of the implementing mechanism constructed
in Section \ref{M}. Section \ref{sec:drm} illustrates two special cases in
which our implementing mechanism can be made into a direct mechanism where
each agent reports a state. In Section \ref{trans}, we discuss the necessity
of domain restrictions in establishing Theorem \ref{Nash}.

\subsection{The Mechanism}

\label{M}

We construct a mechanism $\mathcal{M}=\left( (M_{i},\tau _{i})_{i\in 
\mathcal{I}},g\right) $ which will be used to prove Theorem \ref{Nash}. The
mechanism shares a number of features of the implementing mechanisms in \cite%
{maskin99} and in Abreu and Matsushima (1992, 1994), which we summarize at
the end of the subsection. The construction involves two major building
blocks that we call the \textit{best challenge scheme} and \textit{dictator
lotteries}, respectively. After introducing these building blocks, we will
define the message space, allocation rule, and transfer rule of our
implementing mechanism.

For each agent $i$, as a preliminary step, we define 
\begin{equation*}
\Theta _{i}\equiv \left\{ v_{i}(\cdot ,\theta ):\theta \in \Theta \right\} 
\text{.}
\end{equation*}%
That is, $\Theta _{i}$ is the set of expected utility functions of agent $i$
induced by some state $\theta $. Denote by $\theta _{i}\in \Theta _{i}$ the
expected utility function of agent $i$ obtained at state $\theta \in \Theta $%
, namely, that $\theta _{i}=v_{i}(\cdot ,\theta )$. We call $\theta _{i}$
the type of player $i$ at state $\theta $. We denote by $u_{i}\left( \cdot
,\theta _{i}\right) $ the quasilinear utility function which corresponds to
type $\theta _{i}$, namely, that for each $x=\left( \ell ,\left(
t_{i}\right) _{i\in \mathcal{I}}\right) \in X$, we have $u_{i}\left(
x,\theta _{i}\right) \equiv v_{i}(\ell ,\theta )+t_{i}$. For a
Maskin-monotonic SCF $f$, we have $f\left( \theta \right) =f(\tilde{\theta})$
if states $\theta $ and $\tilde{\theta}$ induce the same type profile (i.e., 
$\theta _{i}=\tilde{\theta}_{i}$ for every $i$). Hence, if a type profile $(%
\tilde{\theta}_{i})_{i\in \mathcal{I}}$ is induced by some state $\theta \in
\Theta $, we may abuse the notation to write $(\tilde{\theta}_{i})_{i\in 
\mathcal{I}}\in \Theta $ and set $f((\tilde{\theta}_{i})_{i\in \mathcal{I}%
})\equiv f\left( \theta \right) $.

\begin{comment}
\label{r1}It is possible no state in $\Theta $ induces a given type profile.
For example, suppose we that have two states $\alpha $ and $\beta $ and two
agents $A$ and $B$ who have an identical expected utility function that
varies across the states, namely $\alpha _{A}=\alpha _{B}\neq \beta
_{A}=\beta _{B}.$ In this example, there are four type profiles: $(\alpha
_{A},\alpha _{B}),(\alpha _{A},\beta _{B}),(\beta _{A},\alpha _{B})$, and $%
(\beta _{A},\beta _{B})$, and yet neither the type profile $(\alpha
_{A},\beta _{B})$ nor $(\beta _{A},\alpha _{B})$ corresponds to a state.
\end{comment}

\subsubsection{Best Challenge Scheme}

\label{B}

For $\left( x,\theta _{i}\right) \in X\times \Theta _{i}$, we use $\mathcal{L%
}_{i}\left( x,\theta _{i}\right) $ to denote the lower-contour set at
allocation $x$ in $X$ for type $\theta _{i}$, i.e.,%
\begin{equation*}
\mathcal{L}_{i}\left( x,\theta _{i}\right) =\left\{ x^{\prime }\in
X:u_{i}\left( x,\theta _{i}\right) \geq u_{i}(x^{\prime },\theta
_{i})\right\} .
\end{equation*}%
We use $\mathcal{SU}_{i}\left( x,\theta _{i}\right) $ to denote the strict
upper-contour set of $x\in X$ for type $\theta _{i}$, i.e.,%
\begin{equation*}
\mathcal{SU}_{i}\left( x,\theta _{i}\right) =\left\{ x^{\prime }\in
X:u_{i}(x^{\prime },\theta _{i})>u_{i}(x,\theta _{i})\right\} \text{.}
\end{equation*}%
Hence, according to Definition \ref{mm}, an SCF $f$ satisfies Maskin
monotonicity if and only if for every pair of states $\tilde{\theta}$ and $%
\theta $ in $\Theta $, 
\begin{equation}
f(\tilde{\theta})\not=f\left( \theta \right) \Rightarrow \exists i\in 
\mathcal{I}\text{ s.t. }\mathcal{L}_{i}(f(\tilde{\theta}),\tilde{\theta}%
_{i})\cap \mathcal{SU}_{i}(f(\tilde{\theta}),\theta _{i})\not=\varnothing .
\label{whistle}
\end{equation}

Agent $i$ in (\ref{whistle}) is called a \textquotedblleft whistle-blower"
or a \textquotedblleft test agent," and an allocation in $\mathcal{L}_{i}(f(%
\tilde{\theta}),\tilde{\theta}_{i})\cap \mathcal{SU}_{i}(f(\tilde{\theta}%
),\theta _{i})$ is called a \textquotedblleft test
allocation\textquotedblright\ for agent $i$ and the ordered pair of states $(%
\tilde{\theta},\theta )$. We now define a notion called \emph{the best
challenge scheme}, which plays a crucial role in proving Theorem \ref{Nash}.
We say that a mapping $x:\Theta \times \Theta _{i}\rightarrow X$ is a \emph{%
challenge scheme} for an SCF $f$ if and only if, for each pair of state $%
\tilde{\theta}\in \Theta $ and type $\theta _{i}\in \Theta _{i}$, 
\begin{equation*}
\left\{ 
\begin{array}{ll}
x(\tilde{\theta},\theta _{i})\in \mathcal{L}_{i}(f(\tilde{\theta}),\tilde{%
\theta}_{i})\cap \mathcal{SU}_{i}(f(\tilde{\theta}),\theta _{i})\text{,} & 
\text{if }\mathcal{L}_{i}(f(\tilde{\theta}),\tilde{\theta}_{i})\cap \mathcal{%
SU}_{i}(f(\tilde{\theta}),\theta _{i})\neq \varnothing \text{;} \\ 
x(\tilde{\theta},\theta _{i})=f(\tilde{\theta})\text{,} & \text{if }\mathcal{%
L}_{i}(f(\tilde{\theta}),\tilde{\theta}_{i})\cap \mathcal{SU}_{i}(f(\tilde{%
\theta}),\theta _{i})=\varnothing \text{.}%
\end{array}%
\right.
\end{equation*}%
We may think of state $\tilde{\theta}$ as an announcement made by one or
more other agents that agent $i$ of type $\theta _{i}$ could
\textquotedblleft challenge\textquotedblright\ (as a whistle-blower). The
following lemma shows that there is a challenge scheme in which each
whistle-blower $i$ facing state announcement $\tilde{\theta}$ finds it
weakly best to challenge $\tilde{\theta}$ by simply reporting his true type $%
\theta _{i}$.

\begin{lemma}
\label{BC}There is a challenge scheme $x(\cdot ,\cdot )$ for an SCF $f$ such
that for every state $\tilde{\theta}$ and type $\theta _{i}$, 
\begin{equation}
u_{i}(x(\tilde{\theta},\theta _{i}),\theta _{i})\geq u_{i}(x(\tilde{\theta}%
,\theta _{i}^{\prime }),\theta _{i}),\forall \theta _{i}^{\prime }\in \Theta
_{i}\text{.}  \label{b}
\end{equation}
\end{lemma}

We relegate its formal proof to Appendix \ref{app:BC}.\footnote{%
We owe special thanks to Phil Reny for suggesting the lemma which simplifies
the implementing mechanism adopted in an earlier version of our paper.} In
defining the implementing mechanism, we shall invoke a challenge scheme
which satisfies (\ref{b}). We call such a challenge scheme \emph{the best
challenge scheme}. In words, under the best challenge scheme, for any state $%
\tilde{\theta}$, agent $i$ of type $\theta _{i}$ weakly prefers the
allocation $x(\tilde{\theta},\theta _{i})$ to any other $x(\tilde{\theta}%
,\theta _{i}^{\prime })$ induced by announcing $\theta _{i}^{\prime }\neq
\theta _{i}$.

\subsubsection{Dictator Lotteries}

\label{Dict}

Let $\tilde{X}\equiv A\cup \bigcup_{i\in \mathcal{I},\theta _{i}\in \Theta
_{i},\tilde{\theta}\in \Theta }x(\tilde{\theta},\theta _{i}).$ We can then
conclude that $\tilde{X}$ is a finite set over which all agents' utilities
are bounded, because $v_{i}(\cdot ,\theta )$ is bounded, $\Theta $ is
finite, and we pre-specify $x(\tilde{\theta},\theta _{i})$ for each $i\in 
\mathcal{I},$ type $\theta _{i}\in \Theta _{i}$, and state $\tilde{\theta}%
\in \Theta .$ Hence, we can choose $\eta ^{\prime }>0$ as an upper bound on
the monetary value of a change of allocation in $\tilde{X}$, that is, 
\begin{equation}
\eta ^{\prime }>\sup_{i\in \mathcal{I},\theta _{i}\in \Theta
_{i},x,x^{\prime }\in \tilde{X}}\left\vert u_{i}(x,\theta
_{i})-u_{i}(x^{\prime },\theta _{i})\right\vert \text{.}  \label{D'}
\end{equation}

We now state a result which ensures the existence of what we call \textit{%
dictator lotteries} for agent $i$. In particular, a collection of lotteries
is called dictator lotteries of agent $i$ if it satisfies Conditions (\ref{d}%
) and (\ref{d-w}) stated in Lemma \ref{AM}. Condition (\ref{d}) shows that
within dictator lotteries, each agent has a strict incentive to reveal his
true type, whereas Condition (\ref{d-w}) says that these dictator lotteries
are strictly less preferred than any allocations in $\tilde{X}$.

\begin{lemma}
\label{AM}For each agent $i\in \mathcal{I}$, there exists a collection of
lotteries $\{y_{i}(\theta _{i})\}_{\theta _{i}\in \Theta _{i}}$ such that
for all types $\theta _{i},\theta _{i}^{\prime }\in \Theta _{i}$ with $%
\theta _{i}\neq \theta _{i}^{\prime }$, we have%
\begin{equation}
u_{i}\left( y_{i}\left( \theta _{i}\right) ,\theta _{i}\right) >u_{i}\left(
y_{i}\left( \theta _{i}^{\prime }\right) ,\theta _{i}\right) \text{;}
\label{d}
\end{equation}%
moreover, for each $j\in \mathcal{I}$ and type $\theta _{j}^{\prime }\in
\Theta _{j}$, we also have that, for every $x\in \tilde{X}$, 
\begin{equation}
u_{i}(y_{j}(\theta _{j}^{\prime }),\theta _{i})<u_{i}(x,\theta _{i})\text{.}
\label{d-w}
\end{equation}
\end{lemma}

Since two distinct types in $\Theta _{i}$ induce different expected utility
functions over $\Delta \left( A\right) $, it follows from \cite[Lemma]{AM92}
that we can prove the existence of lotteries $\left\{ y_{i}^{\prime }\left(
\cdot \right) \right\} \subset \Delta \left( A\right) $ that satisfy
Condition (\ref{d}). To satisfy Condition (\ref{d-w}), we simply add a
penalty of $\eta ^{\prime }$ to each outcome of the lotteries $\left\{
y_{i}^{\prime }\left( \theta _{i}\right) \right\} _{\theta _{i}\in \Theta
_{i}}$. More precisely, for each $\theta _{i}\in \Theta _{i}$, we set 
\begin{equation*}
y_{i}(\theta _{i})=(y_{i}^{\prime }(\theta _{i}),-\eta ^{\prime },\ldots
,-\eta ^{\prime })\in X\text{.}
\end{equation*}

\subsubsection{Message Space}

A generic message of agent $i$ is described as follows: 
\begin{equation*}
m_{i}=\left( m_{i}^{1},m_{i}^{2}\right) \in M_{i}=M_{i}^{1}\times
M_{i}^{2}=\Theta _{i}\times \left[ \times _{j=1}^{I}\Theta _{j}\right] \text{%
.}
\end{equation*}%
That is, agent $i$ is asked to make (1) a report of his own type (which we
denote by $m_{i}^{1}$); and (2) a report of a type profile (which we denote
by $m_{i}^{2}$). To simplify the notation, we write $m_{i,j}^{2}=\tilde{%
\theta}_{j}$ if agent $i$ reports in $m_{i}^{2}$ that agent $j$ is of type $%
\tilde{\theta}_{j}$. Recall that agents have complete information about the
true state. If the true state is $\theta $, we say that agent $i$ sends a
truthful first report if $m_{i}^{1}=\theta _{i}$ and a truthful second
report if $m_{i}^{2}=\left( \theta _{j}\right) _{j\in \mathcal{I}}$. Note
that each agent is asked to report a type profile in $M_{i}^{2}$ instead of
a state. Hence, the mechanism must take care of the difficulties in
identifying the state from a type profile which we explain at the beginning
of Section \ref{M}.

It is useful to compare the message space of our mechanism with that of the
implementing mechanism in \cite{maskin99}. In Maskin's mechanism (see %
\citet[p. 31]{maskin99}), each agent is asked to report a preference profile
and an integer, as well as an allocation. The allocation need not be
specified in the case of SCFs, since there is no ambiguity about the
socially desirable outcome assigned to each state. In contrast, we ask each
agent to report a preference/type profile and a \emph{type}. The type
component of the message space plays the role of an integer in Maskin's
mechanism in knocking out unwanted equilibria, albeit in a different manner.
As the integer game admits no equilibrium when there is disagreement over
most preferred outcomes, it is used to assure that undesirable message
profiles do not form an equilibrium. However, the logic of their argument no
longer works when the goal is to achieve implementation in mixed-strategy
Nash equilibria by a finite mechanism. Indeed, there is no a priori way to
rule out \textit{any} message profile because any of them might be played
with positive probability in a mixed-strategy Nash equilibrium.

By making use of the type component $m_{i}^{1}$ in the message space, we
appeal to the approach of Abreu and Matsushima (1992, 1994)) to resolve the
issue. More precisely, we design the mechanism so that when an unwanted
message profile is triggered in equilibrium, the type report $m_{i}^{1}$
must coincide with agent $i$'s preference under the true state. Through the
cross-checking of the preferences and preference profiles reported by the
agents (in a similar manner to Abreu and Matsushima (1992, 1994)), it
further implies that the unwanted message profile could not have happened.
Unlike Abreu and Matsushima (1992, 1994), however, to ensure that $m_{i}^{1}$
is truthful, we must guarantee that the designer's twin goals of allowing
for whistle-blowing/challenges (as in \cite{maskin99}) and eliciting the
truth (from the dictator lotteries, as in Abreu and Matsushima (1992, 1994))
can be aligned perfectly. It is achieved through Lemmas 1 and 2: since
truth-telling is weakly optimal for the former and strictly optimal for the
latter, we can make the truth-telling in $m_{i}^{1}$ strictly optimal by
taking a convex combination of the best challenge scheme and dictator
lotteries. Hence, Maskin meets Abreu and Matsushima. We formalize the idea
in Section \ref{ar}.

\subsubsection{Allocation Rule}

\label{ar}

For each message profile $m\in M$, the allocation is determined as follows:%
\begin{equation*}
g\left( m\right) =\frac{1}{I(I-1)}\sum_{i\in \mathcal{I}}\sum_{j\neq i}\left[
e_{i,j}\left( m_{i},m_{j}\right) \left( \frac{1}{2}%
\sum_{k=i,j}y_{k}(m_{k}^{1})\right) \oplus \left( 1-e_{i,j}\left(
m_{i},m_{j}\right) \right) x(m_{i}^{2},m_{j}^{1})\right] \text{,}
\end{equation*}%
where $\{y_{k}(\cdot )\}\ $are the dictator lotteries for agent $k$ obtained
from Lemma \ref{AM}, and $\alpha x\oplus \left( 1-\alpha \right) x^{\prime }$
denotes the outcome which corresponds to the compound lottery that outcome $%
x $ occurs with probability $\alpha $, and outcome $x^{\prime }$ occurs with
probability $1-\alpha $;\footnote{%
More precisely, if $x=\left( \ell ,\left( t_{i}\right) _{i\in \mathcal{I}%
}\right) $ and $x^{\prime }=\left( \ell ^{\prime },\left( t_{i}^{\prime
}\right) _{i\in \mathcal{I}}\right) $ are two outcomes in $X$, we identify $%
\alpha x\oplus \left( 1-\alpha \right) x^{\prime }$ with the outcome $\left(
\alpha \ell \oplus \left( 1-\alpha \right) \ell ^{\prime },\left( \alpha
t_{i}+\left( 1-\alpha \right) t_{i}^{\prime }\right) _{i\in \mathcal{I}%
}\right) $. For simplicity, we also write the compound lottery $\frac{1}{2}%
y_{i}\left( m_{i}^{1}\right) \oplus \frac{1}{2}y_{j}\left( m_{j}^{1}\right) $
as $\frac{1}{2}\sum_{k=i,j}y_{k}\left( m_{k}^{1}\right) $.} moreover, we
define 
\begin{equation*}
e_{i,j}(m_{i},m_{j})=\left\{ 
\begin{array}{ll}
0\text{,} & \text{if }m_{i}^{2}\in \Theta \text{, }m_{i}^{2}=m_{j}^{2}\text{%
, and }x(m_{i}^{2},m_{j}^{1})=f(m_{i}^{2})\text{;} \\ 
\varepsilon \text{,} & \text{if }m_{i}^{2}\in \Theta \text{, and [}%
m_{i}^{2}\neq m_{j}^{2}\text{ or }x(m_{i}^{2},m_{j}^{1})\neq f(m_{i}^{2})%
\text{];} \\ 
1\text{,} & \text{if }m_{i}^{2}\notin \Theta \text{.}%
\end{array}%
\right.
\end{equation*}%
In what follows, we say that the second reports of agent $i$ and agent $j$
are \emph{consistent} if $m_{i}^{2}=m_{j}^{2}$ \emph{and} the common type
profile identifies a state in $\Theta $; moreover, we say that agent $j$ 
\emph{does not} \emph{challenge} agent $i$ if $%
x(m_{i}^{2},m_{j}^{1})=f(m_{i}^{2})$.\footnote{%
Observe that we make the first report of both agents $i$ and $j$ effective
(through affecting the compound lottery $\frac{1}{2}\sum_{k=i,j}y_{k}\left(
m_{k}^{1}\right) $), regardless of whether pair $\left( i,j\right) $ or pair 
$\left( j,i\right) $ is picked. The construction will be used in proving
Claim \ref{c0}, which, in turn, is used to prove Claim \ref{pc2}.}

In words, the designer first chooses an ordered pair of distinct agents $%
(i,j)$ with equal probability. The outcome function distinguishes three
cases: (1) if the second reports of agent $i$ and agent $j$ are consistent 
\emph{and} agent $j$ does not challenge agent $i$, then we implement $%
f\left( m_{i}^{2}\right) $; (2) if agent $i$ reports a type profile which
does not identify a state in $\Theta $, then we implement the dictator
lottery $\frac{1}{2}\sum_{k=i,j}y_{k}\left( m_{k}^{1}\right) $; (3)
otherwise, we implement the compound lottery:%
\begin{equation*}
C_{i,j}^{\varepsilon }(m_{i},m_{j})\equiv \varepsilon \left( \frac{1}{2}%
\sum_{k=i,j}y_{k}\left( m_{k}^{1}\right) \right) \oplus \left( 1-\varepsilon
\right) x(m_{i}^{2},m_{j}^{1})\text{.}
\end{equation*}%
That is, $C_{i,j}^{\varepsilon }(m_{i},m_{j})$ is an $\left( \varepsilon
,1-\varepsilon \right) $-combination of (i) the two dictator lotteries--$%
y_{i}\left( m_{i}^{1}\right) $ and $y_{j}(m_{j}^{1})$--which occur with
equal probability; and (ii) the allocation specified by the best challenge
scheme $x(m_{i}^{2},m_{j}^{1})$.

By (\ref{D'}), we can choose $\varepsilon >0$ sufficiently small, and $\eta
>0$ sufficiently large\footnote{%
Instead of using $\eta ^{\prime }$ defined in (\ref{D'}), we choose $\eta $
because the mechanism may produce a strictly larger finite set of
alternatives than those contained in $\tilde{X}$. For instance, allocations
from the dictator lotteries may occur from the mechanism but are not
contained in $\tilde{X}$.} such that firstly we have 
\begin{equation}
\eta >\sup_{i\in \mathcal{I},\theta _{i}\in \Theta _{i},m,m^{\prime }\in
M}\left\vert u_{i}(g\left( m\right) ,\theta _{i})-u_{i}(g\left( m^{\prime
}\right) ,\theta _{i})\right\vert \text{;}  \label{D}
\end{equation}%
secondly it does not disturb the \textquotedblleft
effectiveness\textquotedblright\ of agent $j$'s challenge; put formally,%
\begin{eqnarray}
x(m_{i}^{2},m_{j}^{1}) &\neq &f(m_{i}^{2})\Rightarrow  \notag \\
u_{j}(C_{i,j}^{\varepsilon }(m_{i},m_{j}),m_{i,j}^{2})
&<&u_{j}(f(m_{i}^{2}),m_{i,j}^{2})\text{ and }u_{j}(C_{i,j}^{\varepsilon
}(m_{i},m_{j}),m_{j}^{1})>u_{j}(f(m_{i}^{2}),m_{j}^{1})\text{.}  \label{bw}
\end{eqnarray}%
It means that whenever agent $j$ challenges agent $i$, the lottery $%
C_{i,j}^{\varepsilon }(m_{i},m_{j})$ is strictly worse than $f\left(
m_{i}^{2}\right) $ for agent $j$ when agent $i$ tells the truth about agent $%
j$'s preference in $m_{i}^{2}$; moreover, the lottery $C_{i,j}^{\varepsilon
}(m_{i},m_{j})$ is strictly better than $f\left( m_{i}^{2}\right) $ for
agent $j$ when agent $j$ tells the truth in $m_{j}^{1}$, a fact which
implies that agent $i$ tells a lie about agent $j$'s preference.

\subsubsection{Transfer Rule}

\label{Fmoney}

We now define the transfer rule. For every message profile $m\in M$ and
every agent $i\in \mathcal{I}$, we specify the transfer received by agent $i$
as follows: 
\begin{equation*}
\tau _{i}(m)=\sum_{j\neq i}\left[ \tau _{i,j}^{1}(m_{i},m_{j})+\tau
_{i,j}^{2}(m_{i},m_{j})\right] ,
\end{equation*}%
where for each agent $j\neq i$, we define%
\begin{eqnarray}
\tau _{i,j}^{1}\left( m_{i},m_{j}\right) &=&\left\{ 
\begin{tabular}{ll}
$0$, & if $m_{i,j}^{2}=m_{j,j}^{2}$; \\ 
$-\eta $, & if $m_{i,j}^{2}\not=m_{j,j}^{2}$ and $m_{i,j}^{2}\neq m_{j}^{1}$;
\\ 
$\eta $, & if $m_{i,j}^{2}\not=m_{j,j}^{2}\text{ }$and $%
m_{i,j}^{2}=m_{j}^{1} $.%
\end{tabular}%
\right.  \label{money1} \\
\tau _{i,j}^{2}\left( m_{i},m_{j}\right) &=&\left\{ 
\begin{array}{ll}
0\text{,} & \text{if }m_{i,i}^{2}=m_{j,i}^{2}\text{;} \\ 
-\eta \text{,} & \text{if }m_{i,i}^{2}\neq m_{j,i}^{2}\text{.}%
\end{array}%
\right.  \label{money2}
\end{eqnarray}%
Recall that $\eta >0$ is chosen to be larger than the maximal utility
difference from the outcome function $g\left( \cdot \right) $; see (\ref{D}).

In words, for each pair of agents $(i,j)$, if their second reports on agent $%
j$'s type coincide ($m_{ij}^{2}=m_{jj}^{2}$), then no transfer will be made;
if their second reports on agent $j$'s type differ ($m_{ij}^{2}\neq
m_{jj}^{2}$), then we consider the following two subcases: (i) if agent $i$%
's second report about agent $j$'s type matches agent $j$'s first report ($%
m_{i,j}^{2}=m_{j}^{1}$), then agent $j$ pays $\eta $ to agent $i$; (ii) if
agent $i$'s second report about agent $j$'s type does not match agent $j$'s
first report ($m_{i,j}^{2}\neq m_{j}^{1}$), then both agents pay $\eta $ to
the designer. Note that the first report $m_{i}^{1}$ does not affect the
transfer to agent $i$.


\subsection{Proof of Theorem \protect\ref{Nash}}

\label{proof}

As we argue in Section \ref{sec:mono}, Maskin monotonicity is a necessary
condition for Nash implementation. We therefore focus on the
\textquotedblleft if\textquotedblright\ part of the proof. Fix an arbitrary
true state $\theta $ throughout the proof. Recall that $\theta _{i}$ stands
for agent $i$'s type at state $\theta $ and $\left( \theta _{i}\right)
_{i\in \mathcal{I}}$ denotes the true type profile.

We argue that the truth-telling message profile $m$ (i.e., $m_{i}=(\theta
_{i},\theta )$ for each agent $i$) constitutes a pure-strategy Nash
equilibrium. Since $m$ is truthful, for all agents $i$ and $j$, we have $%
e_{i,j}(m_{i},m_{j})=0$ and $\tau _{i}\left( m\right) =0$ (consistency and
no challenge). Consider a possible deviation $\tilde{m}_{i}$ of agent $i$
from $m$. First, if $\tilde{m}_{i,j}^{2}=\theta _{j}^{\prime }\neq \theta
_{j}$ for some $j\in \mathcal{I}$, then the message profile $(\tilde{m}%
_{i},m_{-i})$ induces the penalty of $\eta $ from rule $\tau
_{i,j}^{1}(\cdot )$ if $j\neq i$, and rule $\tau _{i,j}^{2}(\cdot )$ if $j=i$%
. As a result, $\tilde{m}_{i}$ is strictly worse against $m_{-i}$ than $%
m_{-i}$.

Second, if $\tilde{m}_{i}^{1}\neq \theta _{i}$ and $\tilde{m}_{i}^{2}=\theta 
$, $(\tilde{m}_{i},m_{-i})$ leads either to $x(\theta ,\tilde{m}%
_{i}^{1})=f(\theta )$ and thereby the same payoff, or to $x(\theta ,\tilde{m}%
_{i}^{1})\neq f(\theta )$. In the latter case, the message profile $(\tilde{m%
}_{i},m_{-i})$ results in the outcome $C_{i,j}^{\varepsilon }(\tilde{m}%
_{i},m_{j})$ which, by (\ref{bw}), is strictly worse than $f(\theta )$
induced by $m$. Furthermore, deviating from $m_{i}$ to $\tilde{m}_{i}$ does
not affect the transfer of agent $i$. Therefore, the truth-telling message
profile $m$ constitutes a pure-strategy Nash equilibrium.

We next show that for every Nash equilibrium $\sigma $ of the game $\Gamma (%
\mathcal{M},\theta )$ and every message profile $m$ reported with positive
probability under $\sigma $, we must achieve the socially desirable outcome,
i.e., $g\left( m\right) =f\left( \theta \right) $ and $\tau _{i}\left(
m\right) =0$ for every agent $i$. The proof is divided into three steps.

\noindent \textbf{Step 1}: \textit{Contagion of truth}. If agent $j$
announces his type truthfully in his first report with probability one, then
everyone must also report agent $j$'s type truthfully in their second report;

\noindent \textbf{Step 2}: \textit{Consistency}. Every agent reports the
same state $\tilde{\theta}$ in the second report;

\noindent \textbf{Step 3}: \textit{No challenge}. No agent challenges the
common reported state $\tilde{\theta}$, i.e., $x(\tilde{\theta},m_{j}^{1})=f(%
\tilde{\theta})$ for every agent $j \in \mathcal{I}$.

Consistency implies that $\tau _{i}\left( m\right) =0$ for every agent $i\in 
\mathcal{I}$, whereas no challenge together with Maskin monotonicity of the
SCF $f$ implies that $g\left( m\right) =f(\tilde{\theta})=f\left( \theta
\right) $. It completes the proof of Theorem \ref{Nash}. We now proceed to
establish these three steps. In the rest of the proof, we fix $\sigma $ as
an arbitrary mixed-strategy Nash equilibrium of the game $\Gamma (\mathcal{M}%
,\theta )$.

As a consequence of Lemmas \ref{AM} and \ref{BC}, the mechanism has the
following crucial property which we will make use of in establishing the
implementation.

\begin{claim}
\label{c0}Let $\sigma $ be a Nash equilibrium of the game $\Gamma (\mathcal{M%
},\theta )$. If $m_{i}^{1}\neq \theta _{i}$ for some $m_{i}\in $supp$(\sigma
_{i})$, then for every agent $j\neq i$, we have $e_{i,j}\left(
m_{i},m_{j}\right) =e_{j,i}\left( m_{j},m_{i}\right) =0$ with $\sigma _{j}$%
-probability one.
\end{claim}

The claim essentially follows from Lemmas \ref{BC} and \ref{AM}. Indeed, the
two lemmas together imply that agents must have a strict incentive to tell
the truth in their first report, as long as switching from a lie to truth
affects the allocation with positive probability. A detailed verification of
the claim, however, is tedious, as it involves checking different cases of
the value of functions $e_{i,j}\left( \cdot \right) $ and $e_{j,i}\left(
\cdot \right) $. We relegate its formal proof to Appendix \ref{pc0}.

\subsubsection*{Step 1: Contagion of Truth}

\label{CoT}

\begin{claim}
\label{1st}The following two statements hold:

\noindent (a) If agent $j$ sends a truthful first report with $\sigma _{j}$%
-probability one, then every agent $i\neq j$ must report agent $j$'s type
truthfully in his second report with $\sigma _{i}$-probability one.

\noindent (b) If every agent $i\neq j$ reports the same type $\tilde{\theta}%
_{j}$ of agent $j$ in his second report with $\sigma _{i}$-probability one,
then agent $j$ must also report the type $\tilde{\theta}_{j}$ in his second
report with $\sigma _{j}$-probability one.
\end{claim}

\begin{proof}
We first prove (a). Suppose instead that there exist some agent $i \in 
\mathcal{I}$ and some message $m_{i}$ played with $\sigma _{i}$-positive
probability such that $m_i$ misreports agent $j$'s type in the second
report, i.e., $m_{i,j}^{2}\not=\theta _{j}$. Let $\tilde{m}_{i}$ be a
message that differs from $m_{i}$ only in reporting $j$'s type truthfully $%
\tilde{m}_{i,j}^{2}=\theta _{j}$. Such a change affects only $\tau
_{i,j}^{1}(\cdot )$. For every $m_{-i}$ played with $\sigma _{-i}$-positive
probability, we consider the following two cases.

\noindent \textbf{Case 1}: $m_{j,j}^{2}=\theta _{j}$

Since agent $j$ sends a truthful first report with $\sigma _{j}$-probability
one, due to the construction of $\tau _{i,j}^{1}(\cdot )$, we have $\tau
_{i,j}^{1}\left( m_{i},m_{-i}\right) =-\eta $ whereas $\tau _{i,j}^{1}\left( 
\tilde{m}_{i},m_{-i}\right) =0.$

\noindent \textbf{Case 2}: $m_{j,j}^{2}\neq \theta _{j}$

Since agent $j$ sends a truthful type in the first report with $\sigma _{j}$%
-probability one, according to the construction of $\tau _{i,j}^{1}\left(
\cdot \right) ,$ we have $\tau _{i,j}^{1}\left( m_{i},m_{-i}\right) $ is
either $0$ or $-\eta $ whereas $\tau _{i,j}^{1}\left( \tilde{m}%
_{i},m_{-i}\right) =\eta $.

Thus, in terms of transfers, the gain from reporting $\tilde{m}_{i}$ rather
than $m_{i}$ is at least $\eta $, which is larger than the maximal utility
loss from the outcome function $g\left( \cdot \right) $ by (\ref{D}). Hence, 
$\tilde{m}_{i}$ is a profitable deviation from $m_{i}$ against $\sigma _{-i}$%
. As it contradicts the hypothesis that $m_{i}\in \mbox{supp}(\sigma _{i})$,
we have established (a).

We now prove (b). Suppose, on the contrary, that there exists some message $%
m_{j}$ played with $\sigma _{j}$-positive probability such that $%
m_{j,j}^{2}\neq \tilde{\theta}_{j}$. Let $\tilde{m}_{j}$ be a message that
is identical to $m_{j}$ except that $\tilde{m}_{j,j}^{2}=\tilde{\theta}_{j}$%
. Such a change affects only $\tau _{j,i}^{2}\left( \cdot \right) $.
According to the construction of $\tau _{j,i}^{2}\left( \cdot \right) $ and
since every agent $i\neq j$ reports $\tilde{\theta}_{j}$ in the second
report with $\sigma _{i}$-probability one, agent $j$ saves the penalty of $%
\left( I-1\right) \eta $ from reporting $\tilde{m}_{j}$ instead of $m_{j}$.
Again, since $\eta $ is greater than the maximal utility difference by (\ref%
{D}), we conclude that $\tilde{m}_{j}$ is a profitable deviation from $m_{j}$
against $\sigma _{-i}$. It contradicts the hypothesis that $m_{j}\in %
\mbox{supp}(\sigma _{j})$. Hence, we prove (b).
\end{proof}

\subsubsection*{Step 2: Consistency}

\label{consistency}

Claim \ref{pc1} shows that in equilibrium, all agents must announce the same
state $\tilde{\theta}$ with probability one.

\begin{claim}
\label{pc1}There exists a state $\tilde{\theta}\in \Theta $ such that every
agent announces $\tilde{\theta}$ in their second report with probability one.
\end{claim}

\begin{proof}
We consider the following two cases:

\noindent \textbf{Case 1}: \textit{Everyone tells the truth in the first
report with probability one, i.e., }$m_{i}^{1}=\theta _{i}$\textit{\ with }$%
\sigma _{i}$\textit{-probability one\ for every agent }$i \in \mathcal{I}$.%
\vspace{0.2cm}

It follows directly from Claim \ref{1st} that $m_{i}^{2}=\theta $ with $%
\sigma _{i}$-probability one for every agent $i \in \mathcal{I}$.\vspace{%
0.2cm}

\noindent \textbf{Case 2}: \textit{There exists agent }$i$\textit{\ who
tells a lie in the first report with }$\sigma _{i}$\textit{-positive} 
\textit{probability.\vspace{0.2cm}}

That is, there exists $m_{i}\in \mbox{supp}(\sigma _{i})$ such that $%
m_{i}^{1}\neq \theta _{i}$.\textit{\vspace{0.2cm}} By Claim \ref{c0}, $%
\left( m_{i},m_{-i}\right) $ is consistent with $\sigma _{-i}$-probability
one. In particular, there exists $\tilde{\theta}\in \Theta $ such that every
agent $j\neq i~$must report 
\begin{equation}
m_{j}^{2}=m_{i}^{2}=\tilde{\theta}\text{ with }\sigma _{j}\text{-probability
one.}  \label{k1}
\end{equation}%
Hence, by Claim \ref{1st}(b), for every $\tilde{m}_{i}\in $supp$\left(
\sigma _{i}\right) $, we have 
\begin{equation}
\tilde{m}_{i,i}^{2}=m_{i,i}^{2}=\tilde{\theta}_{i}\text{.}  \label{k2}
\end{equation}%
We now prove that for every $\tilde{m}_{i}\in $supp$\left( \sigma
_{i}\right) $, we have $\tilde{m}_{i}^{2}=m_{i}^{2}=\tilde{\theta}$, which
would complete the proof. We prove it by contradiction, i.e., suppose there
exists $\tilde{m}_{i}\in $supp$\left( \sigma _{i}\right) $ such that%
\begin{equation}
\tilde{m}_{i}^{2}\neq m_{i}^{2}\text{.}  \label{k3}
\end{equation}%
Furthermore, (\ref{k1}) and (\ref{k3}) imply that for every agent $j\neq i$, 
$e_{j,i}\left( m_{j},\tilde{m}_{i}\right) =\varepsilon $ with $\sigma _{j}$%
-probability one. Hence, by Claim \ref{c0}, every agent $j\neq i$ must tell
the truth in the first report, i.e., $m_{j}^{1}=\theta _{j}$ with $\sigma
_{j}$-probability one. As a result, Claim \ref{1st}(a) implies for every
agent $j\neq i$%
\begin{equation}
\tilde{m}_{i,j}^{2}=m_{i,j}^{2}=\theta _{j}\text{ with }\sigma _{i}\text{%
-probability one.}  \label{k4}
\end{equation}%
Finally, (\ref{k2}) and (\ref{k4}) imply $\tilde{m}_{i}^{2}=m_{i}^{2}$,
contradicting (\ref{k3}).
\end{proof}

\subsubsection*{Step 3: No Challenge}

\label{no challenge}

By Claim \ref{pc1}, there exists a common state $\tilde{\theta}\in \Theta $
with $\sigma _{i}$-probability one for every agent $i \in \mathcal{I}$. We
now show in Claim \ref{pc2} that no one challenges the common state $\tilde{%
\theta}$.

\begin{claim}
\label{pc2}No agent challenges with positive probability the common state $%
\tilde{\theta}$ announced in the second report.
\end{claim}

\begin{proof}
Suppose by way of contradiction that $x(\tilde{\theta},m_{i}^{1})\neq f(%
\tilde{\theta})$ for some message $m_{i}\in $supp$\left( \sigma _{i}\right) $%
. By Claim \ref{pc1}, we have $x(m_{j}^{2},m_{i}^{1})\neq f(m_{j}^{2})$ for
every message $m_{j}\in $supp$\left( \sigma _{j}\right) $ and every agent $%
j\neq i$. It implies that $e_{j,i}\left( m_{j},m_{i}\right) =\varepsilon $
with $\sigma _{j}$-probability one for every $j\neq i$ and $m_{j}\in %
\mbox{supp}(\sigma _{j})$. By Claim \ref{c0}, we have $m_{j}^{1}=\theta _{j}$
with $\sigma _{j}$-probability one and $m_{i}^{1}=\theta _{i}$. Thus, we
obtain $x(\tilde{\theta},\theta _{i})\neq f(\tilde{\theta})$. By the
construction of the best challenge scheme, we also have $x(\tilde{\theta}%
,\theta _{i})\in \mathcal{L}_{i}(f(\tilde{\theta}),\tilde{\theta}_{i})\cap 
\mathcal{SU}_{i}(f(\tilde{\theta}),\theta _{i})$. Then, by (\ref{bw}), every
message $\bar{m}_{i}$ with $x(\tilde{\theta},\bar{m}_{i}^{1})=f(\tilde{\theta%
})$ cannot be a best response against $\sigma _{-i}$. Indeed, since $x(%
\tilde{\theta},\theta _{i})\in \mathcal{SU}_{i}(f(\tilde{\theta}),\theta
_{i})$, it is a profitable deviation to replace $\bar{m}_{i}^{1}$ by $\theta
_{i}$. Hence, $x(\tilde{\theta},\tilde{m}_{i})\neq f(\tilde{\theta})$ and $%
e_{j,i}\left( m_{j},\tilde{m}_{i}\right) =\varepsilon $\ for every $\tilde{m}%
_{i}\in $supp$(\sigma _{i})$. Once again, by Claim \ref{c0}, we have $\tilde{%
m}_{i}^{1}=\theta _{i}$ with $\sigma _{i}$-probability one. Therefore, every
agent's first report is truthful with probability one. By Claim \ref{1st},
we conclude that $\tilde{\theta}=\theta $. Since $x(\tilde{\theta},\theta
_{i})\neq f(\tilde{\theta})$, it follows that $x(\tilde{\theta},\theta _{i})$
belongs to the empty intersection $\mathcal{L}_{i}(f(\theta ),\theta
_{i})\cap \mathcal{SU}_{i}(f(\theta ),\theta _{i})$, which is impossible.
\end{proof}

\subsection{Implementation in a Direct Mechanism}

\label{sec:drm}

In this section we present two special cases in which our implementing
mechanism can be made into a direct mechanism. Both cases require three or
more agents. A direct (revelation) mechanism is a mechanism $\left(
(M_{i}),g,(\tau _{i})\right) _{i\in \mathcal{I}}$ in which (i) agents are
asked to report the state (i.e., $M_{i}=\Theta $ for every agent $i$), and
(ii) a unanimous report leads to the socially desirable outcome with no
transfers (i.e., $g\left( \theta ,...,\theta \right) =f\left( \theta \right)
,$ and $\tau _{i}\left( \theta \right) =0,$ for every $i\in \mathcal{I}$ and 
$\theta \in \Theta $). Our notion of direct mechanism is adopted in, for
example, \cite{DS91} and \cite[Definition 179.2]{or} both of which ask each
agent to report a state.

Although direct mechanisms invoke a simpler message space than the augmented
mechanisms used in the full implementation literature, the literature on
partial implementation has attempted to construct mechanisms that are
simpler or easier to implement than direct mechanisms, allowing lotteries
and transfers. See, for example, \cite{DM00} and \cite{PR02}. While our
result complements these papers, our main focus is to study full
implementation in mixed-strategy Nash equilibrium without making use of
integer or modulo games.

The first case shows that every Maskin-monotonic SCF is fully implementable
in pure-strategy Nash equilibria in a direct mechanism. Pure-strategy Nash
implementation means that we only require that each pure-strategy Nash
equilibrium achieve desirable outcomes, i.e., condition (ii) of Definition %
\ref{def-nash} holds only for pure-strategy Nash equilibria. Indeed, one
might expect that by penalizing disagreement with transfers, the designer
can easily obtain a unanimous state announcement without using
integer/modulo games. Once there is a unanimous state announcement in
equilibrium, Maskin monotonicity will ensure implementation, as it does in 
\cite{maskin99}. The following proposition formalizes the idea; see Appendix %
\ref{app:prop:pure} for a proof.

\begin{proposition}
\label{prop:pure}Suppose that there are at least three agents and the SCF $f$
satisfies Maskin monotonicity. Then, $f$ is implementable in pure-strategy
Nash equilibria by a direct mechanism.
\end{proposition}

The idea of \textquotedblleft penalizing disagreement" becomes problematic
once we consider mixed-strategy equilibria. Indeed, the direct mechanism
which we construct in proving Proposition \ref{prop:pure} is reminiscent of
modulo games, which, as is well known, admit unwanted mixed-strategy
equilibria. Thus, it should come at no surprise that the direct mechanism
also admits unwanted mixed-strategy equilibria.

The second case establishes mixed-strategy Nash implementation in direct
mechanisms by considering a state space of a \textquotedblleft product
form,\textquotedblright\ i.e., there is a one-to-one correspondence between $%
\Theta $ and $\times _{i=1}^{I}\Theta _{i}$. We state the following result
and relegate its proof to Appendix \ref{app:prop:drm-product}:

\begin{proposition}
\label{prop:drm-product}Suppose that there are at least three agents, there
is a one-to-one correspondence between $\Theta $ and $\times
_{i=1}^{I}\Theta _{i}$, and the SCF $f$ satisfies Maskin monotonicity. Then, 
$f$ is implementable in mixed-strategy Nash equilibria by a direct mechanism.
\end{proposition}

Proposition \ref{prop:drm-product} represents an extreme case in which
mixed-strategy Nash implementation can be achieved in a direct mechanism.
Product state space naturally arises in a Bayesian setup with a full-support
common prior. While such a full-support prior is precluded by the
complete-information assumption, it is consistent with \textquotedblleft
almost complete information" which we will introduce in Section \ref{perturb}%
.

\subsection{Implementation without Off-the-Equilibrium Transfers\label{trans}%
}

The following example illustrates the fact of that without any domain
restriction such as quasilinear preferences with transfers, some
Maskin-monotonic SCF cannot be implemented by mixed-strategy Nash equilibria
in finite mechanisms.

\begin{example}[Example 4 of Jackson (1992)]
\label{example}Consider the environment with two agents $1$ and $2$. Suppose
that there are four alternatives $a,b,c$, and $d$ and two states $\theta $
and $\theta ^{\prime }$. Suppose that agent 1 has the state-independent
preference $a\succ _{1}b\succ _{1}c\sim _{1}d,$ and agent $2$ has the
preference $a\succ _{2}^{\theta }b\succ _{2}^{\theta }d\succ _{2}^{\theta }c$
at state $\theta $ and preference $b\succ _{2}^{\theta ^{\prime }}a\succ
_{2}^{\theta ^{\prime }}c\sim _{2}^{\theta ^{\prime }}d$ at state $\theta
^{\prime }$. Consider the SCF $f$ such that $f\left( \theta \right) =a$ and $%
f\left( \theta ^{\prime }\right) =c$.
\end{example}

With no restrictions on agents' preferences, \cite{jackson92} shows that for
every finite mechanism which implements $f$ in pure-strategy Nash
equilibria, there must also exist a \textquotedblleft bad" mixed-strategy
Nash equilibrium such that at state $\theta ^{\prime }$ the equilibrium
outcome differs from $c$ with positive probability.\footnote{%
We briefly recap the argument here. Let $\mathcal{M}$ be a finite mechanism
which implements the SCF $f$ in pure-strategy Nash equilibria. Consider a
mechanism which restricts the message space of $\mathcal{M}$ such that,
against any message of agent $i$, the opponent agent $j$ can choose a
message that induces either outcome $a$ or $b$. The restricted set of
messages is nonempty since the equilibrium message profile at state $\theta $
leads to outcome $a$. It follows that at state $\theta ^{\prime }$, the game
induced by the restricted mechanism must have a mixed-strategy Nash
equilibrium. Moreover, the equilibrium outcome must be $a$ or $b$ with
positive probability; otherwise, agent 2 can deviate to induce outcome $a$
or $b$ with positive probability. Since $c$ and $d$ are ranked lowest by
both agents at state $\theta ^{\prime }$, the mixed-strategy equilibrium
must remain an equilibrium at state $\theta ^{\prime }$ in the game induced
by $\mathcal{M}$; moreover, the equilibrium fails to achieve $f\left( \theta
^{\prime }\right) =c$.} Since $f$ satisfies Maskin monotonicity, the example
shows that without imposing any domain restrictions on the environment, it
is impossible to implement any Maskin-monotonic SCF in mixed-strategy
equilibria by a finite mechanism. However, regardless of the cardinal
representation of the preferences in Jackson's example, the SCF $f$ can
actually be implemented in mixed-strategy equilibria with arbitrarily small
transfers off the equilibrium; for more discussion, see Section \ref{small}
and in particular, footnote \ref{small2}.

\section{Extensions}

\label{extensions}

We now establish several extensions of our main result (Theorem \ref{Nash}).
In Section \ref{perturb}, we show that the implementation result is robust
to information perturbations. That is, we establish that our implementation
result remains valid in any incomplete-information environment that is close
to\ our complete-information benchmark. In Section \ref{SCC}, we extend our
result to the case of social choice correspondences (henceforth, SCCs).
Section \ref{small} clarifies how the designer can modify the implementing
mechanism to make the size of transfers arbitrarily small. For the sake of
clarity, we will not discuss any combination of multiple extensions. For
instance, we will study the case of SCCs only in Section \ref{SCC} but focus
entirely on SCFs in the rest of the paper.

The extensions involve more technical details. Thus, we assume, in this
section, that the set $A$ (of pure alternatives) is finite and relegate all
the proofs to the appendix.

\subsection{Robustness to Information Perturbations}

\label{perturb} 

\cite{CE03} and \cite{AFHKT} consider a designer who not only wants all
equilibria of her mechanism to yield a desirable outcome under complete
information, but is also concerned about the possibility that agents may
entertain small doubts about the true state. They argue that such a designer
should insist on implementing the SCF in the closure of a solution concept
as the amount of incomplete information about the state vanishes. \cite{CE03}
adopt undominated Nash equilibrium and \cite{AFHKT} adopt subgame-perfect
equilibrium as a solution concept in studying the robustness issue.

To allow for information perturbations, suppose that the agents do not
observe the state directly but are informed of the state via signals. The
set of agent $i$'s signals is denoted as $S_{i}$, which is identified with $%
\Theta $, i.e., $S_{i}\equiv \Theta $.\footnote{%
We adopt the formulation from \cite{CE03} and \cite{AFHKT}. Our result holds
for any alternative formulation under which the (Bayesian) Nash equilibrium
correspondence has a closed graph.} A signal profile is an element $s=\left(
s_{1},...,s_{I}\right) \in S\equiv \times _{i\in \mathcal{I}}S_{i}.$ When
the realized signal profile is $s$, agent $i$ observes only his own signal $%
s_{i}$. Let $s_{i}^{\theta }$ denote the signal which corresponds to state $%
\theta ,$ and we write $s^{\theta }=\left( s_{i}^{\theta }\right) _{i\in I}$%
. State and signals are drawn from some prior distribution over $\Theta
\times S$. In particular, complete information can be modelled as a prior $%
\mu $ such that $\mu \left( \theta ,s\right) =0$ whenever $s\not=s^{\theta }$%
. Such a $\mu $ will be called a \emph{complete-information prior}. We
assume that for each agent $i$, the marginal distribution on $i$'s signals
places a strictly positive weight on each of $i$'s signals, that is, marg$%
_{S_{i}}\mu \left( s_{i}\right) >0$ for every $s_{i}\in S_{i}$, so that the
posterior belief given every signal is well defined. For every prior $\nu $,
we also write $\nu \left( \cdot |s_{i}\right) $ for the conditional
distribution of $\nu $ on signal $s_{i}.$

The distance between two priors is measured by the uniform metric. That is,
for every two priors $\mu $ and $\nu ,$ we have $d\left( \mu ,\nu \right)
\equiv \max_{\theta ,s}\left\vert \mu \left( \theta ,s\right) -\nu \left(
\theta ,s\right) \right\vert $. Write $\nu ^{\varepsilon }\rightarrow \mu \ $%
if $d\left( \nu ^{\varepsilon },\mu \right) \rightarrow 0$ as $\varepsilon
\rightarrow 0$. A prior $\nu $ together with a mechanism $\mathcal{M}%
=((M_{i},\tau _{i})_{i\in \mathcal{I}},g)$ induces an incomplete-information
game, which we denote by $\Gamma \left( \mathcal{M},\nu \right) $. A
(mixed-)strategy of agent $i$ is now a mapping $\sigma _{i}:S_{i}\rightarrow
\Delta \left( M_{i}\right) $.

The designer may resort to a solution concept $\mathcal{E}$ for the game $%
\Gamma \left( \mathcal{M},\nu \right) $ (such as Bayesian Nash equilibrium)
which induces a set of mappings from $\Theta \times S$ to $X$, which we call 
\emph{acts}, following \cite{CE03}. For instance, each Bayesian Nash
equilibrium $\sigma $ induces the act $\alpha _{\sigma }$ with $\alpha
_{\sigma }\left( \theta ,s\right) \equiv \sigma \left( s\right) \circ \left(
g,\left( \tau _{i}\right) _{i\in \mathcal{I}}\right) ^{-1},$ where we abuse
the notation to identify the finite-support distribution $\sigma \left(
s\right) \circ \left( g,\left( \tau _{i}\right) _{i\in \mathcal{I}}\right)
^{-1}$ on $X$ with an allocation in $X$. We denote the set of acts induced
by the solution concept $\mathcal{E}$ as $\mathcal{E}\left( \mathcal{M},\nu
\right) $. We endow $X$ with a topology with respect to which the utility
function $u_{i}$ is continuous on $X$.\footnote{%
For instance, it is the case if $A$ is a (Hausdorff) topological space, $%
v_{i}\left( a,\theta \right) $ is bounded and continuous in $a$, and $\Delta
\left( A\right) $ is endowed with the weak$^{\ast }$-topology. Then, $%
X\equiv \Delta \left( A\right) \times 
\mathbb{R}
^{I}$, endowed with the product topology, is also a Hausdorff topological
space.} We now define $\overline{\mathcal{E}}$-implementation.

\begin{definition}
\label{defc}An SCF $f$ is $\overline{\mathcal{E}}$-implementable under the
complete-information prior $%
\mu
$ if there exists a mechanism $\mathcal{M}=\left( (M_{i}),g,(\tau
_{i})\right) _{i\in \mathcal{I}}$ such that for every $\left( \theta
,s\right) \in $ supp$\left( \mu \right) $ and every sequence of priors $%
\{\nu ^{n}\}$ converging to $%
\mu
$, the following two requirements hold: (i) there is a sequence of acts $%
\left\{ \alpha _{n}\right\} $ with $\alpha _{n}\in \mathcal{E}\left( 
\mathcal{M},\nu _{n}\right) $ such that $\alpha _{n}\left( \theta ,s\right)
\rightarrow f(\theta )$; and (ii) for every sequence of acts $\left\{ \alpha
_{n}\right\} $ with $\alpha _{n}\in \mathcal{E}\left( \mathcal{M},\nu
_{n}\right) $, we have $\alpha _{n}\left( \theta ,s\right) \rightarrow
f(\theta )$.
\end{definition}

\cite{CE03} and \cite{AFHKT} show that Maskin monotonicity is a necessary
condition for $\overline{UNE}$-implementation and $\overline{SPE}$%
-implementation, respectively.\footnote{\cite{AFHKT} adopt sequential
equilibrium as the solution concept for the incomplete-information game $%
\Gamma \left( \mathcal{M},\nu \right) $.} The result of \cite{CE03} implies
that implementation of a non-Maskin-monotonic SCF in undominated Nash
equilibria such as the result in \cite{AM94} is necessarily vulnerable to
information perturbations. Moreover, both \citet[Theorem 2]{CE03} and \cite%
{AFHKT} establish the sufficiency result by using an infinite mechanism with
an integer game and restricting attention to pure-strategy equilibria. It
raises the question as to whether their robustness test may be too demanding
when it is applied to finite mechanisms where mixed-strategy equilibria have
to be taken seriously, since the implementing mechanism of \cite{JPS94},
that of \cite{AM94}, or the simple mechanism in Section 5 of \cite{MR88} are
considered examples of such finite mechanisms.

The canonical mechanism which we propose in the proof of Theorem \ref{Nash}
is indeed finite, and we show that that finite mechanism implements every
Maskin-monotonic SCF in mixed-strategy Nash equilibria. Since the solution
concept of Bayesian Nash equilibrium, viewed as a correspondence on priors,
has a closed graph, that finite mechanism also achieves $\overline{NE}$%
-implementation. We now obtain the following result as a corollary of
Theorem \ref{Nash} in our setup with lotteries and transfers.

\begin{proposition}
\label{prop:ne-bar}Let $\mathcal{E}$ be a solution concept such that $%
\varnothing \neq \mathcal{E}\left( \mathcal{M},%
\mu
\right) \subseteq NE\left( \mathcal{M},%
\mu
\right) $ for each finite mechanism $\mathcal{M}$ and a complete-information
prior $%
\mu
$. Then, every Maskin-monotonic SCF $f$ is $\overline{\mathcal{E}}$%
-implementable.
\end{proposition}

The condition $\varnothing \neq \mathcal{E}\left( \mathcal{M},%
\mu
\right) \subseteq NE\left( \mathcal{M},%
\mu
\right) $ is satisfied for virtually every refinement of Nash equilibrium,
because we allow for mixed-strategy equilibria and $\Gamma \left( \mathcal{M}%
,%
\mu
\right) $ is a finite game.

\subsection{Social Choice Correspondences}

\label{SCC}

A large portion of the implementation literature strives to deal with social
choice correspondences (hereafter, SCCs), i.e., multi-valued social choice
rules. In this section, we extend our Nash implementation result to cover
the case of SCCs. We suppose that the designer's objective is specified by
an SCC $F:\Theta \rightrightarrows X$; and for simplicity, we assume that $%
F\left( \theta \right) $ is a finite set for each state $\theta \in \Theta $%
. It includes the special case where the co-domain of $F$ is $A$. Following 
\cite{maskin99}, we first define the notion of Nash implementation for an
SCC.

\begin{definition}
\label{Def-implement-SCC}An SCC $F$ is \textbf{implementable in
mixed-strategy Nash equilibria} \textbf{by a finite mechanism} if there
exists a mechanism $\mathcal{M}=\left( (M_{i},\tau _{i})_{i\in \mathcal{I}%
},g\right) $ such that for every state $\theta \in \Theta $, the following
two conditions are satisfied: (i) for every $x\in F(\theta )$, there exists
a pure-strategy Nash equilibrium $m\ $in the game $\Gamma (\mathcal{M}%
,\theta )$ with $g(m)=x$ and $\tau _{i}(m)=0$ for every agent $i\in \mathcal{%
I}$; and (ii) for every $m\in \mbox{supp}\ (NE(\Gamma (\mathcal{M},\theta
))) $, we have $\mbox{supp}(g(m))\subseteq F\left( \theta \right) $ and $%
\tau _{i}\left( m\right) =0$ for every agent $i\in \mathcal{I}$.
\end{definition}

Second, we state the definition of Maskin monotonicity for an SCC.

\begin{definition}
\label{mono-SCC}An SCC $F$ satisfies \textbf{Maskin monotonicity} if for
each pair of states $\tilde{\theta}$ and $\theta $ and $z\in F(\tilde{\theta}%
)\backslash F\left( \theta \right) $, some agent $i\in \mathcal{I}$ and some
allocation $z^{\prime }\in X$ exist such that 
\begin{equation*}
\tilde{u}_{i}(z^{\prime },\tilde{\theta})\leq \tilde{u}_{i}(z,\tilde{\theta})%
\text{ and }\tilde{u}_{i}(z^{\prime },\theta )>\tilde{u}_{i}(z,\theta )\text{%
.}
\end{equation*}
\end{definition}

We now state our Nash implementation result for SCCs and relegate the proof
to Appendix \ref{NashC}.\footnote{%
When there are only two agents, we can still show that every
Maskin-monotonic SCC $F$ is \textit{weakly} implementable in Nash
equilibria, that is, there exists a mechanism which has a pure-strategy Nash
equilibrium and satisfies requirement (ii) in Definition \ref%
{Def-implement-SCC}.}

\begin{theorem}
\label{NashC}Suppose there are at least three agents. An SCC $F$ is
implementable in mixed-strategy Nash equilibria by a finite mechanism if and
only if it satisfies Maskin monotonicity.
\end{theorem}

Compared with Theorem \ref{Nash} for SCFs, Theorem \ref{NashC} needs to
overcome additional difficulties. In the case of SCFs, when the agents'
second reports are consistent at a common state $\tilde{\theta}$, they will
be associated with a single outcome $f(\tilde{\theta})$. Hence, if agent $i$%
's second report is challenged, then \emph{every} second report which is
played with positive probability by any agent must also be challenged in
equilibrium. Together with Claim \ref{c0}, it implies that every agent must
tell the truth in their first and second report, which leads to a
contradiction in the proof of Claim \ref{pc2}.

In the case of SCCs, each allocation $x\in F\left( \theta \right) $ has to
be implemented as the outcome of some pure-strategy equilibrium. Hence, each
agent must also report an allocation to be implemented. It also follows that
a challenge scheme for an SCC must be defined for a type $\theta _{i}$ to
challenge a pair $(\tilde{\theta},x)$ with $x\in F(\tilde{\theta})$. As a
result, even when the agents' second reports are consistent at state $\tilde{%
\theta}$ (which still holds by Claim \ref{pc1}), they might still be
randomizing between two allocations $x$ and $x^{\prime }$ in $F(\tilde{\theta%
})$ such that $(\tilde{\theta},x)$ is challenged and yet $(\tilde{\theta}%
,x^{\prime })$ is not. Hence, we cannot follow a similar argument as in
Claim \ref{c0} to derive a contradiction. Instead, we build on the
implementing mechanism in Section \ref{M} and show that agent $i$ will not
report $(\tilde{\theta},x)$ which can be challenged either by (i) agent $%
j\neq i$ or by (ii) agent $i$ himself. We deal with Case (i) by imposing a
large penalty on agent $i$ conditional on agent $j$'s challenging $(\tilde{%
\theta},x)$, whereas we deal with Case (ii) by allowing agent $i$ to
challenge himself without having to pay the penalty.

\noindent \textbf{Remark.} \cite{MR12finite} also consider deterministic
SCCs in a separable environment studied in \cite{JPS94}. \cite{MR12finite}
identify a condition (which they call top-$D$ inclusiveness) under which an
SCC is implementable in mixed-strategy Nash equilibria in finite mechanisms
if and only if it satisfies \textit{set-monotonicity} (proposed by \cite%
{MR12}). There are several differences between our Theorem \ref{NashC} and
their result. First, \cite{MR12finite} require only the existence of
mixed-strategy equilibria but we follow \cite{maskin99} in requiring the
existence of pure-strategy equilibria in part (i) of Definition \ref%
{Def-implement-SCC}. Second, \cite{MR12finite} consider an ordinal setting,
while we consider a cardinal setting. These two features of \cite{MR12finite}
are the reason why they use set-monotonicity as a necessary condition for
characterizing their ordinal Nash implementation.\footnote{%
In \cite{MAMwp}, we study the concept of ordinal Nash implementation
proposed by \cite{MR12}. The notion requires that the implementing mechanism
achieve mixed-strategy Nash implementation for every cardinal representation
of preferences over lotteries. We show that ordinal almost monotonicity, as
defined in \cite{sanver06}, is a necessary and sufficient condition for
ordinal Nash implementation.} Third, our quasilinear environments with
transfers are more restrictive than the separable environments considered by 
\cite{MR12finite}. Finally, \cite{MR12finite} need \textquotedblleft top $D$%
-inclusiveness\textquotedblright\ as an additional condition, which requires
that there exist at least one agent for whom the SCC contains the agent's
best outcome within the range of the SCC for every state of the world,
whereas we impose no conditions beyond Maskin monotonicity for the SCC.

\subsection{Small Transfers}

\label{small}

One potential drawback of the mechanism we propose for Theorem \ref{Nash} is
that the size of transfers may be large. To tackle the problem, we use the
technique introduced by \cite{AM94} to show that if the SCF satisfies Maskin
monotonicity in the restricted domain without any transfer, then it is
Nash-implementable with arbitrarily small transfers.

We first propose a notion of Nash implementation with bounded transfers off
the equilibrium and still no transfers on the equilibrium.

\begin{definition}
An SCF $f:\Theta \rightarrow \Delta \left( A\right) $ is implementable in
mixed-strategy Nash equilibria by a finite mechanism \textbf{with transfers
bounded by $\bar{\tau}$} if there exists a finite mechanism $\mathcal{M}%
=\left( (M_{i},\tau _{i})_{i\in \mathcal{I}},g\right) $ such that for every
state $\theta \in \Theta $ and $m\in M$, (i) there exists a pure-strategy
Nash equilibrium in the game $\Gamma (\mathcal{M},\theta )$; (ii) for each $%
m\ $in $\mbox{supp}\ (NE(\Gamma (\mathcal{M},\theta )))$, we have $g\left(
m\right) =f\left( \theta \right) $ and $\tau _{i}\left( m\right) =0$ for
every agent $i\in \mathcal{I}$; and (iii) $|\tau _{i}\left( m\right) |\leq 
\bar{\tau}$ for every $m\in M$ and every agent $i\in \mathcal{I}.$
\end{definition}

Next, we propose a notion of Nash implementation in which there are no
transfers on the equilibrium and only arbitrarily small transfers off the
equilibrium.

\begin{definition}
An SCF $f$ is implementable in mixed-strategy Nash equilibria by a finite
mechanism \textbf{with arbitrarily small transfers} if, for every $\bar{\tau}%
>0,$ the SCF $f$ is implementable in Nash equilibria by a finite mechanism
with transfers bounded by $\bar{\tau}.$
\end{definition}

We say that an SCF $f$ satisfies Maskin monotonicity in the restricted
domain $\Delta \left( A\right) $ if $f(\tilde{\theta})\neq f\left( \theta
\right) $ implies that there are an agent $i$ and some lottery $x(\tilde{%
\theta},\theta _{i})$ in $\Delta \left( A\right) $ such that $x(\tilde{\theta%
},\theta _{i})\ $belongs to $\mathcal{L}_{i}(f(\tilde{\theta}),\tilde{\theta}%
_{i})\cap \mathcal{SU}_{i}(f(\tilde{\theta}),\theta _{i})$. Here, for $%
\left( \ell ,\theta _{i}\right) \in \Delta \left( A\right) \times \Theta
_{i} $, we use $\mathcal{L}_{i}\left( \ell ,\theta _{i}\right) $ to denote
the lower-contour set at allocation $\ell $ in $\Delta \left( A\right) $ for
type $\theta _{i}$, i.e.,%
\begin{equation*}
\mathcal{L}_{i}\left( \ell ,\theta _{i}\right) =\left\{ \ell ^{\prime }\in
\Delta (A):v_{i}\left( \ell ,\theta \right) \geq v_{i}(\ell ^{\prime
},\theta )\right\} .
\end{equation*}%
In a similar fashion, $\mathcal{SU}_{i}\ $is defined. Clearly, Maskin
monotonicity in the restricted domain $\Delta \left( A\right) $ is stronger
than Maskin monotonicity in the domain $X$, as the former requires that the
test allocation be a lottery over alternatives without transfer. In Appendix %
\ref{app-small}, we assume there are at least three agents, and prove the
following result.\footnote{\label{small2}\noindent In the case with only two
agents, Theorem 3 still holds if there exists an alternative $w\in A$ which
is the worst alternative for any agent at any state. In that case, we can
simply modify the \textquotedblleft voting rule\textquotedblright\ $\phi $
in the proof of Theorem 3 to be $\phi \left( m^{h}\right) =f(\tilde{\theta})$
if both agents announce a common type profile which identifies a state $%
\tilde{\theta}$ in $m^{h}$; and $\phi \left( m^{h}\right) =w$ otherwise. In
particular, $w=c$ in Example 4 of Jackson (1992) and thus the SCF can be
implemented with arbitrarily small transfers. Moreover, the conclusion holds
regardless of the utility representation of the agents' preferences.
However, note that we assume that agents have quasilinear utilities while
Jackson's example does not make such an assumption.}

\begin{theorem}
\label{Thm:small-transfer}Suppose there are at least three agents. An SCF $%
f_{A}:\Theta \rightarrow \Delta (A)$ is implementable in mixed-strategy Nash
equilibria by a finite mechanism with arbitrarily small transfers if $f_{A}$
satisfies Maskin monotonicity in the restricted domain.
\end{theorem}

\subsection{Infinite State Space}

\label{infinite}

One significant assumption we have made in this paper is that the state
space is finite. In Appendix \ref{Infi}, we extend Theorem \ref{Nash} to an
infinite state space in which the agents' utility functions are continuous.
A similar extension was raised as an open question for virtual
implementation in \cite{AM92} (see their Section 5) and it has not been
answered to our knowledge.

In Appendix \ref{Infi}, we construct an extension of the implementing
mechanism for mixed-strategy Nash implementation which accommodates an
infinite state space. We state the result as follows:

\begin{theorem}
\label{infiNash}Suppose that $\Theta $ is a Polish space. Then, an SCF $f$
satisfies Maskin monotonicity if and only if there exists a mechanism which
implements $f$ in mixed-strategy Nash equilibria. Moreover, if $\Theta $ is
compact and both the utility function $\left\{ v_{i}\left( a,\cdot \right)
\right\} _{a\in A}$ and the SCF are continuous functions on $\Theta $, then
the implementing mechanism has a compact message space together with a
continuous outcome function and continuous transfer rules.
\end{theorem}

One notable feature of this extension is that as long as the setting is the
one with a compact state space, the continuous SCF, and continuous utility
functions, the resulting implementing mechanism will also be compact and
continuous. This feature ensures that best responses are always well defined
in our mechanism; hence, it differentiates our construction from the
traditional one invoking the use of integer games.

The proof of Theorem \ref{infiNash} needs to overcome two difficulties.
First, in a finite state space, the transfer rules $\tau _{i,j}^{1}$ and $%
\tau _{i,j}^{2}$ which we define in (\ref{money1}) and (\ref{money2}) impose
either a large penalty and/or a large reward as long as the designer sees a
discrepancy in the agents' announcements. With a continuum of states/types,
however, such a drastic change in transfer scale is precluded by the
continuity requirement. Hence, our first challenge is to suitably define $%
\tau _{i,j}^{1}$ and $\tau _{i,j}^{2}$ so that they vary continuously yet
still incentivize them to tell the truth.

Second, in an infinite setting, we know of no way to construct a challenge
scheme which pre-selects a test allocation in a continuous manner. As a
result, we cannot have the agents report their type, let alone the true
type, to cast a challenge to state $\tilde{\theta}$. Instead, we will
restore the continuity of the outcome function by asking them to report a
test allocation $x$ directly. Despite this change, we will establish a
counterpart of Condition (\ref{bw}) as Lemmas \ref{infiworse} and \ref%
{infibetter} in Appendix \ref{Infi}.

\subsection{The Ordinal Approach}

\label{ordinal}

We have assumed that the agents are expected utility maximizers. This leaves
open the issue as to whether, and to what extent, our implementation result
depends on the designer's knowledge about the cardinalization of the agents'
preferences over lotteries. To address the issue, we discuss how our result
can accommodate an ordinal setting.

First, we introduce the notion of \emph{ordinal Nash implementation}. The
notion requires that the mixed-strategy Nash implementation be obtained for 
\emph{any} cardinal representation of the ordinal preferences over the
finite set of pure alternatives $A$. Formally, we follow the approach
proposed by \cite{MR12}.

Suppose that at state $\theta $, the agents commonly know only that their
ordinal rankings over the set of pure alternatives $A$. We write the induced
ordinal preference profile at state $\theta $ by $(\succeq _{i}^{\theta
})_{i\in \mathcal{I}}$. We also assume no redundancy, i.e., $\theta \neq
\theta ^{\prime }$ implies $\succeq _{i}^{\theta }\not=$ $\succeq
_{i}^{\theta ^{\prime }}$ for some agent $i$. It is taken for granted that
distinct pair of $\succeq _{i}^{\theta }$and $\succeq _{i}^{\theta ^{\prime
}}$induce different preference orderings over $A,$ and also that no players
are indifferent over all elements of $A.$ For each $i\in \mathcal{I}$, let $%
v_{i}:A\times \Theta \rightarrow \mathbb{R}$ be a \textit{cardinal
representation} of $(\succeq _{i}^{\theta })_{\theta \in \Theta }$ over $A$,
i.e., for each pair of alternatives $a,a^{\prime }\in A$, agent $i\in 
\mathcal{I}$, and state $\theta \in \Theta $, we have 
\begin{equation*}
v_{i}(a,\theta )\geq v_{i}(a^{\prime },\theta )\Leftrightarrow a\succeq
_{i}^{\theta }a^{\prime }\text{.}
\end{equation*}%
We assume that each function $v_{i}$ takes a value in $\left[ 0,1\right] .$
In addition, each cardinal representation $v_{i}$ induces an expected
utility function on $\Delta \left( A\right) $ which, by abuse of notation,
we also denote by $v_{i}:\Delta (A)\times \Theta \rightarrow \mathbb{R}$. We
denote by $V_{i}^{\theta }$ the set of all cardinal representations $%
v_{i}\left( \cdot ,\theta \right) $ of $\succeq _{i}^{\theta }$. Following 
\cite{MR12}, we focus our discussion on the case of a deterministic SCF,
i.e., $f:\Theta \rightarrow A$. We say that an SCF $f$ is ordinally Nash
implementable if it is implementable in mixed-strategy Nash equilibria
independently of the cardinal representation. We formalize this idea in the
following definition.

\begin{definition}
An SCF $f$ is \textit{ordinally} Nash implementable if there exists a
mechanism $\mathcal{M}$ such that, for every $\theta \in \Theta $ and every
profile of cardinal representations $v=(v_{i})_{i\in \mathcal{I}}$ of $%
(\succeq _{i}^{\theta })_{i\in \mathcal{I},\theta \in \Theta }$, the
following two conditions are satisfied: (i) there exists a pure-strategy
Nash equilibrium $m$ in the game $\Gamma (\mathcal{M},\theta ,v)$ such that $%
g(m)=f\left( \theta \right) $ and $\tau _{i}(m)=0$ for every $i\in \mathcal{I%
}$; and (ii) for every $m\in \mbox{supp}(NE(\Gamma (\mathcal{M},\theta ,v)))$%
, we have $\mbox{supp}(g(m))=f(\theta )$ and $\tau _{i}(m)=0$ for every $%
i\in \mathcal{I}$.
\end{definition}

In Appendix \ref{aord}, we introduce the notion of \emph{ordinal almost
monotonicity} proposed by \cite{sanver06}. It roughly says that whenever the
SCF designates different outcomes at states $\theta $ and $\theta ^{\prime }$%
, there must be an agent with a \emph{deterministic} test allocation which
displays a suitable preference reversal with respect to the socially
desirable outcome at the two states. We obtain the following result:

\begin{theorem}
\label{ord}An SCF $f$ is ordinally Nash implementable if and only if it
satisfies ordinal almost monotonicity.
\end{theorem}

It is worth emphasizing that Theorem \ref{infiNash} becomes the key to
proving this result. We show that ordinal almost monotonicity of an SCF
implies Maskin monotonicity of the SCF consistently extended from $\Theta $
(now the set of ordinal preference profiles) to the set of cardinalizations.
Since the set of all cardinalizations extended from $\Theta $ becomes an
infinite state space, the extension of ordinal almost monotonicity renders
Theorem \ref{infiNash} applicable.

Instead of relying on ordinal almost monotonicity, \cite{MR12} propose a
notion called \textit{set-monotonicity} for SCCs. They show that the notion
of set-monotonicity is weaker than Maskin monotonicity and is necessary and
\textquotedblleft almost sufficient\textquotedblright\ in their notion of
implementation in mixed-strategy Nash equilibria. There are three further
differences between the results of \cite{MR12} and ours. First, \cite{MR12}
require only the existence of mixed-strategy equilibria but we follow \cite%
{maskin99} in requiring the existence of a pure-strategy equilibrium. This
difference makes our ordinal implementation notion more demanding than that
of \cite{MR12}. Second, we use monetary transfers, while \cite{MR12} do not.
Although ordinal almost monotonicity is weaker than set-monotonicity, by our
use of transfers, we obtain our stronger notion of ordinal mixed-strategy
Nash implementation \`{a} la \cite{maskin99} for the case of SCFs by means
of ordinal almost monotonicity. Finally, \cite{MR12} also study the case of
SCCs which we omit here.

\section{Concluding Remarks}

\label{conclusion}

Despite its tremendous success, implementation theory has also been
criticized on various fronts. A major critique is that the mechanisms used
to achieve full implementation are not \textquotedblleft
natural\textquotedblright . To address the critique, \cite{jackson92}
proposes to characterize the class of SCFs which can be fully implemented in
\textquotedblleft natural\textquotedblright\ mechanisms, even at the cost of
imposing domain restrictions.

We consider our results as an important step in advancing Jackson's research
program. Specifically, we propose to recast an implementation problem by
requiring that the implementing mechanism be finite/well-behaved and have no
unwanted mixed-strategy equilibria. Such requirements are to be anticipated,
when the setting of interest is indeed finite/well-behaved to start with. We
prove a first set of benchmark results on mixed-strategy Nash implementation
by considering environments with lotteries and transfers. We also show that
our results are robust to information perturbations and amenable to handling
prominent extensions such as SCCs, small transfers, infinite settings, and
ordinal settings.

There have been some past attempts at tackling implementation in Bayesian
(incomplete-information) environments, such as \cite{MR90} and \cite{ACG03}.
Both papers are considered as previous attempts at full Bayesian
implementation by well-behaved mechanisms in incomplete-information
environments with transfers. However, they do not consider mixed-strategy
equilibria and we are aware of no results on mixed-strategy Bayesian
implementation achieved by well-behaved mechanisms. The novelty of our paper
is to take care of mixed-strategy equilibria by a well-behaved mechanism. We
take complete-information environments with a restricted preference domain
as a natural starting point and leave for future research the important yet
more challenging extensions to incomplete-information environments.\footnote{%
Allowing for integer games, \cite{kunimoto2019} and \cite{SV2010}
characterize mixed-strategy Bayesian implementation in more general
environments.}

Our implementation results are obtained by imposing transfers off the
equilibrium. This feature is intimately related to the burgeoning literature
on repeated implementation, such as \cite{LS11} and \cite{MR17}, in which
continuation values can serve as transfers in our construction.\footnote{%
We thank Hamid Sabourian for drawing our attention to this point.} It is an
important future direction to investigate the formal connection.

\appendix

\section{Appendix}

\label{appendix}

In this Appendix, we provide the proofs omitted from the main body of the
paper.

\subsection{Proof of Lemma \protect\ref{BC}}

\label{app:BC}

First, we elaborate the proof of Lemma \ref{BC} here.

\begin{proof}
Consider a challenge scheme $\bar{x}(\cdot ,\cdot )$. First, we show that we
can modify $\bar{x}(\cdot ,\cdot )$ into a new challenge scheme $x(\cdot
,\cdot )$ such that 
\begin{equation}
x(\tilde{\theta},\theta _{i})\neq f(\tilde{\theta})\text{ and }x(\tilde{%
\theta},\theta _{i}^{\prime })\neq f(\tilde{\theta})\Rightarrow u_{i}(x(%
\tilde{\theta},\theta _{i}),\theta _{i})\geq u_{i}(x(\tilde{\theta},\theta
_{i}^{\prime }),\theta _{i})\text{.}  \label{b1}
\end{equation}%
To construct $x(\cdot ,\cdot )$, for each player $i$, we distinguish two
cases: (a) if $\bar{x}(\tilde{\theta},\theta _{i})=f(\tilde{\theta})$ for
all $\theta _{i}\in \Theta _{i}$, then set $x(\tilde{\theta},\theta _{i})=%
\bar{x}(\tilde{\theta},\theta _{i})=f(\tilde{\theta})$; (b) if $\bar{x}(%
\tilde{\theta},\theta _{i})\neq f(\tilde{\theta})$ for some $\theta _{i}\in
\Theta _{i}$, then define $x(\tilde{\theta},\theta _{i})\ $as the most
preferred allocation of type $\theta _{i}$ in the finite set 
\begin{equation*}
X(\tilde{\theta})=\left\{ \bar{x}(\tilde{\theta},\theta _{i}^{\prime
}):\theta _{i}^{\prime }\in \Theta _{i}\text{ and }\bar{x}(\tilde{\theta}%
,\theta _{i}^{\prime })\neq f(\tilde{\theta})\right\} .
\end{equation*}%
Since $\bar{x}(\tilde{\theta},\theta _{i}^{\prime })\in \mathcal{L}_{i}(f((%
\tilde{\theta}),\tilde{\theta}_{i})$, we have $u_{i}(x(\tilde{\theta},\theta
_{i}),\tilde{\theta}_{i})\leq u_{i}(f(\tilde{\theta}),\tilde{\theta}_{i})$;
moreover, since $x(\tilde{\theta},\theta _{i})\ $as the most preferred
allocation of type $\theta _{i}$ in $X(\tilde{\theta})$ and $\bar{x}(\tilde{%
\theta},\theta _{i})\in \mathcal{SU}_{i}(f(\tilde{\theta}),\theta _{i})$, it
follows that $u_{i}(x(\tilde{\theta},\theta _{i}),\theta _{i})>u_{i}(f\left( 
\tilde{\theta}\right) ,\theta _{i})$. In other words, $x(\cdot ,\cdot )$
remains a challenge scheme. Moreover, $x(\cdot ,\cdot )$ satisfies (\ref{b1}%
) by construction.

Next, for each state $\tilde{\theta}$ and type $\theta _{i}$, we show that $%
x(\cdot ,\cdot )$ satisfies (\ref{b}). We proceed by considering the
following two cases. First, suppose that $x(\tilde{\theta},\theta _{i})\neq
f(\tilde{\theta})$. Then, by (\ref{b1}), it suffices to consider type $%
\theta _{i}^{\prime }$ with $x(\tilde{\theta},\theta _{i}^{\prime })=f(%
\tilde{\theta})$. Since $x(\tilde{\theta},\theta _{i}^{\prime })=f(\tilde{%
\theta})$ and $x(\tilde{\theta},\theta _{i})\neq f(\tilde{\theta})$, then it
follows from $x(\tilde{\theta},\theta _{i})\in \mathcal{SU}_{i}(f(\tilde{%
\theta}),\theta _{i})$ that $u_{i}(x(\tilde{\theta},\theta _{i}),\theta
_{i})>u_{i}(x(\tilde{\theta},\theta _{i}^{\prime }),\theta _{i})$. Hence, (%
\ref{b}) holds. Second, suppose that $x(\tilde{\theta},\theta _{i})=f(\tilde{%
\theta})$. Then, it suffices to consider type $\theta _{i}^{\prime }$ with $%
x(\tilde{\theta},\theta _{i}^{\prime })\neq f(\tilde{\theta})$. Since $x(%
\tilde{\theta},\theta _{i})=f(\tilde{\theta})$, we have $\mathcal{L}_{i}(f(%
\tilde{\theta}),\tilde{\theta}_{i})\cap \mathcal{SU}_{i}(f(\tilde{\theta}%
),\theta _{i})=\varnothing $. Moreover, $x(\tilde{\theta},\theta
_{i}^{\prime })\neq f(\tilde{\theta})$ implies that $x(\tilde{\theta},\theta
_{i}^{\prime })\in \mathcal{L}_{i}(f(\tilde{\theta}),\tilde{\theta}_{i})$.
Hence, we must have $x(\tilde{\theta},\theta _{i}^{\prime })\notin $ $%
\mathcal{SU}_{i}(f(\tilde{\theta}),\theta _{i})$. That is, $u_{i}(x(\tilde{%
\theta},\theta _{i}),\theta _{i})\geq u_{i}(x(\tilde{\theta},\theta
_{i}^{\prime }),\theta _{i}),$ i.e., (\ref{b}) holds.
\end{proof}

\subsection{Proof of Proposition \protect\ref{prop:pure}}

\label{app:prop:pure}

To facilitate the comparison with \cite{maskin99}, we assume that there are
three or more agents and define the following direct mechanism, denoted by $%
\mathcal{M}^{\text{D}}$, according to three rules:\vspace{0pt}

\noindent \textbf{Rule 1.} If there exists state $\tilde{\theta}$ such that
every agent announces $\tilde{\theta}$, then implement the outcome $f(\tilde{%
\theta})$.

\noindent \textbf{Rule 2.} If there exists state $\tilde{\theta}$ such that
everyone except agent $i$ announces $\tilde{\theta}$ and agent $i$ announces 
$\tilde{\theta}^{\prime }$, then implement a test allocation $x(\tilde{\theta%
},\tilde{\theta}_{i}^{\prime })$ for agent $i$ and the ordered pair of
states $(\tilde{\theta},\tilde{\theta}^{\prime })$; and if there is no such
test allocation, implement $f(\tilde{\theta})$. Moreover, charge agent $i+1$
(mod $I$) a large penalty $2\eta $, where the scale $\eta $ dominates any
difference in utility from allocation.

\noindent \textbf{Rule 3.} Otherwise, implement $f(m_{1})$. Moreover, charge
each agent $i$ a penalty of $\eta $ if $i$ reports a state which is not
reported by the unique majority (i.e., $\left\{ m_{i}\right\} \neq \arg
\max_{\tilde{\theta}}|\{j\in \mathcal{I}:m_{j}=\tilde{\theta}\}|$).\footnote{%
Note that Rule 3 penalizes every agent by $\eta ,$ if each of them reports a
different state.}

Now let the true state be $\theta .$

It follows from Rule 2 that since $\theta $ is the true state, $x(\theta ,%
\tilde{\theta}_{i}^{\prime })\neq f(\theta )$ implies that $x(\theta ,\tilde{%
\theta}_{i}^{\prime })\in \mathcal{L}_{i}(f(\theta ),\theta _{i})$. Hence,
everyone reporting the true state constitutes a pure-strategy Nash
equilibrium.

Now fix an arbitrary pure-strategy Nash equilibrium $m$. First, we claim
that $m$ cannot trigger Rule 2. Suppose that Rule 2 is triggered, and let
agent $i$ be the odd man out. Then, agent $i+1$ finds it strictly profitable
to deviate to announce $m_{i}$. After such a deviation, since $I\geq 3$,
either Rule 3 is triggered or it remains in Rule 2, but agent $i$ is no
longer the odd man out. Thus, agent $i+1$ saves at least $\eta $ (from
paying $2\eta $ to paying $\eta $ or $0$). Such a deviation may also change
the allocation selected by the outcome function $g\left( \cdot \right) $,
which induces utility change less than $\eta .$ Hence, agent $i+1$ strictly
prefers deviating to announce $m_{i}$, which contradicts the hypothesis that 
$m$ is a Nash equilibrium.

Second, we claim that $m$ cannot trigger Rule 3 either. Suppose that Rule 3
is triggered. Pick an arbitrary state reported by some (not necessarily
unique) majority of agents, i.e., $\hat{\theta}\in \arg \max_{\tilde{\theta}%
}|\{j\in I:m_{j}=\tilde{\theta}\}|$. Let $\mathcal{I}_{\hat{\theta}}$ be the
set of agents who report $\hat{\theta}$. Clearly, $\mathcal{I}_{\hat{\theta}%
}\subsetneq \mathcal{I}$, because Rule 3 (rather than Rule 1) is triggered.
Then, we can find an agent $i^{\ast }\in \mathcal{I}_{\hat{\theta}}$ such
that agent $i^{\ast }+1$ (mod $I$) is not in $I_{\hat{\theta}}$. Since agent 
$i^{\ast }+1$ does not belong to the unique majority, he must pay $\eta $
under $m$. Then, agent $i^{\ast }+1$ will strictly prefer deviating to
announce $m_{i^{\ast }}=\hat{\theta}$. After such a deviation, either Rule 3
is triggered, and agent $i^{\ast }+1$ falls in the unique majority who
reports $\hat{\theta}$; or Rule 2 is triggered, but agent $i^{\ast }$ cannot
be the odd man out. Thus, agent $i^{\ast }+1$ saves $\eta $ (from paying $%
\eta $ to paying $0$) and $\eta ^{\prime }$ is larger than the maximal
utility change induced by different allocations in $g\left( \cdot \right) $.
The existence of profitable deviation of agent $i^{\ast }+1$ contradicts the
hypothesis that $m$ is a Nash equilibrium.

Hence, we conclude that $m$ must trigger Rule 1. It follows that $f(\tilde{%
\theta})=f\left( \theta \right) $. Otherwise, by Maskin monotonicity, a
whistle blower can deviate to trigger Rule 2.

\subsection{Proof of Proposition \protect\ref{prop:drm-product}}

\label{app:prop:drm-product}

The proof is based on modifying the implementing mechanism and the proof of
Theorem \ref{Nash}. We only provide a sketch here. Set $M_{i}=M_{i}^{1}%
\times M_{i}^{2}$ where $M_{i}^{1}=\Theta _{i}$ and $M_{i}^{2}=\times
_{j\neq i}\Theta _{j}$. Since $I\geq 3$, the type of each agent is reported
by at least two agents in their second report. For each message profile $%
m=\left( m_{i}\right) _{i=1}^{I}$, denote by $\tilde{\Theta}\left( m\right) $
the set of state induced from the agents' second report, namely that $\tilde{%
\theta}\in \tilde{\Theta}\left( m\right) $ iff for every $i\in \mathcal{I}$,
we have $\tilde{\theta}_{i}=m_{j,i}^{2}$ for some agent $j\in \mathcal{I}$
(possibly $j=i$). Then, we modify the outcome function: 
\begin{equation*}
g\left( m\right) =\frac{1}{I\left\vert \tilde{\Theta}\left( m\right)
\right\vert }\sum_{i\in \mathcal{I}}\sum_{\tilde{\theta}\in \tilde{\Theta}%
\left( m\right) }\left[ e\left( m\right) \frac{1}{I}\sum_{j\in \mathcal{I}%
}y_{j}(m_{j}^{1})\oplus (1-e\left( m\right) )x(\tilde{\theta},m_{i}^{1})%
\right]
\end{equation*}%
where $e\left( m\right) =0$ if (i) $\tilde{\Theta}\left( m\right) $ contains
a unique state (consistency); and (ii) $x(\tilde{\theta},m_{i}^{1})=f(\tilde{%
\theta})$ for every agent $i$ and every $\tilde{\theta}\in \tilde{\Theta}%
\left( m\right) $ (no challenge); otherwise, $e\left( m\right) =\varepsilon $%
.\footnote{%
Here we do not have the case with $e\left( m\right) =1$ since $\Theta
=\times _{i=1}^{I}\Theta _{i}$ implies that $\tilde{\Theta}\left( m\right)
\subseteq \Theta $.} For the transfer rule, we define 
\begin{equation*}
\hat{\tau}_{i,j}^{1}\left( m_{i},m_{-i}\right) =\left\{ 
\begin{tabular}{ll}
$0$ & if $m_{i,j}^{2}=m_{k,j}^{2}$ for all $k\in \mathcal{I}\backslash
\left\{ i,j\right\} $; \\ 
$-\eta $ & if $m_{i,j}^{2}\not=m_{k,j}^{2}$ for some $k\in \mathcal{I}%
\backslash \left\{ i,j\right\} $ and $m_{i,j}^{2}\neq m_{j}^{1}$; \\ 
$\eta $ & if $m_{i,j}^{2}\not=m_{k,j}^{2}\text{ }$for some $k\in \mathcal{I}%
\backslash \left\{ i,j\right\} $ and $m_{i,j}^{2}=m_{j}^{1}$.%
\end{tabular}%
\right.
\end{equation*}%
Set $\tau _{i}(m)=\sum_{j\neq i}\hat{\tau}_{i,j}^{1}(m)$. As the agents no
longer report their own type in the second report, we do not need to define $%
\tau _{i,j}^{2}\left( \cdot \right) $.

The proof of implementation follows the same steps as the proof of Theorem %
\ref{Nash} and we only highlight the difference. First, for the contagion of
truth argument, we can only establish Claim \ref{1st}(a) because in the
modified mechanism, the agents no longer report their own type in the second
report so that we do not have rule $\tau _{i,j}^{2}\left( \cdot \right) $.
For the consistency argument, it turns out that Claim \ref{1st}(a) suffices.
Specifically, consider an arbitrary message $m_{i}\in $supp$\left( \sigma
_{i}\right) $ such that $m_{i}^{1}\neq \theta _{i}$.\textit{\vspace{0.2cm} }%
The same argument as in the proof of Claim \ref{pc1} implies that $\left(
m_{i},m_{-i}\right) $ is consistent for every $m_{-i}\in $supp$\left( \sigma
_{-i}\right) $. To show that $(\tilde{m}_{i},m_{-i})$ is consistent for any
other $\tilde{m}_{i}\in $supp$\left( \sigma _{i}\right) ,$ we make use of
the assumption that we have three or more agents. In particular, since $%
\left( m_{i},m_{-i}\right) $ is consistent for every $m_{-i}\in $supp$\left(
\sigma _{-i}\right) $, if $(\tilde{m}_{i},m_{-i})$ is inconsistent, it must
be $\tilde{m}_{i,k}^{2}\not=m_{j,k}^{2}$ for some $j\not=i$, $k\not=i$, and $%
k\not=j$. By Claim \ref{c0} agent $k$ must report his true type with
probability one. Then, it follows from Claim \ref{1st}(a) that $\tilde{m}%
_{i,k}^{2}=m_{j,k}^{2}$ with probability one and we have reached a
contradiction. The argument for no challenge remains the same.

\subsection{Proof of Claim \protect\ref{c0}}

\label{pc0}

Suppose that $m_{i}^{1}\neq \theta _{i}$ for some $m_{i}\in $supp$\left(
\sigma _{i}\right) $. Consider a message $\tilde{m}_{i}$ which differs from $%
m_{i}$ only in sending a truthful first report, i.e., $\tilde{m}%
_{i}^{1}=\theta _{i}$ and $\tilde{m}_{i}^{2}=m_{i}^{2}$. We prove the claim
by showing that $\tilde{m}_{i}$ is not a strictly better response than $%
m_{i} $ against $m_{j}$ only when $e_{i,j}\left( m_{i},m_{j}\right)
=e_{j,i}\left( m_{j},m_{i}\right) =0$. Recall that the first report of agent 
$i$ has no effect on his own transfer.

We consider first the case that the designer uses agent $j$'s report to
check agent $i$'s report. In that situation, the first report of agent $i$
has no effect on the function $e_{i,j}\left( \cdot ,m_{j}\right) $ for every 
$m_{j}$. Hence, we have $e_{i,j}\left( \tilde{m}_{i},m_{j}\right)
=e_{i,j}\left( m_{i},m_{j}\right) $. Moreover, if $m_{i}^{2}\notin \Theta $,
then $e_{i,j}\left( \tilde{m}_{i},m_{j}\right) =e_{i,j}\left(
m_{i},m_{j}\right) =1$; thus, by Lemma \ref{AM}, $\tilde{m}_{i}$ is a
strictly better response than $m_{i}$ against $m_{j}$. Hence, we may assume $%
m_{i}^{2}\in \Theta $ and consider the following two cases:

\noindent \emph{Case 1.1. }$e_{i,j}\left( \tilde{m}_{i},m_{j}\right)
=e_{i,j}\left( m_{i},m_{j}\right) =\varepsilon $.

\bigskip It follows from Lemmas \ref{AM} and \ref{BC} that 
\begin{equation*}
u_{i}(C_{i,j}^{\varepsilon }(\tilde{m}_{i},m_{j}),\theta
_{i})-u_{i}(C_{i,j}^{\varepsilon }(m_{i},m_{j}),\theta _{i})>0\text{.}
\end{equation*}%
Hence, $\tilde{m}_{i}$ is a strictly better response than $m_{i}$ against $%
m_{j}$.

\noindent \emph{Case 1.2. }$e_{i,j}\left( \tilde{m}_{i},m_{j}\right)
=e_{i,j}\left( m_{i},m_{j}\right) =0$.

Since $m_{i}^{2}=\tilde{m}_{i}^{2}$, both $\left( m_{i},m_{j}\right) $ and $%
\left( \tilde{m}_{i},m_{j}\right) $ lead to the same outcome $%
x(m_{i}^{2},m_{j}^{1})=x(\tilde{m}_{i}^{2},m_{j}^{1})=f\left(
m_{i}^{2}\right) $.

Next, suppose that the designer uses agent $i$'s report to check agent $j$'s
report. Again, if $m_{j}^{2}\notin \Theta $, then $e_{j,i}\left( m_{j},%
\tilde{m}_{i}\right) =e_{j,i}\left( m_{j},m_{i}\right) =1$; thus, by Lemma %
\ref{AM}, $\tilde{m}_{i}$ is a strictly better response than $m_{i}$ against 
$m_{j}$. Hence, we may assume $m_{j}^{2}\in \Theta $ and consider the
following four cases:

\noindent \emph{Case 2.1.} $e_{j,i}\left( m_{j},m_{i}\right) =\varepsilon $
and $e_{j,i}\left( m_{j},\tilde{m}_{i}\right) =0$.

It follows from (\ref{d-w}) and Lemma \ref{BC} that 
\begin{equation*}
u_{i}(f(m_{j}^{2}),\theta _{i})-u_{i}(C_{j,i}^{\varepsilon
}(m_{j},m_{i}),\theta _{i})>0,
\end{equation*}%
where $f(m_{j}^{2})$ is the outcome induced by $\left( m_{j},\tilde{m}%
_{i}\right) $.

\noindent \emph{Case 2.2.} $e_{j,i}\left( m_{j},m_{i}\right) =0$ and $%
e_{j,i}\left( m_{j},\tilde{m}_{i}\right) =\varepsilon $.

Since $e_{j,i}\left( m_{j},m_{i}\right) =0$, we have $m_{i}^{2}=\tilde{m}%
_{i}^{2}=m_{j}^{2}$. Hence, $e_{j,i}\left( m_{j},\tilde{m}_{i}\right)
=\varepsilon $ implies that $x(m_{j}^{2},\tilde{m}_{i}^{1})=x(m_{j}^{2},%
\theta _{i})\neq f(m_{j}^{2})$. Thus, it follows from (\ref{bw}) that%
\begin{equation*}
u_{i}(C_{j,i}^{\varepsilon }(m_{j},\tilde{m}_{i}),\theta
_{i})-u_{i}(f(m_{j}^{2}),\theta _{i})>0,
\end{equation*}%
where $f(m_{j}^{2})$ is the outcome induced by $\left( m_{j},m_{i}\right) $.

\noindent \emph{Case 2.3.} $e_{j,i}\left( m_{j},m_{i}\right) =e_{j,i}\left( 
\tilde{m}_{j},m_{i}\right) =\varepsilon $.

It follows from Lemmas \ref{AM} and \ref{BC} that 
\begin{equation*}
u_{i}(C_{j,i}^{\varepsilon }(m_{j},\tilde{m}_{i}),\theta
_{i})-u_{i}(C_{j,i}^{\varepsilon }(m_{j},m_{i}),\theta _{i})>0\text{.}
\end{equation*}%
\emph{Case 2.4. }$e_{j,i}\left( m_{j},m_{i}\right) =e_{j,i}\left( \tilde{m}%
_{j},m_{i}\right) =0$.

Both $\left( m_{j},m_{i}\right) $ and $\left( m_{j},\tilde{m}_{i}\right) $
lead to the same outcome $x(m_{j}^{2},\tilde{m}%
_{i}^{1})=x(m_{j}^{2},m_{i}^{1})=f(m_{j}^{2})$.

In sum, as long as $e_{i,j}\left( m_{i},m_{j}\right) =\varepsilon $ or $%
e_{j,i}\left( m_{j},m_{i}\right) =\varepsilon $ (Case 1.1 and Cases
2.1-2.3), $\tilde{m}_{i}$ is a strictly better response than $m_{i}$ against 
$m_{j}$. Hence, in order for $\tilde{m}_{i}$ not to be a profitable
deviation, we must have $e_{i,j}\left( m_{i},m_{j}\right) =e_{j,i}\left(
m_{j},m_{i}\right) =0$.

\subsection{Proof of Theorem \protect\ref{NashC}}

\label{proof: NashC}

We first extend the notion of\emph{\ }a\emph{\ challenge scheme} for an SCC.
Fix agent $i$ of type $\theta _{i}$. For each state $\tilde{\theta}\in
\Theta $ and $z\in F(\tilde{\theta}),$ if $\mathcal{L}_{i}(z,\tilde{\theta}%
_{i})\cap \hspace{0.1cm}\mathcal{SU}_{i}(z,\theta _{i})\neq \varnothing $,
we select some $x(\tilde{\theta},z,\theta _{i})\in \mathcal{L}_{i}(z,\tilde{%
\theta}_{i})\cap \mathcal{SU}_{i}(z,\theta _{i})$; otherwise, we set $x(%
\tilde{\theta},z,\theta _{i})=z$. In the sequel, we define $F\left( \Theta
\right) \equiv \bigcup_{\theta \in \Theta }F\left( \theta \right) $. Observe
that $F\left( \Theta \right) $ is a finite set, since each $F\left( \theta
\right) $ is assumed to be finite.

As in the case of SCFs, the following lemma shows that there is a challenge
scheme under which truth-telling induces the best allocation. In addition,
we choose the challenge scheme in such a way that for every agent $i$, type $%
\theta _{i}$, and state $\tilde{\theta}$ under which the challenge is
effective (i.e. $x(\tilde{\theta},z,\theta _{i})\not=z$), no type $\theta
_{i}^{\prime \prime }\in \Theta _{i}$ is indifferent between $x(\tilde{\theta%
},z,\theta _{i})$ and any allocation $z^{\prime }$ in $F(\Theta )$.

\begin{lemma}
\label{BCC}For any SCC $F$, there is a challenge scheme $\{x(\tilde{\theta}%
,z,\theta _{i})\}_{i\in \mathcal{I},\tilde{\theta}\in \Theta ,z\in F(\tilde{%
\theta}),\theta _{i}\in \Theta _{i}}$ such that for every $i\in \mathcal{I}%
,\ \tilde{\theta}\in \Theta ,\ z\in F(\tilde{\theta})$, and $\theta _{i}\in
\Theta _{i}$, 
\begin{equation}
u_{i}(x(\tilde{\theta},z,\theta _{i}),\theta _{i})\geq u_{i}(x(\tilde{\theta}%
,z,\theta _{i}^{\prime }),\theta _{i}),\forall \theta _{i}^{\prime }\in
\Theta _{i}\text{;}  \label{best-C}
\end{equation}

moreover, whenever, $x(\tilde{\theta},z,\theta _{i})\not=z$, we have 
\begin{equation}
u_{i}(x(\tilde{\theta},z,\theta _{i}),\theta _{i}^{\prime \prime
})\not=u_{i}(z^{\prime },\theta _{i}^{\prime \prime }),\forall \theta
_{i}^{\prime \prime }\in \Theta _{i},\forall z^{\prime }\in F(\Theta )\text{.%
}  \label{gC}
\end{equation}
\end{lemma}

\begin{proof}
We first prove (\ref{gC}) by constructing a challenge scheme $\{x(\tilde{%
\theta},z,\theta _{i})\}$. Fix agent $i$ of type $\theta _{i}$. For each
state $\tilde{\theta}\in \Theta $ and $z\in F(\tilde{\theta})$, if $\mathcal{%
L}_{i}(z,\tilde{\theta}_{i})\cap \hspace{0.1cm}\mathcal{SU}_{i}(z,\theta
_{i})=\varnothing $, we let $x(\tilde{\theta},z,\theta _{i})=z$; otherwise,
we define 
\begin{equation*}
\mathcal{S}(i,z,\tilde{\theta},\theta )=\left\{ z^{\prime \prime }\in
X:u_{i}(z^{\prime \prime },\tilde{\theta}_{i})<u_{i}(z,\tilde{\theta}_{i})%
\text{ and }u_{i}(z^{\prime \prime },\theta _{i})>u_{i}(z,\theta
_{i})\right\} .
\end{equation*}%
Observe that $\mathcal{S}(i,z,\tilde{\theta},\theta )$ is a nonempty open
set, since we can add a small penalty to agent $i$ with an allocation in $%
\mathcal{L}_{i}(z,\tilde{\theta}_{i})\cap \hspace{0.1cm}\mathcal{SU}%
_{i}(z,\theta _{i})$.$\,$ Now consider 
\begin{eqnarray*}
&&\mathcal{S}^{\ast }(i,z,\tilde{\theta},\theta ) \\
&\equiv &\mathcal{S}(i,z,\tilde{\theta},\theta )\diagdown \bigcup_{\theta
_{i}^{\prime \prime }\in \Theta _{i}}\bigcup_{z^{\prime }\in F(\Theta
)}\left\{ z^{\prime \prime }\in X:u_{i}(z^{\prime \prime },\theta
_{i}^{\prime \prime })=u_{i}(z^{\prime },\theta _{i}^{\prime \prime
})\right\} \text{.}
\end{eqnarray*}%
Thanks to the finiteness of $F(\Theta )$ and $\Theta _{i}$, $\mathcal{S}%
^{\ast }(i,z,\tilde{\theta},\theta )$ remains a nonempty open set after we
delete finitely many closed sets $\left\{ z^{\prime \prime }\in
X:u_{i}(z^{\prime \prime },\theta _{i}^{\prime \prime })=u_{i}(z^{\prime
},\theta _{i}^{\prime \prime })\right\} $, one for each $\theta _{i}^{\prime
\prime }\in \Theta _{i}$ and $z^{\prime }\in F(\Theta )$. Now we choose an
element $x(\tilde{\theta},z,\theta _{i})\in \mathcal{S}^{\ast }(i,z,\tilde{%
\theta},\theta )$. Hence, we obtain (\ref{gC}). The proof of (\ref{best-C})
is completed once we apply the proof of Lemma \ref{BC} to the challenge
scheme $\{x(\tilde{\theta},z,\theta _{i})\}_{i\in \mathcal{I},\tilde{\theta}%
\in \Theta ,z\in F(\tilde{\theta}),\theta _{i}\in \Theta _{i}}$.
\end{proof}

Next, we propose a mechanism $\mathcal{M}=\left( (M_{i}),g,(\tau
_{i})\right) _{i\in \mathcal{I}}$ which will be used to prove the if-part of
Theorem \ref{NashC}. First, a generic message of agent $i$ is described as
follows: 
\begin{gather*}
m_{i}=\left( m_{i}^{1},m_{i}^{2},m_{i}^{3}\right) \in M_{i}=M_{i}^{1}\times
M_{i}^{2}\times M_{i}^{3}=\Theta _{i}\times \left[ \times _{j=1}^{I}\Theta
_{j}\right] \times F(\Theta )\text{ s.t.} \\
m_{i}^{2}\in \Theta \Rightarrow m_{i}^{3}\in F(m_{i}^{2})\text{. }
\end{gather*}%
That is, agent $i$ is asked to announce (1) agent $i$'s own type (which we
denote by $m_{i}^{1}$); (2) a type profile (which we denote by $m_{i}^{2}$);
(3) an allocation $m_{i}^{3}$ such that $m_{i}^{3}\in F(m_{i}^{2})$ if $%
m_{i}^{2}$ is a state. As we do in the case of SCFs, we write $m_{i,j}^{2}=%
\tilde{\theta}_{j}$ if agent $i$ reports in $m_{i}^{2}$ that agent $j$'s
type is $\tilde{\theta}_{j}$. Likewise, since $F$ is Maskin-monotonic, we
have $F\left( \theta \right) =F\left( \tilde{\theta}\right) $ if $\tilde{%
\theta}_{i}=\theta _{i}$ for every $i$; hence, for $m_{i}^{2}\in \Theta $, $%
F(m_{i}^{2})$ is uniquely defined as~$F\left( \tilde{\theta}\right) $ such
that $\tilde{\theta}_{j}=m_{i,j}^{2}$ for all $j$.

We define $\phi (m)$ as follows: for each $m\in M$, 
\begin{equation*}
\phi (m)=\left\{ 
\begin{tabular}{l}
$x$, \\ 
$m_{1}^{3}$,%
\end{tabular}%
\begin{tabular}{l}
if $\left\vert \left\{ i\in \mathcal{I}:m_{i}^{3}=x\right\} \right\vert \geq
I-1$; \\ 
otherwise.%
\end{tabular}%
\right.
\end{equation*}%
We say that $\phi (m)$ is an \emph{effective allocation} under $m$. In
words, the effective allocation is $x$, if there are $I-1$ players who agree
on allocation $x$; otherwise, the effective allocation is the allocation
announced by agent 1.

The allocation rule $g$ is defined as follows: for each $m\in M$, 
\begin{equation*}
g(m)=\frac{1}{I^{2}}\sum_{i\in \mathcal{I}}\sum_{j\in \mathcal{I}}\left[
e_{i,j}\left( m\right) \frac{1}{I}\sum_{k\in \mathcal{I}}y_{k}\left(
m_{k}^{1}\right) \oplus \left( 1-e_{i,j}\left( m\right) \right) x(\tilde{%
\theta},\phi (m),m_{j}^{1})\right] .
\end{equation*}%
where $\{y_{k}(\theta _{k})\}_{\theta _{k}\in \Theta _{k}}$ are the dictator
lotteries for agent $k$ as defined in Lemma \ref{AM}. Given a message
profile $m,$ and a pair of agents $i$ and $j,$ we say that agent $j$ \emph{%
challenges agent} $i$ if and only if $m_{i}^{3}=\phi (m)$ and $%
x(m_{i}^{2},\phi (m),m_{j}^{1})\neq \phi (m)$, i.e., agent $i$'s reported
allocation is an effective one and agent $j$ challenges this effective
allocation. We define the $e_{i,j}$-function as follows: for each $m\in M$, 
\begin{equation*}
e_{i,j}\left( m\right) =\left\{ 
\begin{array}{ll}
0\text{,} & \text{if }m_{i}^{2}\in \Theta \text{, }m_{i}^{2}=m_{j}^{2}\text{%
, and agent }j\text{ does not challenge agent }i\text{;} \\ 
\varepsilon \text{,} & \text{if }m_{i}^{2}\in \Theta \text{, and [}%
m_{i}^{2}\neq m_{j}^{2}\text{ or agent }j\text{ challenges agent }i\text{];}
\\ 
1\text{,} & \text{if }m_{i}^{2}\notin \Theta \text{.}%
\end{array}%
\right.
\end{equation*}%
Recall that the $e_{i,j}$-function in Section \ref{ar} for the case of SCFs
only depends on $m_{i}$ and $m_{j}$. In contrast, the $e_{i,j}$-function
here depends on the entire message profile, as the nature of the challenge
also depends on whether the allocation reported by agent $i$ is an effective
allocation or not.

Fix $i,j\in \mathcal{I},\varepsilon \in (0,1)$, and $m\in M$. Then, we
define 
\begin{equation*}
C_{i,j}^{\varepsilon }(m)\equiv \varepsilon \times \frac{1}{I}\sum_{k\in 
\mathcal{I}}y_{k}\left( m_{k}^{1}\right) \oplus \left( 1-\varepsilon \right)
\times x(m_{i}^{2},\phi (m),m_{j}^{1})\text{.}
\end{equation*}%
For every message profile $m$ and agent $j$, we can choose $\varepsilon >0$
sufficiently small such that (i) $C_{i,j}^{\varepsilon }(m)$ does not
disturb the \textquotedblleft effectiveness\textquotedblright\ of agent $j$%
's challenge, i.e.,

\begin{eqnarray}
x(m_{i}^{2},\phi (m),m_{j}^{1}) &\neq &\phi (m)\Rightarrow  \notag \\
u_{j}(C_{i,j}^{\varepsilon }(m),m_{i,j}^{2}) &<&u_{j}(\phi (m),m_{i,j}^{2})%
\text{ and }u_{j}(C_{i,j}^{\varepsilon }(m),m_{j}^{1})>u_{j}(\phi
(m),m_{j}^{1})\text{;}  \label{Cbw}
\end{eqnarray}%
moreover, (ii) an \textquotedblleft effective\
self-challenge\textquotedblright\ of agent $j\ $induces a generic outcome
such that at each state, no agent is indifferent between the resulting
outcome and any outcome in $F\left( \Theta \right) $, i.e., 
\begin{eqnarray}
x(m_{j}^{2},\phi (m),m_{j}^{1}) &\neq &\phi (m)\Rightarrow  \notag \\
u_{j}(C_{j,j}^{\varepsilon }(m),\theta _{j}) &\not=&u_{j}(x,\theta _{j})%
\text{ }  \label{Cg} \\
\text{for any }\theta \text{ and any }x &\in &F\left( \Theta \right) \text{.}
\notag
\end{eqnarray}%
Observe that property (ii) can be made satisfied because inequality (\ref{gC}%
) holds in Lemma \ref{BCC}; moreover, by (\ref{d-w}) in Lemma \ref{AM}, $%
u_{j}(C_{j,j}^{\varepsilon }(m),\theta _{j})$ is a strictly decreasing
function in $\varepsilon $.

The transfer to agent $i$ is specified as follows: for each $m\in M$, 
\begin{equation*}
\tau _{i}(m)=\sum_{j\neq i}\left[ \bar{\tau}_{i,j}^{1}(m)+\bar{\tau}%
_{i,j}^{2}(m)+\tau _{i,j}^{3}\left( m\right) \right]
\end{equation*}%
where we set $\bar{\tau}_{i,j}^{1}(m)=2\tau _{i,j}^{1}(m)$ and $\bar{\tau}%
_{i,j}^{2}(m)=2\tau _{i,j}^{2}(m)$, while $\tau _{i,j}^{1}(m)$ and $\tau
_{i,j}^{2}(m)$ are defined as in Section \ref{Fmoney}; moreover, we specify $%
\tau _{i,j}^{3}(m)$ as follows: for each $m\in M$,

\begin{equation*}
\tau _{i,j}^{3}\left( m\right) =\left\{ 
\begin{array}{ll}
-\eta \text{,} & \text{if agent }j\text{ challenges agent }i\text{,} \\ 
0, & \text{otherwise.}%
\end{array}%
\right.
\end{equation*}%
That is, agent $i$ is asked to pay $\eta $ if his reported outcome $%
m_{i}^{3} $ is challenged by agent $j\not=i$. Note that we still require
that $\eta $ be greater than the maximal payoff difference, which is
guaranteed by (\ref{D}) in Section \ref{Fmoney}.

In the rest of the proof of Theorem \ref{NashC}, we fix $\theta $ as the
true state and $\sigma $ as a (possibly mixed strategy) Nash equilibrium of
the game $\Gamma (\mathcal{M},\theta )$ throughout.

To prove Theorem \ref{NashC}, we use a stronger statement than Claim \ref{c0}
since each agent $i$'s dictator lotteries are triggered whenever there is a
pair of agents $\left( j,k\right) $ such that $e_{j,k}\left(
m_{j},m_{k}\right) =\varepsilon $ where $j,k\in \mathcal{I}$ and we possibly
have $j\neq i$ or $k\neq i$. The proof of this stronger claim is identical
to the proof of Case 1.1 in Claim \ref{c0}.

\begin{claim}
\label{cc0}If $m_{i}^{1}\neq \theta _{i}$ for some $m_{i}\in $supp$(\sigma
_{i})$, then we have $e_{j,k}\left( m_{i},m_{-i}\right) =0$ for every $%
m_{-i}\in $supp$\left( \sigma _{-i}\right) $ and every pair of agent(s) $%
j,k\in \mathcal{I}$.
\end{claim}

We now observe that Claims \ref{1st} and \ref{pc1} used in the proof of
Theorem \ref{Nash} hold with exactly the same proof. As we did in the proof
of Theorem \ref{Nash}, by Claim \ref{pc1}, we denote the common state
announced in the agents' second report by $\tilde{\theta}$. In the
following, we establish Claim \ref{pc2C}\ as the counterpart of Claim \ref%
{pc2} used in the proof of Theorem \ref{Nash} in the modified mechanism
introduced above.

For each allocation $x\in F\left( \Theta \right) $, we define the following
set of agents: 
\begin{equation*}
\mathcal{J}\left( x\right) \equiv \left\{ j\in \mathcal{I}:\mathcal{L}_{j}(x,%
\tilde{\theta}_{j})\cap \mathcal{SU}_{j}(x,\theta _{j})=\varnothing \right\} 
\text{.}
\end{equation*}%
The following preliminary claim will be used in proving Claims \ref{pc2C}
and \ref{pc2Cs}.

\begin{claim}
\label{blowthewhistle}For any pair of agent $i$ and $j$ (whether $i=j$ or $%
i\neq j$) and message profile $m\in $supp$\left( \sigma \right) $ such that $%
m_{i}^{3}=\phi (m)$, we have $x(\tilde{\theta},m_{i}^{3},m_{j}^{1})%
\not=m_{i}^{3}\ $if and only if $j\not\in \mathcal{J}\left( m_{i}^{3}\right) 
$.
\end{claim}

\begin{proof}
Fix agent $i\in \mathcal{I}$ and a message profile $m\in $supp$(\sigma )$.
We first prove the if-part. Suppose, on the contrary, that there exists some
agent $j\not\in \mathcal{J}\left( m_{i}^{3}\right) $ such that $x(\tilde{%
\theta},m_{i}^{3},m_{j}^{1})=m_{i}^{3}$. Then, by (\ref{Cbw}), the deviation
from $m_{j}$ to $\tilde{m}_{j}=(\theta _{j},m_{j}^{2},m_{j}^{3})$ delivers a
strictly better payoff for agent $j$ against $m_{-j},$ while, by Lemmas \ref%
{AM} and \ref{BCC}, the deviation from $m_{j}$ to $\tilde{m}_{j}$ generates
no payoff loss for agent $j$ against any $m_{-j}^{\prime }\not=m_{-j}$.
Hence, the deviation $\tilde{m}_{j}$ is profitable, which contradicts the
hypothesis that $\sigma $ is a Nash equilibrium of the game $\Gamma (%
\mathcal{M},\theta )$.

Next, we prove the only-if-part. Suppose, on the contrary, that there exists
some agent $j\in \mathcal{J}\left( m_{i}^{3}\right) $ such that $x(\tilde{%
\theta},m_{i}^{3},m_{j}^{1})\not=m_{i}^{3}.$ Since $j\in \mathcal{J}\left(
m_{i}^{3}\right) $, we must have $m_{j}^{1}\neq \theta _{j}$. Define $\tilde{%
m}_{j}$ as a deviation which is identical to $m_{j}$ except that $\tilde{m}%
_{j}^{3}=\theta _{j}\neq m_{j}^{3}$. Then we have $x(\tilde{\theta}%
,m_{i}^{3},\theta _{j})=m_{i}^{3}$ since $j\in \mathcal{J}\left(
m_{i}^{3}\right) .$ By (\ref{Cbw}), $\tilde{m}_{j}$ generates a strictly
better payoff for agent $j$ than $m_{j}$ against $m_{-j}$. By Lemmas \ref{AM}
and \ref{BCC}, we also know that agent $j$'s payoff generated by $\tilde{m}%
_{j}$ is at least as good as that generated by $m_{j}$ against any $%
m_{-j}^{\prime }\neq m_{-j}$. Hence, $\tilde{m}_{j}$ constitutes a
profitable deviation, which contradicts the hypothesis that $\sigma $ is a
Nash equilibrium of the game $\Gamma (\mathcal{M},\theta )$.
\end{proof}

\begin{claim}
\label{pc2C}No one challenges an allocation announced in the third report of
any other agent, i.e., for any pair of agents $i,j\in \mathcal{I}$ with $%
i\neq j$ and any $m\in $supp$\left( \sigma \right) $, if $m_{i}^{3}=\phi (m)$%
, then $x(\tilde{\theta},m_{i}^{3},m_{j}^{1})=m_{i}^{3}$.
\end{claim}

\begin{proof}
Suppose to the contrary that there exist $i,j\in \mathcal{I}$ with $i\neq j$%
, $m\in \mbox{supp}(\sigma )$ such that $m_{i}^{3}=\phi (m)$, and $x(\tilde{%
\theta},m_{i}^{3},m_{j}^{1})\neq m_{i}^{3}$. By Claim \ref{blowthewhistle}, $%
j\not\in \mathcal{J}\left( m_{i}^{3}\right) .$ We now derive a contradiction
in each of the following two cases.\vspace{0.2cm}

\noindent \textbf{Case (i)}: There is some $\tilde{m}\in $supp$\left( \sigma
\right) $ such that $\mathcal{J}\left( \phi \left( \tilde{m}\right) \right) =%
\mathcal{I}$.

Define $\tilde{m}_i$ as the same as $m_i$ except that $\tilde{m}%
_{i}^{3}=\phi \left( \tilde{m}\right) $. Fix $\hat{m}_{-i}\in $supp$\left(
\sigma _{-i}\right)$. We distinguish two subcases:\vspace{0.2cm}

\noindent \textbf{Case (i.1)}: $\phi \left( m_{i},\hat{m}_{-i}\right)
=m_{i}^{3}$.\vspace{0.2cm}

By Claim \ref{blowthewhistle} and the fact that $j\not\in \mathcal{J}\left(
m_{i}^{3}\right) ,$ we have $x(\tilde{\theta},\phi \left( m_{i},\hat{m}%
_{-i}\right) ,\hat{m}_{j}^{1})\neq \phi \left( m_{i},\hat{m}_{-i}\right) $,
i.e., $\phi \left( m_{i},\hat{m}_{-i}\right) $ must be challenged by $\hat{m}%
_{j}^{1}$ and agent $i$ is penalized by $\eta $ according to $\tau
_{i,j}^{3} $. In comparison, if $\phi \left( \tilde{m}_{i},\hat{m}%
_{-i}\right) =\phi \left( \tilde{m}\right) $, since $\mathcal{J}\left( \phi
\left( \tilde{m}\right) \right) =\mathcal{I}$, it follows from Claim \ref%
{blowthewhistle} that no agent challenges $\phi \left( \tilde{m}_{i},\hat{m}%
_{-i}\right) $; if $\phi \left( \tilde{m}_{i},\hat{m}_{-i}\right) \not=\phi
\left( \tilde{m}\right) ,$ then $\tilde{m}_{i}$ is not effective; hence,
agent $i$ avoids paying the penalty $\eta $ for being challenged.

\noindent \textbf{Case (i.2)}: $\phi \left( m_{i},\hat{m}_{-i}\right)
\not=m_{i}^{3}$.\vspace{0.2cm} Then, the deviation does not change
allocation or transfers from agent $i$'s perspective.

We know that Case (i.1) happens with positive probability from our
hypothesis. Hence, by (\ref{D}), it is a profitable deviation.\vspace{0.2cm}

\noindent \textbf{Case (ii)}: for every $\tilde{m}\in $supp$\left( \sigma
\right) $, $\mathcal{J}\left( \phi \left( \tilde{m}\right) \right) \not=%
\mathcal{I}$.

Fix $\tilde{m}\in \mbox{supp}(\sigma )$. If $\tilde{m}_{k}^{3}=\phi \left( 
\tilde{m}\right) $ for some agent $k\in \mathcal{I}$, then Claim \ref%
{blowthewhistle} implies that $x(\tilde{\theta},\tilde{m}_{k}^{3},\tilde{m}%
_{k^{\prime }}^{1})\not=\tilde{m}_{k}^{3}$ for some agent $k^{\prime }$,
namely that agent $k^{\prime }$ challenges agent $k$ at $\tilde{m}$. Thus,
we know that $e_{k,k^{\prime }}\left( \tilde{m}\right) =\varepsilon .$
Hence, by Claim \ref{cc0}, every agent reports the true type in their first
reports under any $\tilde{m}\in $supp$\left( \sigma \right) .$ Hence, by
Claim \ref{1st}, we conclude that $\tilde{\theta}=\theta $. It implies that $%
\phi \left( \tilde{m}\right) \in F\left( \theta \right) $ and $\mathcal{J}%
\left( \phi \left( \tilde{m}\right) \right) =\mathcal{I}$. Thus, it
contradicts that $\mathcal{J}\left( \phi \left( \tilde{m}\right) \right)
\not=\mathcal{I}$.
\end{proof}

\begin{claim}
\label{pc2Cs}No one challenges an allocation announced in his own third
report, i.e., for every agent $i$, $m\in $supp$\left( \sigma \right) $ and $%
m_{i}^{3}=\phi (m)$ we have $x(\tilde{\theta},m_{i}^{3},m_{i}^{1})=m_{i}^{3}$%
.
\end{claim}

\begin{proof}
Suppose to the contrary that there exist agent $i$ and some message $m\in $%
supp$\left( \sigma \right) $ such that $x(\tilde{\theta}%
,m_{i}^{3},m_{i}^{1})\neq m_{i}^{3}$. By Claim \ref{pc2C} and the
construction of $\phi $, the agent $i$ must be agent $1.$ Moreover, $\phi
(m)=m_{1}^{3}$ and $\phi (m)\neq m_{j}^{3}$ for every $j\neq 1$. Hence, 
\begin{equation}
\phi (\tilde{m}_{1},m_{-1})=\tilde{m}_{1}^{3}\text{ for every }\tilde{m}_{1}
\label{alwayseffective}
\end{equation}%
. By Claim \ref{blowthewhistle}, we know that $1\not\in \mathcal{J}\left(
m_{1}^{3}\right) ,$ i.e., 
\begin{equation}
\mathcal{L}_{1}(m_{1}^{3},\tilde{\theta}_{1})\cap \mathcal{S}\mathcal{U}%
_{1}(m_{1}^{3},\theta _{1})\neq \varnothing \text{;}  \label{self}
\end{equation}%
moreover, for every $\tilde{m}\in $supp$\left( \sigma \right) $ with $\phi (%
\tilde{m})=m_{1}^{3}$, we have $x(\tilde{\theta},m_{1}^{3},\tilde{m}%
_{1}^{1})\not=m_{1}^{3}.$ Next, we shall show that 
\begin{equation}
\tilde{m}\in \text{supp}\left( \sigma \right) \text{ and }\phi (\tilde{m})=%
\tilde{m}_{1}^{3}\Rightarrow x(\tilde{\theta},\tilde{m}_{1}^{3},\tilde{m}%
_{1}^{1})\not=\tilde{m}_{1}^{3}\text{.}  \label{self-1}
\end{equation}

To establish (\ref{self-1}), suppose on the contrary that $x(\tilde{\theta},%
\tilde{m}_{1}^{3},\tilde{m}_{1}^{1})=\tilde{m}_{1}^{3}$ for some $\tilde{m}%
\in $supp$\left( \sigma \right) $ with $\tilde{m}_{1}^{3}=\phi (\tilde{m})$.
We now compare the payoff difference between $m_{i}$ and $\tilde{m}_{i}$
against $\hat{m}_{-1}$ by considering the following two different situations.

\noindent \textbf{Case (i)}: $\phi \left( m_{1},\hat{m}_{-1}\right)
=m_{1}^{3}$ and $\phi \left( \tilde{m}_{1},\hat{m}_{-1}\right) =\tilde{m}%
_{1}^{3}.$

In this case, we know that $x(\tilde{\theta},m_{1}^{3},m_{1}^{1})%
\not=m_{1}^{3}$ and $x(\tilde{\theta},\tilde{m}_{1}^{3},\tilde{m}_{1}^{1})=%
\tilde{m}_{1}^{3}.$ By Claim \ref{pc2C}, no agent challenges any of the
other agents. Thus, the allocation difference occurs only when agent $1$ is
chosen to challenge himself.

\noindent \textbf{Case (ii)}: $\phi \left( m_{1},\hat{m}_{-1}\right)
\not=m_{1}^{3}$ or $\phi \left( \tilde{m}_{1},\hat{m}_{-1}\right) \not=%
\tilde{m}_{1}^{3}$.

By the construction of $\phi $, there exists $z\in F(\tilde{\theta})$ such
that every agent $j\neq 1$ reports $\hat{m}_{j}^{3}=z$. Once again, by the
construction of $\phi $, we also have $\phi \left( m_{1},\hat{m}_{-1}\right)
=$ $\phi \left( \tilde{m}_{1},\hat{m}_{-1}\right) =z$. Moreover, by Claim %
\ref{pc2C}, no agent challenges $z$, i.e., $x(\tilde{\theta},z,\hat{m}%
_{k}^{1})=z$ for every $k\in \mathcal{I}$. Hence, $\left( m_{1},\hat{m}%
_{-1}\right) $ and $\left( \tilde{m}_{1},\hat{m}_{-1}\right) $ deliver the
same allocation and transfer to agent $1$.

In summary, against $\sigma _{-i},$ the payoff difference between $m_{i}$
and $\tilde{m}_{i}$ lies only in case (i) and is equal to 
\begin{equation*}
\frac{1}{I^{2}}u_{i}(C_{i,i}^{\varepsilon }(m_{i},\hat{m}_{-1}),\theta _{i})-%
\frac{1}{I^{2}}u_{i}(\tilde{m}_{i}^{3},\theta _{i})\text{.}
\end{equation*}%
The payoff difference must be zero because both $m_{i}$ and $\tilde{m}_{i}$
are played with positive probability in equilibrium. However, it contradicts
(\ref{gC}). Hence, (\ref{self-1}) holds.

Finally, it follows from (\ref{alwayseffective}) and (\ref{self-1}) that
agent 1 must challenge himself with probability one. By Claim \ref{cc0},
every agent reports the true type in their first reports under any $\tilde{m}%
\in $supp$\left( \sigma \right) $. Moreover, by Claim \ref{1st}, we have $%
\tilde{\theta}=\theta $ which is a contradiction to (\ref{self}).
\end{proof}

It only remains to prove the existence of pure-strategy Nash equilibrium.

\begin{claim}
\label{cc2}For every $x\in F\left( \theta \right) $, there exists a
pure-strategy Nash equilibrium $m\in M$ of the game $\Gamma (\mathcal{M}%
,\theta )$ such that $g(m)=x$ and $\tau _{i}(m)=0$ for every $i\in \mathcal{I%
}$.
\end{claim}

\begin{proof}
Fix an arbitrary allocation $x\in F(\theta )$. We argue that truth-telling
(i.e., $m_{i}=(\theta _{i},\theta ,x)$ for each $i$) constitutes a
pure-strategy equilibrium of the game $\Gamma (\mathcal{M},\theta )$. Note
that reporting $\tilde{m}_{i}$ with $\tilde{m}_{i}^{1}=\theta _{i}$, $\tilde{%
m}_{i}^{2}=\theta $, and $\tilde{m}_{i}^{3}\neq x$ instead of $m_{i}$
affects neither the allocation nor the transfer. The argument for proving
that either $\tilde{m}_{i}^{1}\not=\theta _{i}$ or $\tilde{m}_{i}^{2}\neq
\theta $ cannot be a profitable unilateral deviation for every agent $i$ is
identical to the relevant portion of the proof of Theorem \ref{Nash}.
\end{proof}


\subsection{Proof of Theorem \protect\ref{Thm:small-transfer}}

\label{app-small}

Recall that in the mechanism which we use to prove Theorem \ref{Nash}, agent 
$i$'s generic message is $m_{i}=(m_{i}^{1},m_{i}^{2})\in \Theta _{i}\times %
\left[ \times _{j=1}^{I}\Theta _{j}\right] $. We expand $m_{i}^{2}$ into $H$
copies of $\left[ \times _{j=1}^{I}\Theta _{j}\right] $ and define 
\begin{equation*}
m_{i}=(m_{i}^{1},m_{i}^{2},\ldots ,m_{i}^{H+1})\in \Theta _{i}\times 
\underbrace{\left[ \times _{j=1}^{I}\Theta _{j}\right] \times \cdots \times %
\left[ \times _{j=1}^{I}\Theta _{j}\right] }_{\mbox{$H$
terms}}
\end{equation*}%
where $H$ is a positive integer to be chosen later. For each message profile 
$m\in M$, the allocation is defined as follows:

\begin{equation*}
g(m)=\frac{1}{I(I-1)}\sum_{i\in \mathcal{I}}\sum_{j\neq i}\left[
e_{i,j}\left( m_{i},m_{j}\right) \frac{1}{2}\sum_{k=i,j}y_{k}\left(
m_{k}^{1}\right) \oplus \frac{1-e_{i,j}(m_{i},m_{j})}{H}\left[ x\left(
m_{i}^{2},m_{j}^{H+2}\right) \oplus \sum_{h=3}^{H+1}\phi \left( m^{h}\right) %
\right] \right]
\end{equation*}%
where $\{y_{k}(\cdot )\}\ $are the dictator lotteries\footnote{%
Although the dictator lotteries may contain transfers, we do not take into
account the scale of transfers in it. To dispense with the transfers in the
dictator lotteries, we can use an arbitrarily small amount of money to make
the best challenge schemes and social choice function generic, i.e., any two
resulting outcomes are distinct from each other.} for agent $k$ defined in
Lemma \ref{AM}, $\phi \left( \cdot \right) $ is an outcome function such
that 
\begin{equation*}
\phi \left( m^{h}\right) =\left\{ 
\begin{array}{ll}
f(\tilde{\theta})\text{,} & \text{if }m_{i}^{h}=\tilde{\theta}\in \Theta 
\text{ for at least }I-1\text{ agents;} \\ 
b\text{,} & \text{otherwise, where }b\text{ is an arbitrary outcome in }A,%
\end{array}%
\right.
\end{equation*}%
and 
\begin{equation*}
e_{i,j}(m_{i},m_{j})=\left\{ 
\begin{array}{ll}
0\text{,} & \text{if }m_{i}^{2}\in \Theta \text{, }%
m_{i}^{2}=m_{j}^{2}=m_{i}^{h}=m_{j}^{h}\text{ and }x\left(
m_{i}^{2},m_{j}^{H+2}\right) =f(m_{i}^{2})\text{,\ }\forall h\in \{3,\ldots
,H+1\}\text{;} \\ 
1\text{,} & \text{if }m_{i}^{2}\notin \Theta \text{;} \\ 
\varepsilon \text{,} & \text{otherwise.}%
\end{array}%
\right.
\end{equation*}

We now define the transfer rule. For every message profile $m\in M$ and
agent $i\in \mathcal{I}$, we specify the transfer to agent $i$ as follows: 
\begin{equation*}
\tau _{i}(m)=\sum_{j\neq i}\left[ \tau _{i,j}^{1,2}(m)+\tau _{i,j}^{2,2}(m)%
\right] +\sum_{h=3}^{H+1}\tau _{i}^{h}(m)+d_{i}(m^{2},\ldots ,m^{H+1})
\end{equation*}%
where $\gamma ,\kappa ,\xi >0$ (their size are determined later)

\begin{eqnarray*}
\tau _{i,j}^{1,2}\left( m\right) &=&\left\{ 
\begin{tabular}{ll}
$0$, & if $m_{i,j}^{2}=m_{j,j}^{2}$; \\ 
$-\gamma $ & if $m_{i,j}^{2}\not=m_{j,j}^{2}$ and $m_{i,j}^{2}\neq m_{j}^{1}$%
; \\ 
$\gamma $ & if $m_{i,j}^{2}\not=m_{j,j}^{2}\text{ }$and $%
m_{i,j}^{2}=m_{j}^{1}$.%
\end{tabular}%
\right. \\
\tau _{i,j}^{2,2}\left( m\right) &=&\left\{ 
\begin{array}{ll}
0\text{,} & \text{if }m_{i,i}^{2}=m_{j,i}^{2}\text{;} \\ 
-\gamma \text{,} & \text{if }m_{i,i}^{2}\neq m_{j,i}^{2}\text{;}%
\end{array}%
\right.
\end{eqnarray*}%
moreover, for every $h\in \{3,\ldots ,H+1\}$, 
\begin{equation*}
\tau _{i}^{h}\left( m\right) =\left\{ 
\begin{array}{ll}
-\kappa \text{,} & \text{if there exists }\tilde{\theta}\text{ such that }%
m_{i}^{h}\neq \tilde{\theta}\text{ but }m_{j}^{h}=\tilde{\theta}\text{ for
all }j\not=i\text{;} \\ 
0\text{,} & \text{otherwise,}%
\end{array}%
\right.
\end{equation*}%
and 
\begin{equation*}
d_{i}(m^{2},\ldots ,m^{H+1})=\left\{ 
\begin{array}{ll}
-\xi \text{,} & 
\begin{array}{l}
\text{if there exists }h\in \{3,\ldots ,H+1\}\text{ such that }%
m_{i}^{h}\not=m_{i}^{2}\text{ and }m_{j}^{h^{\prime }}=m_{j}^{2}, \\ 
\text{for all }h^{\prime }\in \{2,\ldots ,h-1\}\text{ and all }j\not=i\text{;%
}%
\end{array}
\\ 
0\text{,} & \text{otherwise.}%
\end{array}%
\right.
\end{equation*}

Finally, we choose positive numbers $\gamma ,\xi ,H,\kappa ,$ and $%
\varepsilon $ such that 
\begin{eqnarray*}
\bar{\tau} &>&\gamma +\left( H-1\right) \kappa +\xi \\
\gamma &>&\xi +\varepsilon \eta \\
\kappa &>&\varepsilon \eta \\
\xi &>&\frac{1}{H}\eta +\kappa .
\end{eqnarray*}%
More precisely, we first fix $\bar{\tau}$ and choose $\gamma <$ $\frac{1}{3}%
\bar{\tau}$ and $\xi <\min \left\{ \frac{1}{3}\bar{\tau},\gamma \right\} $.
Second, we choose $H$ large enough so that $\xi >\frac{1}{H}\eta $. Third,
we choose $\kappa $ small enough such that $\left( H-1\right) \kappa <\frac{1%
}{3}\bar{\tau}$ and $\xi >\frac{1}{H}\eta +\kappa $. Fourth, we choose $%
\varepsilon $ small enough such that $\gamma >\xi +\varepsilon \eta $ and $%
\kappa >\varepsilon \eta $. We can now prove Theorem \ref{Thm:small-transfer}
following the three steps as in the proof of Theorem \ref{Nash}.

\subsubsection{Contagion of Truth}

First, note that Claims \ref{c0} and \ref{1st} hold. The proof of Claim \ref%
{1st} applies with only one minor difference: Here $m_{i}^{2}$ may affect
agent $i$'s payoff through $d_{i}\left( \cdot \right) $. However, a similar
argument follows, since we have $\gamma >\xi +\varepsilon \eta $.\footnote{%
It corresponds to Property (b) in \cite{AM94}.} Let $\theta $ denote the
true state.

\begin{claim}
\label{s1st}If every agent $j$ reports the truth in his first report $\sigma
_{j}$-probability one, then every agent $j$ reports the truth in his $2$%
nd,...,$\left( H+1\right) $th report. That is, $m_{j}^{h}=\theta $ for $%
h=2,\ldots ,H+1.$
\end{claim}

By Claims \ref{c0} and \ref{1st}, every agent $j$ reports the state
truthfully in his $2$nd report. Then, we can follow verbatim the argument on
p. 12 of \cite{AM94} which shows that every agent $j$ reports the state
truthfully in his $h$th report for every $h=2,...,H+1$.

\subsubsection{Consistency}

\begin{claim}
\label{spc1}There exists a state $\tilde{\theta}$ such that every agent
announces $\tilde{\theta}$ in the second report all the way to the last/$%
\left( H+1\right) $th report with probability one.
\end{claim}

\begin{proof}
We prove consistency by considering the two cases as in the proof of Claim %
\ref{pc1}. The proof for the first case remains the same. For the second
case, suppose that one agent, say $i,$\ tells a lie in the first report. As
agent $i$ believes that all the other agents report the same state $\tilde{%
\theta}$ in their second all the way to the last report. By the same
argument in the second case in the proof of Claim \ref{pc1}, we can show
that agent $i$ announces $\tilde{\theta}$ in the second report with
probability one. In addition, for every $h=2,\ldots ,H+2$, as agent $i$
believes that all the other agent report the same state $\tilde{\theta}$, by
the rule $\phi \left( m^{h}\right) $ and $\tau _{i}^{h}\left( m^{h}\right) ,$
we know $m_{i}^{h}=\tilde{\theta}.$
\end{proof}

\subsubsection{No Challenge}

\begin{claim}
\label{spc2}No agent challenges with positive probability the common state $%
\tilde{\theta}$ announced in the second report.
\end{claim}

\begin{proof}
The argument is the same as the proof of Claim \ref{pc2}.
\end{proof}

\subsection{Proof of Theorem \protect\ref{infiNash}}

\label{Infi}

Recall that we assume that set of alternatives $A$ is a finite set, and the
state space $\Theta $ is a Polish (i.e., separable and complete metric)
space. Recall the definition of $\Theta _{i}$ introduced in Section \ref{M}.
For every $\ell \in \Delta \left( A\right) $, we write $v_{i}\left( \ell
,\theta \right) =\ell \cdot \bar{v}_{i}\left( \theta _{i}\right) $ where $%
\bar{v}_{i}\left( \theta _{i}\right) \in \left[ 0,1\right] ^{\left\vert
A\right\vert }$ is a vector of utilities over $A$ induced by $v_{i}\left(
\cdot ,\theta \right) $. Assume that agents' utilities remain bounded; if $f$
specifies some transfers in outcome, the transfers are also bounded in money
and $\eta >0$ still satisfies condition (\ref{D'}). Let $\bar{X}\equiv $ $%
\Delta \left( A\right) \times \left[ -2\eta ,2\eta \right] ^{I}$ and
identify $\bar{X}$ with a compact subset of $%
\mathbb{R}
^{I+\left\vert A\right\vert }$ (endowed with the Euclidean topology). Let $d$
denote the metric on $\Theta $, $d_{i}$ the metric on $\Theta _{i}$, and $%
\rho :\bar{X}\times \bar{X}\rightarrow 
\mathbb{R}
$ a metric on the outcome space. We endow $\times _{j=1}^{I}\Theta _{j}$
with the product topology and $\times _{j=1}^{I}\Theta _{j}$ and $\bar{X}$
with the Borel $\sigma $-algebra. We say that the setting is \emph{compact
and continuous} if $\Theta $ is compact and $\left( v_{i}\left( a,\cdot
\right) \right) _{a\in A,i\in \mathcal{I}}$ and the SCFs are all continuous
functions on $\Theta $.

We introduce the following version of challenge scheme. First, for $\left(
x,\theta _{i}\right) \in X\times \Theta _{i}$, we use $\mathcal{SL}%
_{i}\left( x,\theta _{i}\right) $ to denote the strict lower-contour set at
allocation $x$ for type $\theta _{i}$, i.e.,%
\begin{equation*}
\mathcal{SL}_{i}\left( x,\theta _{i}\right) =\left\{ x^{\prime }\in
X:u_{i}(x,\theta _{i})>u_{i}(x^{\prime },\theta _{i})\right\} \text{.}
\end{equation*}%
For agent $i$ of type $\theta _{i}$, allocation $x\in \bar{X},$ and state $%
\tilde{\theta}\in \Theta $, we construct the following compound lottery: 
\begin{equation*}
\ell (x,\tilde{\theta})=\frac{\rho (x,f(\tilde{\theta}))}{1+\rho (x,f(\tilde{%
\theta}))}x\oplus \frac{1}{1+\rho (x,f(\tilde{\theta}))}f(\tilde{\theta}).
\end{equation*}

Define 
\begin{equation*}
x(\tilde{\theta},x)=\left\{ 
\begin{array}{ll}
\ell (x,\tilde{\theta})\text{,} & \text{if }x\in \mathcal{SL}_{i}(f(\tilde{%
\theta}),\tilde{\theta}_{i})\text{;} \\ 
f(\tilde{\theta})\text{,} & \text{otherwise.}%
\end{array}%
\right.
\end{equation*}%
As we mentioned in the main text, in the infinite setting we know of no way
to construct a challenge scheme by pre-selecting a test allocation
(depending on type $\theta _{i}$) in a continuous manner. As a result, we
cannot have the agents report their type (let alone the true type) to cast a
challenge to state $\tilde{\theta}$. Instead, we will restore the continuity
of the outcome function by asking them to report an allocation $x$ directly
(see Section \ref{mechanism-infinite}). The definition of strict
lower-contour sets can be found in Section \ref{proof: NashC}.

\begin{claim}
\label{Bcontinuous}$x(\tilde{\theta},x)$ is a continuous function on $\bar{X}%
\times \Theta $.
\end{claim}

\begin{proof}
Since $\rho (\cdot ,\cdot )$ is continuous, we have that $\ell (\cdot ,\cdot
)$ is continuous. We show that 
\begin{equation*}
x(\tilde{\theta}\left[ n\right] ,x\left[ n\right] )\rightarrow x(\tilde{%
\theta},x)
\end{equation*}
in each of the following two cases.

\noindent \emph{\noindent \textbf{Case 1.}} $x\in \mathcal{SL}_{i}(f(\tilde{%
\theta}),\tilde{\theta}_{i})$.

In this case, $x(\tilde{\theta},x)=\ell (x,f(\tilde{\theta}))$. Since $f$
and $u_{i}$ are both continuous, it follows that $x\left[ n\right] \in 
\mathcal{SL}_{i}(f(\tilde{\theta}\left[ n\right] ),\tilde{\theta}_{i}\left[ n%
\right] )$ for each $n$ large enough. Thus, $x(\tilde{\theta}\left[ n\right]
,x\left[ n\right] )=$ $\ell (x\left[ n\right] ,\tilde{\theta}\left[ n\right]
)$. Hence, $x(\tilde{\theta}\left[ n\right] ,x\left[ n\right] )\rightarrow $ 
$\ell (x,f(\tilde{\theta}))$ as $(x\left[ n\right] ,\tilde{\theta}\left[ n%
\right] )\rightarrow (x,\tilde{\theta})$.

\noindent \emph{\noindent \textbf{Case 2.} }$x\not\in \mathcal{SL}_{i}(f(%
\tilde{\theta}),\tilde{\theta}_{i})$.

In this case, $x(\tilde{\theta},x)=f(\tilde{\theta})$. If there is some
integer $\overline{n}$ such that $x\left[ n\right] \notin \mathcal{SL}_{i}(f(%
\tilde{\theta}\left[ n\right] ),\tilde{\theta}_{i}\left[ n\right] )$ for
every $n\geq \overline{n}$, then $x(\tilde{\theta}\left[ n\right] ,x\left[ n%
\right] )=f(\tilde{\theta}\left[ n\right] )$. Since $f$ is continuous and $%
\tilde{\theta}\left[ n\right] \rightarrow \tilde{\theta}$, it follows that $%
x(\tilde{\theta}\left[ n\right] ,x\left[ n\right] )\rightarrow f(\tilde{%
\theta})$. Now suppose that there is a subsequence of {$x$}${\left[ n\right]
,\tilde{\theta}\left[ n\right] }$, say itself, such that $x\left[ n\right]
\in \mathcal{SL}_{i}(f(\tilde{\theta}\left[ n\right] ),\tilde{\theta}_{i}%
\left[ n\right] )$ for every $n$. Then, we have $x(\tilde{\theta}\left[ n%
\right] ,x\left[ n\right] )=\ell (x\left[ n\right] ,f(\tilde{\theta}\left[ n%
\right] ))$. Since $\rho $ is jointly continuous, we must have $\rho (x\left[
n\right] ,f(\tilde{\theta}\left[ n\right] ))\rightarrow \rho (x,f(\tilde{%
\theta})).$ Since $x\not\in \mathcal{SL}_{i}(f(\tilde{\theta}),\tilde{\theta}%
_{i})$, it follows that $\rho (x,f(\tilde{\theta}))=0$. By construction of $%
\ell (x\left[ n\right] ,f(\tilde{\theta}\left[ n\right] ))$, we must have $%
\ell (x\left[ n\right] ,f(\tilde{\theta}\left[ n\right] ))\rightarrow f(%
\tilde{\theta})$. Hence, $x(\tilde{\theta}\left[ n\right] ,x\left[ n\right]
)\rightarrow f(\tilde{\theta})$.
\end{proof}

\begin{lemma}
\label{infic0}For each $i\in \mathcal{I}$, there exists a continuous
function $y_{i}:\Theta _{i}\rightarrow X$ such that for all types $\theta
_{i}$ and $\theta _{i}^{\prime }$ of agent $i$ with $\theta _{i}\neq \theta
_{i}^{\prime }$, we have%
\begin{equation}
u_{i}\left( y_{i}\left( \theta _{i}\right) ,\theta _{i}\right) >u_{i}\left(
y_{i}\left( \theta _{i}^{\prime }\right) ,\theta _{i}\right) \text{;}
\label{d1}
\end{equation}%
and for each type $\theta _{j}^{\prime }$ of agent $j\in \mathcal{I}$, we
also have that for every $x\in f\left( \Theta \right) $ 
\begin{equation}
u_{i}(y_{j}(\theta _{j}^{\prime }),\theta _{i})<u_{i}(x,\theta _{i})\text{.}
\label{d2}
\end{equation}%
Moreover, $y_{i}\left( \cdot \right) $ is continuous on $\Theta _{i}$.
\end{lemma}

\begin{proof}
We construct the dictator lotteries in the infinite state space. We first
construct $\ell _{i}\left( \theta _{i}\right) \in \Delta \left( A\right) $
for each $\theta _{i}\in \Theta _{i}$ and let 
\begin{equation*}
y_{i}\left( \theta _{i}\right) =\left( \ell _{i}\left( \theta _{i}\right)
,-2\eta ,...,-2\eta \right) .
\end{equation*}%
Hence, we obtain $u_{i}\left( a,\theta _{i}\right) >u_{i}\left( y_{k}\left(
\theta _{k}^{\prime }\right) ,\theta _{i}\right) $ for all type $\theta _{i}$
and type $\theta _{k}^{\prime }$ of agents $i$ and $k$. Let $\ell ^{\ast }$
be the uniform lottery over $A$, i.e., $\ell ^{\ast }\left[ a\right]
=1/\left\vert A\right\vert $. Pick $r<1/\left\vert A\right\vert $. Consider
the maximization problem as follows: 
\begin{equation*}
\max_{l\in \Delta \left( A\right) }l\cdot \bar{v}_{i}\left( \theta
_{i}\right)
\end{equation*}%
\begin{equation*}
\text{s.t. }\left\Vert \ell -\ell ^{\ast }\right\Vert \leq r
\end{equation*}%
The Kuhn-Tucker condition for $l_{i}\left( \theta _{i}\right) $ to be the
solution is%
\begin{equation*}
\bar{v}_{i}\left( \theta _{i}\right) -2\lambda _{i}\left( \theta _{i}\right)
\left( \ell _{i}\left( \theta _{i}\right) -\ell ^{\ast }\right) =0
\end{equation*}%
We know that $\bar{v}_{i}\left( \theta _{i}\right) $ is not a zero vector.
Hence, $\lambda _{i}\left( \theta _{i}\right) >0$ and $l_{i}\left( \theta
_{i}\right) $ is equal to the normalization of $\frac{1}{2}\left( \frac{\bar{%
v}_{i}\left( \theta _{i}\right) }{\lambda _{i}\left( \theta _{i}\right) }%
+\ell ^{\ast }\right) $ as a lottery. For every $\theta _{i}\neq \theta
_{i}^{\prime }$, since $\bar{v}_{i}\left( \theta _{i}\right) \ $is not an
affine transform of $\bar{v}_{i}\left( \theta _{i}^{\prime }\right) $, it
follows that $\ell _{i}\left( \theta _{i}\right) \not=\ell _{i}\left( \theta
_{i}^{\prime }\right) $. Moreover, by Theorem of the maximum, $\ell
_{i}\left( \cdot \right) $ is a continuous function on $\Theta _{i}$.
\end{proof}

\subsubsection{The Mechanism}

\label{mechanism-infinite}

A generic message of agent $i$ is described as follows: 
\begin{equation*}
m_{i}=\left( m_{i}^{1},m_{i}^{2},m_{i}^{3}\right) \in M_{i}=M_{i}^{1}\times
M_{i}^{2}\times M_{i}^{3}=\Theta _{i}\times \left[ \times _{j=1}^{I}\Theta
_{j}\right] \times \bar{X}.
\end{equation*}%
That is, agent $i$ is asked to make (1) one announcement of agent $i$'s type
(i.e., $m_{i}^{1}$); and (2) one announcement of a type profile (i.e., $%
m_{i}^{2}$); and (3) one announcement of an allocation (i.e., $m_{i}^{3}$).
As in the main text, we write $m_{i,j}^{2}=\tilde{\theta}_{j}$ if agent $i$
reports in $m_{i}^{2}$ that agent $j$'s type is $\tilde{\theta}_{j}$.

\paragraph{Allocation Rule}

For each message profile $m\in M$, the allocation is defined as follows: 
\begin{equation*}
g\left( m\right) =\frac{1}{I(I-1)}\sum_{i\in \mathcal{I}}\sum_{j\neq i}\left[
e_{i,j}\left( m_{i},m_{j}\right) \frac{1}{2}\sum_{k=i,j}y_{k}\left(
m_{k}^{1}\right) \oplus \left( 1-e_{i,j}\left( m_{i},m_{j}\right) \right)
x(m_{i}^{2},m_{j}^{3})\right]
\end{equation*}%
where $y_{k}\left( \theta _{k}\right) =\left( l_{k}\left( \theta _{k}\right)
,t_{1}\left( \theta _{k}\right) ,...,t_{I}\left( \theta _{k}\right) \right) $
is the dictator lottery for agent $k$ with type $\theta _{k}$ defined in
Lemma \ref{infic0} and we define 
\begin{equation*}
e_{i,j}\left( m_{i},m_{j}\right) \equiv \min \left\{ \max \left\{ \tilde{d}%
\left( m_{i}^{2},m_{j}^{2}\right) ,d\left( m_{i}^{2},m_{j}^{2}\right) ,\rho
(m_{j}^{3},f(m_{i}^{2}))^{3}\right\} ,1\right\} \text{,}
\end{equation*}%
where\footnote{%
In comparison with the $e_{i,j}\left( \cdot \right) $ defined in the proof
of Theorem \ref{Nash}, here the terms of $\tilde{d}$ and $d$ correspond to
the consistency check and the term $\rho $ corresponds to the no challenge
check.}%
\begin{equation}
\tilde{d}\left( m_{i}^{2},m_{j}^{2}\right) =\inf_{\theta \in \Theta }d\left(
m_{i}^{2},\theta \right) +\inf_{\theta \in \Theta }d\left( m_{j}^{2},\theta
\right) \text{.}  \label{h2}
\end{equation}%
For each message profile $\left( m_{i},m_{j}\right) $ of agents $i$ and $j$,
let 
\begin{equation*}
C_{i,j}(m_{i},m_{j})\equiv e_{i,j}\left( m_{i},m_{j}\right) \frac{1}{2}%
\sum_{k=i,j}y_{k}\left( m_{k}^{1}\right) \oplus \left( 1-e_{i,j}\left(
m_{i},m_{j}\right) \right) x(m_{i}^{2},m_{j}^{3})\text{.}
\end{equation*}%
In sum, with probability $\frac{1}{I(I-1)}$ an ordered pair $\left(
i,j\right) $ is chosen, then $C_{i,j}(m_{i},m_{j})$ is implemented.

\begin{claim}
\label{gcontinuous} The outcome function $g$ is continuous.
\end{claim}

\begin{proof}
It follows from Claim \ref{Bcontinuous} and Lemma \ref{infic0}.
\end{proof}

\paragraph{Transfer Rule\label{mmm}}

We now define the transfer rule. For every message profile $m\in M$ and
agent $i\in \mathcal{I}$, we specify the transfer to agent $i$ as 
\begin{equation*}
\tau _{i}(m)=\sum_{j\neq i}\left[ \tau _{i,j}^{1}(m)+\tau _{i,j}^{2}(m)%
\right] ,
\end{equation*}%
where $\tau _{i,j}^{1}$ and $\tau _{i,j}^{2}$ will be defined as follows:
Given a message profile $m$ and agent $j$, let $\tilde{m}_{i}^{m}=\left(
m_{i}^{1},\left( m_{j}^{1},m_{i,-j}^{2}\right) ,m_{i}^{3}\right) $ (which
replaces $m_{i,j}^{2}$ in $m_{i}$ by $m_{j}^{1}$), $\hat{m}_{i}^{m}=\left(
m_{i}^{1},\left( m_{j,j}^{2},m_{i,-j}^{2}\right) ,m_{i}^{3}\right) $ (which
replaces $m_{i,j}^{2}$ in $m_{i}$ by $m_{j,j}^{2}$), and $\bar{m}%
_{i}^{m}=\left( m_{i}^{1},\left( m_{j,i}^{2},m_{i,-i}^{2}\right)
,m_{i}^{3}\right) $ (which replaces $m_{i,i}^{2}$ by $m_{j,i}^{2}$). We
define $\tau _{i,j}^{1}$ as follows:

\begin{eqnarray}
\tau _{i,j}^{1}\left( m\right) &=&-\sup_{\theta _{i}^{\prime }}\left\vert
u_{i}(g\left( m\right) ,\theta _{i}^{\prime })-u_{i}(g\left( \tilde{m}%
_{i}^{m},m_{-i}\right) ,\theta _{i}^{\prime })\right\vert  \notag \\
&&+\sup_{\theta _{i}^{\prime }}\left\vert u_{i}(g\left( \hat{m}%
_{i}^{m},m_{-i}\right) ,\theta _{i}^{\prime })-u_{i}(g\left( \tilde{m}%
_{i}^{m},m_{-i}\right) ,\theta _{i}^{\prime })\right\vert  \label{infimoney1}
\\
&&+d_{j}\left( m_{j,j}^{2},m_{j}^{1}\right) -d_{j}(m_{i,j}^{2},m_{j}^{1})%
\text{.}  \notag
\end{eqnarray}%
Observe that $\tau _{i,j}^{1}$ satisfies two important properties:

\begin{enumerate}
\item neither $u_{i}(g\left( \hat{m}_{i}^{m},m_{-i}\right) ,\theta
_{i}^{\prime })-u_{i}(g\left( \tilde{m}_{i}^{m},m_{-i}\right) ,\theta
_{i}^{\prime })$ nor $d_{j}\left( m_{j,j}^{2},m_{j}^{1}\right) $ depends on
agent $i$'s choice of $m_{i,j}^{2}$, since $m_{i,j}^{2}$ has been replaced
by agent $j$'s announcements in both $\hat{m}_{i}^{m}$ and $\tilde{m}%
_{i}^{m} $;

\item $\tau _{i,j}^{1}\left( m\right) =0$ if $m_{i,j}^{2}=m_{j,j}^{2}$.
\end{enumerate}

We next define $\tau _{i,j}^{2}$ as follows:

\begin{equation}
\tau _{i,j}^{2}\left( m\right) =-\sup_{\theta _{i}^{\prime }}\left\vert
u_{i}(g\left( m\right) ,\theta _{i}^{\prime })-u_{i}(g\left( \bar{m}%
_{i}^{m},m_{-i}\right) ,\theta _{i}^{\prime })\right\vert
-d_{i}(m_{i,i}^{2},m_{j,i}^{2})  \label{infmoney2}
\end{equation}

We say that a function $\alpha \left( \cdot \right) $ between two metric
spaces $S$ and $Y$, both endowed with the Borel $\sigma $-algebra, is \emph{%
analytic} if its pre-image of every open set on $Y$ is an analytic set.
Since every analytic set is universally measurable, an analytic function is
\textquotedblleft almost" a measurable function (see pp. 498-499 of \cite%
{stinchcombe-white}). We show below that the mechanism which we are about to
construct has an analytic outcome function and transfer rule. Hence,
whenever we fix a mixed-strategy Nash equilibrium which is a Borel
probability measure on $M$, we can work with the $\sigma $-completion of the
Borel $\sigma $-algebra on $M$ to make all the expected payoffs well defined.%
\footnote{%
As will be clear in the argument later, the outcome function and transfer
rule which we construct is a value function of some optimization problem.
Such value functions are not necessarily continuous when $\Theta $ is not
compact (which is the case when we apply Theorem \ref{infiNash} to prove
Theorem \ref{ord} later).}

\begin{claim}
\label{mcontinuous}The transfer rule $\tau _{i}:M\rightarrow 
\mathbb{R}
$ is an analytic function. Moreover, if the setting is compact and
continuous, then $\tau _{i}\left( \cdot \right) $ is a continuous function.
\end{claim}

\begin{proof}
It follows from Theorem 2.17 of \cite{stinchcombe-white} that $\tau
_{i}(\cdot )$ is an analytic function. Suppose that the setting is compact
and continuous. Then, by Claim \ref{gcontinuous}, $g$ is also continuous on $%
M$. Moreover, by the theorem of maximum, $\tau _{ij}^{1}\left( \cdot \right) 
$ and $\tau _{ij}^{2}\left( \cdot \right) $ are continuous on $M.$ Hence $%
\tau _{ij}^{1}\left( \cdot \right) $ and $\tau _{ij}^{2}\left( \cdot \right) 
$ are both continuous.
\end{proof}

With the claims above, the implementing mechanism $\mathcal{M}=\left(
(M_{i}),g,(\tau _{i})\right) _{i\in \mathcal{I}}$ has been defined;
moreover, when the setting is compact and continuous, $\mathcal{M}$ is a
mechanism with compact sets of message and continuous outcome function and
transfer rule. To show that implementation is achieved by the constructed
mechanism, we only emphasize arguments different from those in a finite
state space. Before we provide the main argument, we establish two lemmas
which play an important role in the proof of Theorem \ref{infiNash}.

Throughout the proof, we denote by $\theta $ the true state. First, we show
that it is strictly worse for every agent to challenge the truth.

\begin{lemma}
\label{infiworse}Let $(m_{i},m_{j})$ be a message profile of agents $i$ and $%
j$ with $x(m_{i}^{2},m_{j}^{3})\neq f(m_{i}^{2})$. Then, $%
u_{j}(C_{i,j}(m_{i},m_{j}),m_{i,j}^{2})<u_{j}(f(m_{i}^{2}),m_{i,j}^{2})$.
\end{lemma}

\begin{proof}
Since $x(m_{i}^{2},m_{j}^{3})\neq f(m_{i}^{2})$, we have $%
u_{j}(x(m_{i}^{2},m_{j}^{3}),m_{i,j}^{2})<u_{j}(f\left( m_{i}^{2}\right)
,m_{i,j}^{2})$. Moreover, by (\ref{d2}) for every $x\in f\left( \Theta
\right) \ $we have $u_{i}\left( x,\theta _{i}\right) >u_{i}\left(
y_{k}\left( \theta _{k}^{\prime }\right) ,\theta _{i}\right) $ for each type 
$\theta _{i}$ and type $\theta _{k}^{\prime }$ of agents $i$ and $k$, we
conclude that 
\begin{equation}
u_{j}(C_{i,j}(m_{i},m_{j}),m_{i,j}^{2})<u_{j}(f(m_{i}^{2}),m_{i,j}^{2})\text{%
.}  \label{infworse}
\end{equation}%
This completes the proof.
\end{proof}

Second, whenever $\mathcal{SL}_{j}(f(\tilde{\theta}),\tilde{\theta}_{i})\cap 
\mathcal{SU}_{j}(f(\tilde{\theta}),\theta _{i})\not=\varnothing $, we show
that it is strictly better for agent $j$ to challenge.

\begin{lemma}
\label{infibetter}Let $(m_{i},m_{j})$ be a message profile of agents $i$ and 
$j$ with $m_{i}^{2}=m_{j}^{2}=\tilde{\theta}$ with $\tilde{\theta}\in \Theta 
$ and $\mathcal{SL}_{j}(f(\tilde{\theta}),\tilde{\theta}_{j})\cap \mathcal{SU%
}_{j}(f(\tilde{\theta}),m_{j}^{1})\not=\varnothing $. Then, we can choose $%
m_{j}^{3}\in \bar{X}$ such that $%
u_{j}(C_{i,j}(m_{i}.m_{j}),m_{j}^{1})>u_{j}(f(m_{i}^{2}),m_{j}^{1})$.
\end{lemma}

\begin{proof}
First, we fix an arbitrary $x\in \mathcal{SL}_{j}(f(\tilde{\theta}),\tilde{%
\theta}_{j})\cap \mathcal{SU}_{j}(f(\tilde{\theta}),m_{j}^{1})$. Let 
\begin{equation*}
m_{j}^{3}=\alpha x\oplus \left( 1-\alpha \right) f(\tilde{\theta}),
\end{equation*}%
where $\alpha \in \left( 0,1\right) .$ Note that $m_{j}^{3}\in \mathcal{SL}%
_{j}(f(\tilde{\theta}),\tilde{\theta}_{j})\cap \mathcal{SU}_{j}(f(\tilde{%
\theta}),m_{j}^{1})$ for every $\alpha \in \left( 0,1\right) $. As $\alpha
\rightarrow 0,$ we have 
\begin{equation*}
\rho (m_{j}^{3},f(\tilde{\theta}))\rightarrow 0\text{ and }%
u_{j}(m_{j}^{3},m_{j}^{1})\rightarrow u_{j}(f(\tilde{\theta}),m_{j}^{1})%
\text{.}
\end{equation*}%
Observe that $\rho (m_{j}^{3},f(\tilde{\theta}))=\alpha \left\Vert x-f(%
\tilde{\theta})\right\Vert $ (recall that $\bar{X}$ is a compact subset of $%
\mathbb{R}
^{I+\left\vert A\right\vert }$). Hence, we can choose $\alpha >0\ $small
enough such that 
\begin{gather}
\rho (m_{j}^{3},f(\tilde{\theta}))<1\text{;}  \label{ep4} \\
\rho (m_{j}^{3},f(\tilde{\theta}))^{3}\left( -3\eta \right) +\alpha \rho
(m_{j}^{3},f(\tilde{\theta}))(u_{j}(x,m_{j}^{1})-u_{j}(f(\tilde{\theta}%
),m_{j}^{1}))>0\text{.}  \label{ep5}
\end{gather}

Now, we have%
\begin{eqnarray*}
&&u_{j}(C_{i,j}(m_{i}.m_{j}),m_{j}^{1})-u_{j}(f(\tilde{\theta}),m_{j}^{1}) \\
&=&e_{i,j}\left( m_{i},m_{j}\right) u_{j}(\frac{1}{2}%
\sum_{k=i,j}y_{k}(m_{k}^{1}),m_{j}^{1})+\left( 1-e_{i,j}\left(
m_{i},m_{j}\right) \right) u_{j}\left(
x(m_{i}^{2},m_{j}^{3}),m_{j}^{1}\right) -u_{j}(f(\tilde{\theta}),m_{j}^{1})
\\
&=&\rho (m_{j}^{3},f(\tilde{\theta}))^{3}u_{j}(\frac{1}{2}%
\sum_{k=i,j}y_{k}(m_{k}^{1}),m_{j}^{1})+(1-\rho (m_{j}^{3},f(\tilde{\theta}%
))^{3})u_{j}(\ell (m_{j}^{3},f(\tilde{\theta})),m_{j}^{1})-u_{j}(f(\tilde{%
\theta}),m_{j}^{1}) \\
&=&\rho (m_{j}^{3},f(\tilde{\theta}))^{3}\left[ u_{j}(\frac{1}{2}%
\sum_{k=i,j}y_{k}(m_{k}^{1}),m_{j}^{1})-u_{j}(f(\tilde{\theta}),m_{j}^{1})%
\right] \\
&&+(1-\rho (m_{j}^{3},f(\tilde{\theta}))^{3})\left[ u_{j}\left( \frac{\rho
(m_{j}^{3},f(\tilde{\theta}))}{1+\rho (m_{j}^{3},f(\tilde{\theta}))}%
m_{j}^{3}\oplus \frac{1}{1+\rho (m_{j}^{3},f(\tilde{\theta}))}f(\tilde{\theta%
}),m_{j}^{1}\right) -u_{j}(f(\tilde{\theta}),m_{j}^{1})\right] \\
&>&\rho (m_{j}^{3},f(\tilde{\theta}))^{3}\left( -3\eta \right) +(1-\rho
(m_{j}^{3},f(\tilde{\theta}))^{3})\left[ \frac{\rho (m_{j}^{3},f(\tilde{%
\theta}))}{1+\rho (m_{j}^{3},f(\tilde{\theta}))}\left( u_{j}\left(
m_{j}^{3},m_{j}^{1}\right) -u_{j}(f(\tilde{\theta}),m_{j}^{1})\right) \right]
\\
&=&\rho (m_{j}^{3},f(\tilde{\theta}))^{3}\left( -3\eta \right) +(1-\rho
(m_{j}^{3},f(\tilde{\theta}))^{3})\left[ \frac{\alpha \rho (m_{j}^{3},f(%
\tilde{\theta}))}{1+\rho (m_{j}^{3},f(\tilde{\theta}))}\left( u_{j}\left(
x,m_{j}^{1}\right) -u_{j}(f(\tilde{\theta}),m_{j}^{1})\right) \right] \\
&>&\rho (m_{j}^{3},f(\tilde{\theta}))^{3}\left( -3\eta \right) +\frac{1}{2}%
\alpha \rho (m_{j}^{3},f(\tilde{\theta}))\left( u_{j}\left(
x,m_{j}^{1}\right) -u_{j}(f(\tilde{\theta}),m_{j}^{1})\right)
\end{eqnarray*}%
where the second equality follows because $m_{i}^{2}=m_{j}^{2}=\tilde{\theta}
$ with $\tilde{\theta}\in \Theta $ and $\rho (m_{j}^{3},f(\tilde{\theta}%
))^{3}<1$ (by (\ref{ep4})); the third equality follows from the definition
of $\ell (m_{j}^{3},f(\tilde{\theta}))$; the fourth equality follows from
the linearity of $u_{j}$ in allocation; the first inequality follows
because\ 
\begin{equation*}
u_{j}\left( \frac{1}{2}\sum_{k=i,j}y_{k}(m_{k}^{1}),m_{j}^{1}\right)
-u_{j}(f(\tilde{\theta}),m_{j}^{1})>-3\eta ;
\end{equation*}%
the last inequality follows from (\ref{ep4}). Hence, it follows from (\ref%
{ep5}) that 
\begin{equation*}
u_{j}(C_{i,j}(m_{i}.m_{j}),m_{j}^{1})-u_{j}(f(\tilde{\theta}),m_{j}^{1})>0.
\end{equation*}%
This completes the proof.
\end{proof}

\subsubsection{Existence of Good Equilibrium}

Consider an arbitrary true state $\theta $. The proof consists of two parts.
In the first part, we argue that truth-telling $m$ where $m_{i}=(\theta
_{i},\theta ,x)$ for each $i\in \mathcal{I}$ constitutes a pure-strategy
equilibrium, where $x\left( \theta ,x\right) =f\left( \theta \right) $.
Under this message profile $m$, $e_{i,j}(m_{i},m_{j})=0$. Firstly, reporting 
$\tilde{m}_{i}$ with either $\tilde{m}_{i,i}^{2}\neq \theta _{i}$ or $\tilde{%
m}_{i,j}^{2}\neq \theta _{j}$ suffers the penalty of $\tau _{i,j}^{2}\left(
m\right) $ or $\tau _{i,j}^{1}\left( m\right) $ and hence cannot be a
profitable deviation by Claim \ref{infi1st}. Secondly, reporting $\tilde{m}%
_{i}$ with $\tilde{m}_{i}^{2}=\theta $ and $\tilde{m}_{i}^{3}=x^{\prime
}\not=x$ either leads to $x\left( \theta ,x^{\prime }\right) =f(\theta )$
and results in no change in payoff or $x\left( \theta ,x^{\prime }\right)
\neq f(\theta )$ which is strictly worse than $f(\theta )$. By Lemma \ref%
{infiworse}, this is not a profitable deviation. Finally, reporting $\tilde{m%
}_{i}$ with $\tilde{m}_{i}^{2}=\theta $, $\tilde{m}_{i}^{3}=\theta _{i}$,
and $\tilde{m}_{i}^{1}\neq \theta _{i}$ does not affect the allocation or
transfer, since we still have $\tau _{i}\left( m\right) =0$ and $%
e_{j,k}(m_{j},m_{k})=0$ for every $j$ and $k$.

In the second part, we show that for every Nash equilibrium $\sigma $ of the
game $\Gamma (\mathcal{M},\theta )$ and every $m\in $supp$(\sigma ),$ we
have that $g\left( m\right) =f\left( \theta \right) $ and $\tau _{i}\left(
m\right) =0$ for every $i\in \mathcal{I}$. The proof of the second part is
divided into three steps: (Step 1) \textit{contagion of truth}: if agent $j$
announces his type truthfully in his first report, then every agent must
also report agent $j$'s type truthfully in their second report; (Step 2) 
\textit{consistency}: every agent reports the same state $\tilde{\theta}$ in
the second report; and (Step 3) \textit{no challenge}: no agent challenges
the common reported state $\tilde{\theta}$, i.e., $x(\tilde{\theta}%
,m_{j}^{3})=f(\tilde{\theta})$ for every $j\in \mathcal{I}$. Then,
consistency implies that $\tau _{i}\left( m\right) =0$ for every $i\in 
\mathcal{I}$, whereas no challenge is invoked so that monotonicity of $f$
together with Lemma \ref{infibetter} implies that $g\left( m\right) =f(%
\tilde{\theta})=f\left( \theta \right) $.

As we do not make use of the notion of best challenge scheme in the infinite
setting, the proof of Claim \ref{c0} is more straightforward. Here $%
m_{i}^{1} $ only affects agent $i$'s own payoff through controlling the
dictator lottery $y_{i}$; in particular, both $e_{i,j}\left(
m_{i},m_{j}\right) $ and $e_{j,i}\left( m_{j},m_{i}\right) $ are not
determined by $m_{i}^{1}$ or $m_{j}^{1}$. Hence, if $e_{i,j}\left(
m_{i},m_{j}\right) =\varepsilon $ or $e_{j,i}\left( m_{j},m_{i}\right)
=\varepsilon $, then $m_{i}^{1}=\theta _{i}$ by (\ref{d1}). We now establish
Steps 1-3.

\subsubsection{Contagion of Truth}

\begin{claim}
\label{infi1st} We establish two results:

\noindent (a) If agent $j$ sends a truthful first report with $\sigma _{j}$%
-probability one, then every agent $i\neq j$ must report agent $j$'s type
truthfully in his second report with $\sigma _{i}$-probability one.

\noindent (b) If every agent $i\neq j$ reports a type $\tilde{\theta}_{j}$
of agent $j$ in his second report with $\sigma _{i}$-probability one, then
agent $j$ must also report $\tilde{\theta}_{j}$ in his second report with $%
\sigma _{j}$-probability one.
\end{claim}

\begin{proof}
First, we prove part (a). Suppose instead that there exists some message $%
m_{i}$ played with $\sigma _{i}$-positive probability such that agent $i$
misreports agent $j$'s type in his second report, i.e., $m_{i,j}^{2}\not=%
\theta _{j}.$ Let $\tilde{m}_{i}$ be a message which differs from $m_{i}$
only in reporting $j$'s type truthfully $\tilde{m}_{i,j}^{2}=\theta _{j}$.
Then, for each $m_{-i}$ played with $\sigma _{-i}$-positive probability,
since agent $j$ reports his type truthfully ($m_{j}^{1}=\theta _{j})$, we
have $\tilde{m}_{i}=\tilde{m}_{i}^{m}$ where $m=\left( m_{i}.m_{-i}\right) $%
. Recall the definition of $\tilde{m}_{i}^{m}$ in Section \ref{mmm}. Hence,
we reach a contradiction if we show $\tilde{m}_{i}$ is a profitable
deviation, i.e., for each $m_{-i}$ played with $\sigma _{-i}$-positive
probability, we have 
\begin{equation*}
u_{i}\left( g\left( \tilde{m}_{i}^{m},m_{-i}\right) ,\theta _{i}\right)
+\tau _{i}\left( \tilde{m}_{i}^{m},m_{-i}\right) >u_{i}\left( g\left(
m\right) ,\theta _{i}\right) +\tau _{i}\left( m\right) \text{.}
\end{equation*}%
First, observe that 
\begin{eqnarray*}
\tau _{i}\left( \tilde{m}_{i}^{m},m_{-i}\right) -\tau _{i}\left( m\right)
&=&\tau _{i,j}^{1}\left( \tilde{m}_{i}^{m},m_{-i}\right) -\tau
_{i,j}^{1}\left( m\right) \\
&=&\sup_{\theta _{i}^{\prime }}\left\vert u_{i}\left( g\left( m\right)
,\theta _{i}^{\prime }\right) -u_{i}\left( g\left( \tilde{m}%
_{i}^{m},m_{-i}\right) ,\theta _{i}^{\prime }\right) \right\vert
+d_{j}\left( m_{i,j}^{2},m_{j}^{1}\right)
\end{eqnarray*}%
where the first equality follows because $\tilde{m}_{i}^{m}$ only differs
from $m_{i}$ in agent $i$'s second report of agent $j$'s type. Thus we have 
\begin{eqnarray*}
&&\left[ u_{i}\left( g\left( \tilde{m}_{i}^{m},m_{-i}\right) ,\theta
_{i}\right) +\tau _{i}\left( \tilde{m}_{i}^{m},m_{-i}\right) \right] -\left[
u_{i}\left( g\left( m\right) ,\theta _{i}\right) +\tau _{i}\left( m\right) %
\right] \\
&=&u_{i}\left( g\left( \tilde{m}_{i}^{m},m_{-i}\right) ,\theta _{i}\right)
-u_{i}\left( g\left( m\right) ,\theta _{i}\right) \\
&&+\sup_{\theta _{i}^{\prime }}\left\vert u_{i}\left( g\left( m\right)
,\theta _{i}^{\prime }\right) -u_{i}\left( g\left( \tilde{m}%
_{i}^{m},m_{-i}\right) ,\theta _{i}^{\prime }\right) \right\vert
+d_{j}\left( m_{i,j}^{2},m_{j}^{1}\right) \\
&>&0.
\end{eqnarray*}%
where the last inequality follows because $m_{i,j}^{2}\not=\theta
_{j}=m_{j}^{1}$.

Second, we prove part (b). Suppose, on the contrary, that there exists some
message $m_{j}$ played with $\sigma _{j}$-positive probability such that
agent $j$ misreports agent $i$'s type in his second report, i.e., $%
m_{j,j}^{2}\neq \tilde{\theta}_{j}$. Let $\bar{m}_{j}$ be a message that is
identical to $m_{j}$ except that $\bar{m}_{j,j}^{2}=\tilde{\theta}_{j}$.
Then, for each $m_{-j}$ played with $\sigma _{-j}$-positive probability,
since every agent $i\neq j$ reports $\tilde{\theta}_{j}$ in agent $i$'s
second report, we have $\bar{m}_{j}=\bar{m}_{j}^{m}$ where $m=\left(
m_{j},m_{-j}\right) $. Recall the definition of $\bar{m}_{i}^{m}$ in Section %
\ref{mmm}. Hence, we reach a contradiction if we show that $\bar{m}_{j}$ is
a profitable deviation, i.e., for each $m_{-j}$ played with $\sigma _{-j}$%
-positive probability, we have 
\begin{equation*}
u_{j}\left( g\left( \bar{m}_{j}^{m},m_{-j}\right) ,\theta _{j}\right) +\tau
_{j}\left( \bar{m}_{j}^{m},m_{-j}\right) >u_{j}\left( g\left( m\right)
,\theta _{j}\right) +\tau _{j}\left( m\right) .
\end{equation*}%
Notice that $\tau _{j,i}^{1}(\bar{m}_{j}^{m},m_{-j})=\tau _{j,i}^{1}(m)$. 
\begin{eqnarray*}
\tau _{j}\left( \bar{m}_{j}^{m},m_{-j}\right) -\tau _{j}\left( m\right)
&=&\sum_{i\not=j}\left\{ \tau _{j,i}^{2}\left( \bar{m}_{j}^{m},m_{-j}\right)
-\tau _{j,i}^{2}\left( m\right) \right\} \\
&=&\sum_{i\not=j}\left\{ 
\begin{array}{c}
\sup_{\theta _{j}^{\prime }}\left\vert u_{j}\left( g\left( m\right) ,\theta
_{j}^{\prime }\right) -u_{j}\left( g\left( \bar{m}_{j}^{m},m_{-j}\right)
,\theta _{j}^{\prime }\right) \right\vert \\ 
+d_{j}\left( m_{j,j}^{2},m_{i,j}^{2}\right)%
\end{array}%
\right\} \text{.}
\end{eqnarray*}%
Thus, we have 
\begin{eqnarray*}
&&\left[ u_{j}\left( g\left( \bar{m}_{j}^{m},m_{-j}\right) ,\theta
_{j}\right) +\tau _{j}\left( \bar{m}_{j}^{m},m_{-j}\right) \right] -\left[
u_{j}\left( g\left( m\right) ,\theta _{j}\right) +\tau _{j}\left( m\right) %
\right] \\
&=&u_{j}\left( g\left( \bar{m}_{j}^{m},m_{-j}\right) ,\theta _{j}\right)
-u_{j}\left( g\left( m\right) ,\theta _{j}\right) \\
&&+\sum_{i\not=j}\left\{ 
\begin{array}{c}
\sup_{\theta _{j}^{\prime }}\left\vert u_{j}\left( g\left( m\right) ,\theta
_{j}^{\prime }\right) -u_{j}\left( g\left( \bar{m}_{j}^{m},m_{-j}\right)
,\theta _{j}^{\prime }\right) \right\vert \\ 
+d_{j}\left( m_{j,j}^{2},m_{i,j}^{2}\right)%
\end{array}%
\right\} \\
&>&0.
\end{eqnarray*}%
where the last inequality follows because $m_{j,j}^{2}\neq \tilde{\theta}%
_{j}=m_{i,j}^{2}$. This completes the proof.
\end{proof}

\subsubsection{Consistency}

The argument for consistency follows verbatim the proof of Claim \ref{pc1}
in the proof of Theorem \ref{Nash}.

\subsubsection{No Challenge}

\begin{claim}
\label{infinoch}No agent challenges with positive probability the common
state $\tilde{\theta}$ announced in the second report.
\end{claim}

\begin{proof}
Provided that a consistent report of $\tilde{\theta}$ is achieved, it
suffices to show that $\mathcal{SL}_{j}(f(\tilde{\theta}),\tilde{\theta}%
_{j})\cap \mathcal{S}\mathcal{U}_{j}(f(\tilde{\theta}),\theta
_{j})=\varnothing $ for every agent $j\in \mathcal{I}$. Suppose to the
contrary that $\mathcal{SL}_{j}(f(\tilde{\theta}),\tilde{\theta}_{j})\cap 
\mathcal{S}\mathcal{U}_{j}(f(\tilde{\theta}),\theta _{j})\neq \varnothing $
for some agent $j$. Then, we first show that $x(\tilde{\theta}%
,m_{j}^{3})\neq f(\tilde{\theta})$ for every $m_{j}\in $supp$\left( \sigma
_{j}\right) $. By Lemma \ref{infibetter}, there exists $x\in \mathcal{SL}%
_{j}(f(\tilde{\theta}),\tilde{\theta}_{j})\cap \mathcal{S}\mathcal{U}_{j}(f(%
\tilde{\theta}),\theta _{j}).$ If $x(\tilde{\theta},m_{j}^{3})=f(\tilde{%
\theta}),$ then $\tilde{m}_{j}=\left( m_{j}^{1},m_{j}^{2},x\right) $ is a
strictly profitable deviation from announcing $m_{j}$. This deviation
results in a better allocation $x(\tilde{\theta},x)\in \mathcal{S}\mathcal{U}%
_{j}(f(\tilde{\theta}),\theta _{j})$. Hence, we have $x(\tilde{\theta}%
,m_{j}^{3})\neq f(\tilde{\theta})$ for every $m_{j}\in $supp$\left( \sigma
_{j}\right) $. It follows that the dictator lottery is triggered with
positive probability. Thus, by (\ref{d1}), each agent $i$ has a strict
incentive to announce the true type in his first report (i.e., $%
m_{i}^{1}=\theta _{i}$) with $\sigma _{i}$-probability one. By Claim \ref%
{infi1st}, we conclude that $\tilde{\theta}=\theta $ and hence $\mathcal{SL}%
_{j}(f(\theta ),\theta _{j})\cap \mathcal{S}\mathcal{U}_{j}(f(\theta
),\theta _{j})\neq \varnothing $, which is impossible.
\end{proof}

\subsection{Proof of Theorem \protect\ref{ord}}

\label{aord}

First, we obtain a stronger version of Lemma \ref{AM}, namely, there is a
set of dictator lotteries which can used to elicit the agents' true types,
regardless of their cardinal representation. This follows from the same
proof of Lemma in \cite{AM92}; moreover, the dictator lottery constructed
remains valid (in the sense of (\ref{am-ord})) as long as the preferences
exhibit monotonicity with respect to first-order stochastic dominance.
Recall that for each $x=\left( \ell ,\left( t_{i}\right) _{i\in \mathcal{I}%
}\right) \in X$, $u_{i}\left( x,\theta _{i}\right) \equiv v_{i}(\ell ,\theta
)+t_{i}$.

\begin{lemma}
\label{ordinal-dictator}For each $i\in \mathcal{I}$, there exists a function 
$y_{i}:\Theta _{i}\rightarrow X$ such that for every $\theta _{i},\theta
_{i}^{\prime }\in \Theta _{i}$ with $\theta _{i}\neq \theta _{i}^{\prime }$
and every cardinal representation $v_{i}(\cdot )$ of $(\succeq _{i}^{\theta
})_{\theta \in \Theta }$, 
\begin{equation}
u_{i}(y_{i}(\theta _{i}),\theta _{i})>u_{i}(y_{i}(\theta _{i}^{\prime
}),\theta _{i})\text{;}  \label{am-ord}
\end{equation}%
moreover, for each type $\theta _{j}^{\prime }$ of agent $j\in \mathcal{I}$,
we also have that for every $x\in f\left( \Theta \right) $%
\begin{equation}
u_{i}(y_{j}(\theta _{j}^{\prime }),\theta _{i})<u_{i}(x,\theta _{i})\text{.}
\label{ord2}
\end{equation}
\end{lemma}

To satisfy condition (\ref{ord2}), we simply add a penalty of $\eta $ to
each outcome of the lotteries $\left\{ y_{i}^{\prime }\left( \theta
_{i}\right) \right\} $ where $\eta $ is chosen in the same fashion as in (%
\ref{D'}) given each cardinal representation $v_{i}(\cdot )$. Moreover,
since each $v_{i}(\cdot )$ takes values in $\left[ 0,1\right] $, we also
save the notation and take $\eta $ to be independent of $v_{i}(\cdot )$.

We first introduce the following definitions of contour sets under ordinal
preferences. For $\left( a,\theta _{i}\right) \in A\times \Theta _{i}$,
under ordinal preference $\succeq _{i}^{\theta }$, we write the
upper-contour set, the lower-contour set, the strict upper-contour set, and
the strict lower-contour set as follows:%
\begin{eqnarray*}
U_{i}\left( a,\theta _{i}\right) &=&\left\{ a^{\prime }\in A:a^{\prime
}\succeq _{i}^{\theta }a\right\} \text{;} \\
L_{i}\left( a,\theta _{i}\right) &=&\left\{ a^{\prime }\in A:a\succeq
_{i}^{\theta }a^{\prime }\right\} \text{;} \\
SU_{i}\left( a,\theta _{i}\right) &=&\left\{ a^{\prime }\in A:a^{\prime
}\succ _{i}^{\theta }a\right\} \text{;} \\
SL_{i}\left( a,\theta _{i}\right) &=&\left\{ a^{\prime }\in A:a\succ
_{i}^{\theta }a^{\prime }\right\} \text{;}
\end{eqnarray*}%
where $\succ _{i}^{\theta }$ denotes the strict preference induced by $%
\succeq _{i}^{\theta }$. We now introduce the notion of ordinal almost
monotonicity proposed by \cite{sanver06} as the key condition for Theorem %
\ref{ord}.

\begin{definition}
An SCF $f$ satisfies \textbf{ordinal} \textbf{almost monotonicity} if, for
every pair of states $\theta $ and $\tilde{\theta}$, with $f(\tilde{\theta}%
)\not=f\left( \theta \right) $, there is some agent $i\in \mathcal{I}$ such
that either 
\begin{equation*}
L_{i}(f(\tilde{\theta}),\tilde{\theta}_{i})\cap SU_{i}(f(\tilde{\theta}%
),\theta _{i})\not=\varnothing \text{,}
\end{equation*}%
or 
\begin{equation*}
SL_{i}(f(\tilde{\theta}),\tilde{\theta}_{i})\cap U_{i}(f(\tilde{\theta}%
),\theta _{i})\not=\varnothing \text{.}
\end{equation*}
\end{definition}

\subsubsection{Proof of the Only-If Part}

In the proof, we make use of Claim C from \cite{MR12} which is reproduced as
follows:

\begin{claim}
\label{C}Suppose that $L_{i}\left( f\left( \theta \right) ,\theta
_{i}\right) \subseteq L_{i}(f\left( \theta \right) ,\tilde{\theta}_{i})$ and 
$SL_{i}\left( f\left( \theta \right) ,\theta _{i}\right) \subseteq
SL_{i}(f\left( \theta \right) ,\tilde{\theta}_{i}).$ Then, given every
cardinal representation $v_{i}\left( \cdot ,\theta \right) $ of $%
\succcurlyeq _{i}^{\theta },$ there exists a cardinal representation $%
v_{i}(\cdot ,\tilde{\theta})$ of $\succcurlyeq _{i}^{\tilde{\theta}}$such
that $v_{i}(a,\tilde{\theta})\leq $ $v_{i}\left( a,\theta \right) $ for all $%
a\in A$ and $v_{i}(f\left( \theta \right) ,\tilde{\theta})=$ $v_{i}\left(
f\left( \theta \right) ,\theta \right) $
\end{claim}

Suppose $f$ is ordinally Nash implementable in a mechanism $\mathcal{M}%
=\left( (M_{i}),g,(\tau _{i})\right) _{i\in \mathcal{I}}$ but not almost
monotonic. That is, there exists $\theta ,\tilde{\theta}\in \Theta $ such
that for each agent $i\in \mathcal{I}$, we have $L_{i}\left( f\left( \theta
\right) ,\theta _{i}\right) \subseteq L_{i}(f\left( \theta \right) ,\tilde{%
\theta}_{i})$ and $SL_{i}\left( f\left( \theta \right) ,\theta _{i}\right)
\subseteq SL_{i}(f\left( \theta \right) ,\tilde{\theta}_{i})$, but $f\left(
\theta \right) \not=f(\tilde{\theta}).$ Since $\mathcal{M}$ ordinally Nash
implements $f$, we know that for every cardinal representation $\left(
v_{i}\right) _{i\in \mathcal{I}}$, there exists a pure-strategy Nash
equilibrium $m^{\ast }$ in the game $\Gamma (\mathcal{M},\theta ,v)$\ such
that $g\left( m^{\ast }\right) =f\left( \theta \right) $. Since $f\left(
\theta \right) \not=f(\tilde{\theta})$, the message profile $m^{\ast }$
cannot be a Nash equilibrium at state $\tilde{\theta}$ for every cardinal
representation $v_{i}$. Then, there exist agent $i$ and message $m_{i}\in
M_{i}$ such that 
\begin{eqnarray*}
v_{i}\left( g\left( m_{i}^{\ast },m_{-i}^{\ast }\right) ,\theta \right)
+\tau _{i}\left( m_{i}^{\ast },m_{-i}^{\ast }\right) &\geq &v_{i}\left(
g\left( m_{i},m_{-i}^{\ast }\right) ,\theta \right) +\tau _{i}\left(
m_{i},m_{-i}^{\ast }\right) \text{;} \\
v_{i}(g\left( m_{i}^{\ast },m_{-i}^{\ast }\right) ,\tilde{\theta})+\tau
_{i}\left( m_{i}^{\ast },m_{-i}^{\ast }\right) &<&v_{i}(g\left(
m_{i},m_{-i}^{\ast }\right) ,\tilde{\theta})+\tau _{i}\left(
m_{i},m_{-i}^{\ast }\right) \text{.}
\end{eqnarray*}

Summing up the two inequalities, we obtain that for every cardinal
representation $v_{i}$ 
\begin{equation}
v_{i}\left( g\left( m_{i}^{\ast },m_{-i}^{\ast }\right) ,\theta \right)
-v_{i}(g\left( m_{i}^{\ast },m_{-i}^{\ast }\right) ,\tilde{\theta}%
)>v_{i}\left( g\left( m_{i},m_{-i}^{\ast }\right) ,\theta \right)
-v_{i}(g\left( m_{i},m_{-i}^{\ast }\right) ,\tilde{\theta})\text{.}
\label{card}
\end{equation}%
Note that $g\left( m_{i}^{\ast },m_{-i}^{\ast }\right) =f\left( \theta
\right) $. By Claim \ref{C}, however, we can construct a cardinal utility
representation $v_{i}(\cdot ,\tilde{\theta})$ of $\succcurlyeq _{i}^{\tilde{%
\theta}}$ such that $v_{i}(a,\tilde{\theta})\leq $ $v_{i}\left( a,\theta
\right) $ for all $a\in A$ and $v_{i}(f\left( \theta \right) ,\tilde{\theta}%
)=$ $v_{i}\left( f\left( \theta \right) ,\theta \right) .$ Therefore, the
left-hand side of (\ref{card}) is zero, while the right-hand side is
non-negative. This is a contradiction.

\subsubsection{Proof of the If Part}

Let $f$ be an SCF which is ordinally almost monotonic on $\Theta $. Recall
that $V_{i}^{\theta }$ denotes the set of all cardinal representations $%
v_{i}\left( \cdot ,\theta \right) $ of $\succeq _{i}^{\theta }$. Define $%
V^{\theta }=\times _{i\in \mathcal{I}}V_{i}^{\theta }$ with a generic
element $v^{\theta }$. By no redundancy of $\Theta $, $\theta \neq \theta
^{\prime }$ implies that $\succeq _{i}^{\theta }\neq \succeq _{i}^{\theta
^{\prime }}$ for some agent $i\in \mathcal{I}$. Hence, $\{V^{\theta }:\theta
\in \Theta \}$ forms a partition of $\Theta ^{\ast }\equiv \bigcup_{\theta
\in \Theta }V^{\theta }$ which is the set of all cardinal utility profiles
of agent $i$ induced by $\Theta $. Observe that $\Theta ^{\ast }$ is a
Polish space.\footnote{%
Since any product or disjoint union of a countable family of Polish spaces
remains a Polish space (see Proposition A.1(b) in p. 550 of \cite{treves}),
it suffices to argue that $V_{i}^{\theta }$ is a Polish space. Let $V=\left[
0,1\right] ^{\left\vert A\right\vert }$ be the set of possible
cardinalizations. We may write $V_{i}^{\theta }=\bigcap_{a\in
A}V_{i,a}^{\theta }$ where for each $a\in A$, we set 
\begin{equation*}
V_{i,a}^{\theta }\equiv \bigcap_{\left\{ b\in A:a\succ _{i}^{\theta
}b\right\} }\left\{ v\in V:v\left( a\right) >v\left( b\right) \right\}
\bigcap \bigcap_{\left\{ b\in A:a\sim _{i}^{\theta }b\right\} }\left\{ v\in
V:v\left( a\right) =v\left( b\right) \right\} \text{.}
\end{equation*}%
It follows that $V_{i}^{\theta }$ is a finite intersection of open subsets
and closed subsets of the Polish space $V$ and hence remains a Polish space
(see Proposition A.1(a)(c)(e) in p. 550 of \cite{treves}).} For the sake of
notational simplicity, we write $\theta _{i}^{\ast }$ as a generic element
in $\Theta _{i}^{\ast }$ and $\theta ^{\ast }=\left( \theta _{i}^{\ast
}\right) _{i\in \mathcal{I}}.$ Let $f^{\ast }:\Theta ^{\ast }\rightarrow A$
be the SCF on $\Theta ^{\ast }\,$induced by $f:\Theta \rightarrow A$ such
that $f^{\ast }\left( \theta ^{\ast }\right) =f\left( \theta \right) $ if
and only if $\theta ^{\ast }\in V^{\theta }$.

We prove the if-part by establishing two claims: First, we show that $%
f^{\ast }$ is Maskin-monotonic in Claim \ref{c-m}. Hence, Theorem \ref%
{infiNash} implies that $f^{\ast }$ is implementable in Nash equilibria on $%
\Theta ^{\ast }$. Second, it follows from Claim \ref{c-i} that $f$ is
ordinally Nash implementable on $\Theta $.

\begin{claim}
\label{c-m}If $f$ is ordinally almost monotonic on $\Theta $, then $f^{\ast
} $ is strictly Maskin-monotonic on $\Theta ^{\ast }$.
\end{claim}

\begin{proof}
Let $\theta ^{\ast }$ and $\tilde{\theta}^{\ast }$ be in $\Theta ^{\ast }$
such that $f^{\ast }\left( \theta ^{\ast }\right) \not=f^{\ast }(\tilde{%
\theta}^{\ast }).$ Since $f^{\ast }\left( \theta ^{\ast }\right) =f\left(
\theta \right) $ if and only if $\theta ^{\ast }\in V^{\theta }$, we must
have two states $\theta $ and $\tilde{\theta}$ $\in \Theta $ such that $%
\theta ^{\ast }\in V^{\theta }$ and $\tilde{\theta}^{\ast }\in V^{\tilde{%
\theta}}$, and $f\left( \theta \right) \not=f(\tilde{\theta})$. Since $f$
satisfies ordinal almost monotonicity, there exists agent $i\in \mathcal{I}$
and outcomes $a$, $a^{\prime }\in A$ such that either $a\in L_{i}(f(\tilde{%
\theta}),\tilde{\theta}_{i})\cap SU_{i}(f(\tilde{\theta}),\theta _{i})$ or $%
a^{\prime }\in SL_{i}(f(\tilde{\theta}),\tilde{\theta}_{i})\cap U_{i}(f(%
\tilde{\theta}),\theta _{i}).$ Then, either choose $t_{i}<0$ such that $%
\left( a,\left( t_{i},\mathbf{0}\right) \right) $ $\in \mathcal{SL}%
_{i}(f^{\ast }(\tilde{\theta}^{\ast }),\tilde{\theta}_{i}^{\ast })\cap 
\mathcal{SU}_{i}(f^{\ast }(\tilde{\theta}^{\ast }),\theta _{i}^{\ast })$ or $%
t_{i}^{\prime }>0$ such that $\left( a^{\prime },\left( t_{i}^{\prime },%
\mathbf{0}\right) \right) \in \mathcal{SL}_{i}(f^{\ast }(\tilde{\theta}%
^{\ast }),\tilde{\theta}_{i}^{\ast })\cap \mathcal{SU}_{i}(f^{\ast }(\tilde{%
\theta}^{\ast }),\theta _{i}^{\ast })$ where $\mathbf{0\in \mathbb{R}}^{I-1}$
means that every agent $j\not=i$ incurs no transfer. Hence, $f^{\ast }$ is
strictly Maskin-monotonic on $\Theta ^{\ast }$.
\end{proof}

\begin{claim}
\label{c-i}If $f^{\ast }$ is implementable in Nash equilibria, then $f$ is 
\textit{ordinally} Nash implementable.
\end{claim}

\begin{proof}
Suppose the SCF $f^{\ast }$ is implementable in Nash equilibria on $\Theta
^{\ast }$.\textbf{\ }Then, there exists a mechanism $\mathcal{M}=\left(
(M_{i}),g,(\tau _{i})\right) _{i\in \mathcal{I}}$ such that for every state $%
\theta ^{\ast }\in \Theta ^{\ast }$ and message profile $m\in M$, (i) there
exists a pure-strategy Nash equilibrium in the game $\Gamma (\mathcal{M}%
,\theta ^{\ast })$; and (ii) $m\in \mbox{supp}\ (NE(\Gamma (\mathcal{M}%
,\theta ^{\ast })))\Rightarrow g\left( m\right) =f^{\ast }\left( \theta
^{\ast }\right) $ and $\tau _{i}\left( m\right) =0$ for every $i\in \mathcal{%
I}$. Thus, for every state $\theta ^{\ast }\in V^{\theta }$, we must have
(i) there exists a pure-strategy Nash equilibrium in the game $\Gamma (%
\mathcal{M},\theta ,v^{\theta ^{\ast }});$ and (ii) $m\in \mbox{supp}\
(NE(\Gamma (\mathcal{M},\theta ,v^{\theta ^{\ast }})))\Rightarrow g\left(
m\right) =f\left( \theta \right) $ and $\tau _{i}\left( m\right) =0$ for
every $i\in \mathcal{I}$. Hence, $f$ is ordinally Nash implementable on $%
\Theta .$
\end{proof}

\newpage 
\bibliographystyle{econometrica}
\bibliography{mam}

\end{document}